\documentclass[acmlarge]{acmart}

\AtBeginDocument{
  \providecommand\BibTeX{{\normalfont B\kern-0.5em{\scshape i\kern-0.25em b}\kern-0.8em\TeX}}
}

\setcopyright{acmcopyright}
\copyrightyear{2025}
\acmYear{2025}
\acmDOI{}

\acmJournal{CSUR}
\acmVolume{1}
\acmNumber{01}
\acmArticle{01}
\acmMonth{01}

\usepackage{subcaption}
\usepackage{amsmath}
\PassOptionsToPackage{usenames,dvipsnames}{xcolor}

\usepackage{tikz}
\usetikzlibrary{arrows, arrows.meta}

\newcommand{\circlefill}[1]{
  \begin{tikzpicture}
    \draw (0,0) circle (4pt);
    \ifnum#1=25
      \fill[black] (0,0) -- (90:4pt) arc (90:180:4pt) -- cycle;
    \fi
    \ifnum#1=50
      \fill[black] (0,0) -- (0:4pt) arc (0:180:4pt) -- cycle;
    \fi
    \ifnum#1=75
      \fill[black] (0,0) -- (-90:4pt) arc (-90:180:4pt) -- cycle;
    \fi
    \ifnum#1=100
      \fill[black] (0,0) circle (4pt);
    \fi
  \end{tikzpicture}
}

\usepackage{algorithmic}
\usepackage{longtable}
\usepackage{array}
\usepackage{paralist}
\usepackage{enumitem}
\usepackage{pdflscape}
\usepackage{acro}
\usepackage{rotating}
\usepackage{colortbl}
\usepackage{ulem}
\usepackage{ragged2e}
\usepackage{graphicx}
\usepackage{multirow}
\usepackage{booktabs}
\usepackage{hhline}
\usepackage{balance}
\usepackage{subcaption}

\providecommand{\Description}[1]{}

\begin{document}

\title[
{Energy-Efficient Resource Management in Microservices-Based Fog and Edge Computing: State-of-the-Art and Future Directions}]
{Energy-Efficient Resource Management in Microservices-Based Fog and Edge Computing: State-of-the-Art and Future Directions}

\author{Ali~Akbar~Vali}
\affiliation{%
  \institution{Department of Computer Engineering and IT, University of Kurdistan}
  \city{Sanandaj}
  \country{Iran}}
\email{akbar.vali@uok.ac.ir}

\author{Sadoon~Azizi}
\authornote{This is the corresponding author}
\affiliation{%
  \institution{Department of Computer Engineering and IT, University of Kurdistan}
  \city{Sanandaj}
  \country{Iran}}
\email{s.azizi@uok.ac.ir}

\author{Mohammad Shojafar}
\affiliation{%
  \institution{5G/6GIC, Institute for Communication Systems, University of Surrey}
  \city{Guildford}
  \country{UK}}
\email{m.shojafar@surrey.ac.uk}

\author{Rajkumar Buyya}
\affiliation{%
  \institution{Quantum Cloud Computing and Distributed Systems (qCLOUDS) Laboratory, School of Computing and Information Systems, The University of Melbourne}
  \city{Melbourne}
  \country{Australia}}
\email{rbuyya@unimelb.edu.au}

\renewcommand{\shortauthors}{Vali, Azizi, Shojafar, and Buyya}

\begin{abstract}
The exponential growth of Internet of Things (IoT) devices has intensified the demand for efficient and responsive services. To address this demand, fog and edge computing have emerged as distributed paradigms that bring computational resources closer to end users, reducing latency, bandwidth limitations, and energy consumption. However, these paradigms present challenges in resource management due to resource constraints, computational heterogeneity, dynamic workloads, and diverse Quality of Service (QoS) requirements. This paper presents a comprehensive survey of state-of-the-art resource management strategies in microservices-based fog and edge computing, focusing on energy-efficient solutions. We systematically review and classify over 136 studies (2020–2024) into five key subdomains: service placement, resource provisioning, task scheduling and offloading, resource allocation, and instance selection. Our categorization is based on optimization techniques, targeted objectives, and the strengths and limitations of each approach. Additionally, we examine existing surveys, identifying unresolved challenges and gaps in the literature. By highlighting the lack of synergy among fundamental resource management components, we outline promising research directions leveraging AI-driven optimization, quantum computing, and serverless computing. This survey serves as a comprehensive reference for researchers and practitioners, providing a unified, energy-aware perspective on resource management in microservices-based fog and edge computing, and paving the way for more integrated, efficient, and sustainable future solutions.
\end{abstract}

\ccsdesc[500]{General and reference~Surveys and overviews}

\keywords{Fog Computing, Edge Computing, Internet of Things, Resource Management, Energy Efficiency, Microservices Architecture, Optimization Techniques}

\maketitle
\section{Introduction}
\label{sec:Introduction}
The rapid expansion of devices on the IoT has fundamentally transformed computational paradigms, driving an increasing demand for efficient, low-latency services that are essential for real-time applications. To address these demands, fog and edge computing have emerged as distributed architectures that process data closer to its source, thereby reducing latency, alleviating bandwidth constraints, and mitigating the energy inefficiencies inherent in centralized cloud computing. Fog computing extends cloud capabilities to intermediate nodes between devices and the cloud, while edge computing processes data directly at or near the source (e.g., IoT devices). Mobile Edge Computing (MEC) specifically leverages mobile network infrastructure for edge processing. This survey uses ‘fog’ and ‘edge’ as broad paradigms, with MEC as a specialized subset. However, the decentralized nature of these environments introduces critical resource management challenges, including resource constraints, computational heterogeneity, and dynamic workload variations. Addressing these challenges requires a systematic and adaptive approach to resource management. This survey offers a comprehensive analysis of existing strategies in microservices-based fog and edge computing environments, with a particular focus on energy-efficient solutions that enhance scalability, reliability, and sustainability. By classifying and evaluating more than 136 studies, this work serves as a structured reference for researchers and practitioners aiming to optimize resource management in these distributed computing systems~\cite{premsankar2022energy}.\\
Fog and edge computing have emerged as decentralized paradigms that extend computational capabilities toward data sources, thereby enabling low-latency, bandwidth-efficient, and energy-aware processing for time-critical applications such as autonomous driving, smart cities, and industrial automation. In contrast to conventional cloud-centric architectures, these paradigms mitigate network congestion and reduce service response time, while offering substantial potential for improving energy efficiency in resource-constrained edge environments~\cite{sharma2024intelligent,tzenetopoulos2021fade}. However, the increasing adoption of microservices-based architectures (MSAs) in fog and edge ecosystems has amplified the complexity of resource management, as the modular, distributed, and fine-grained nature of microservices intensifies the interactions among service placement, resource provisioning, task scheduling and offloading, resource allocation, and instance selection. When combined with the stringent energy constraints of fog and edge devices, diverse QoS requirements, fluctuating workloads, and unstable network conditions, the orchestration challenge becomes even more critical. Ensuring energy-efficient operation while maintaining reliable service delivery across heterogeneous, geo-distributed, and resource-limited infrastructures demands a holistic understanding of all core resource management dimensions. These factors highlight the importance of a comprehensive survey that systematically analyzes energy-aware resource management strategies for microservices-based fog and edge computing, offering a unified perspective that can guide future research and system design~\cite{premsankar2022energy}.

Over the past few years, resource management in fog and edge computing has gained significant traction, leading to a growing body of research and several survey studies. However, existing surveys typically address only a subset of the problem space—often focusing on individual aspects such as service placement, task offloading, or autoscaling—without examining all core resource management components in an integrated manner. Moreover, prior works rarely consider energy efficiency and microservices-driven system models as first-class concerns, despite their increasing relevance in real deployments. This fragmented coverage leaves a clear gap in the literature and underscores the need for a unified, energy-aware survey that holistically examines the five fundamental dimensions of resource management within microservices-based fog and edge environments. Motivated by this gap, the present survey provides the first consolidated taxonomy and critical analysis that collectively bring these dimensions together.

Overall, this survey provides a comprehensive and energy-aware reference on resource management in microservices-based fog and edge computing. By systematically covering all five fundamental dimensions—service placement, resource provisioning, task scheduling and offloading, resource allocation, and instance selection—we deliver a unified taxonomy from an algorithmic perspective, highlighting the advantages and limitations of existing solutions. Furthermore, by identifying open research challenges and outlining future directions, this work offers a holistic foundation to guide the development of more intelligent, efficient, and sustainable fog/edge computing systems.

To conduct this survey, we adopted a systematic multi-stage research methodology inspired by evidence-based review practices. A comprehensive literature search was performed across the major academic databases — IEEE Xplore, ACM Digital Library, SpringerLink, ScienceDirect, and Wiley Online Library — covering studies published between January 2020 and December 2024. Early-access 2025 papers were included if their online version appeared during the review period.

The search queries were designed to capture five fundamental resource management domains — service placement, resource provisioning, task scheduling and offloading, resource allocation, and instance (or replica) selection — combined with architectural and environmental terms such as microservice, fog, edge, and mobile edge computing. Each domain-specific query used logical combinations of these keywords (e.g., "fog" OR "edge" OR "MEC" with "microservice" and the corresponding management term). This initial search yielded 1,171 publications. After removing duplicates, non-English records, and short or non-peer-reviewed works, we screened the titles and abstracts to retain studies explicitly addressing energy-aware objectives within microservices-based fog or edge architectures. A full-text assessment of the remaining papers resulted in a final corpus of 136 studies spanning the five target dimensions.

\begin{figure}[t]
    \centering
    \begin{minipage}[c]{0.48\textwidth}
        \centering
        \includegraphics[width=\linewidth]{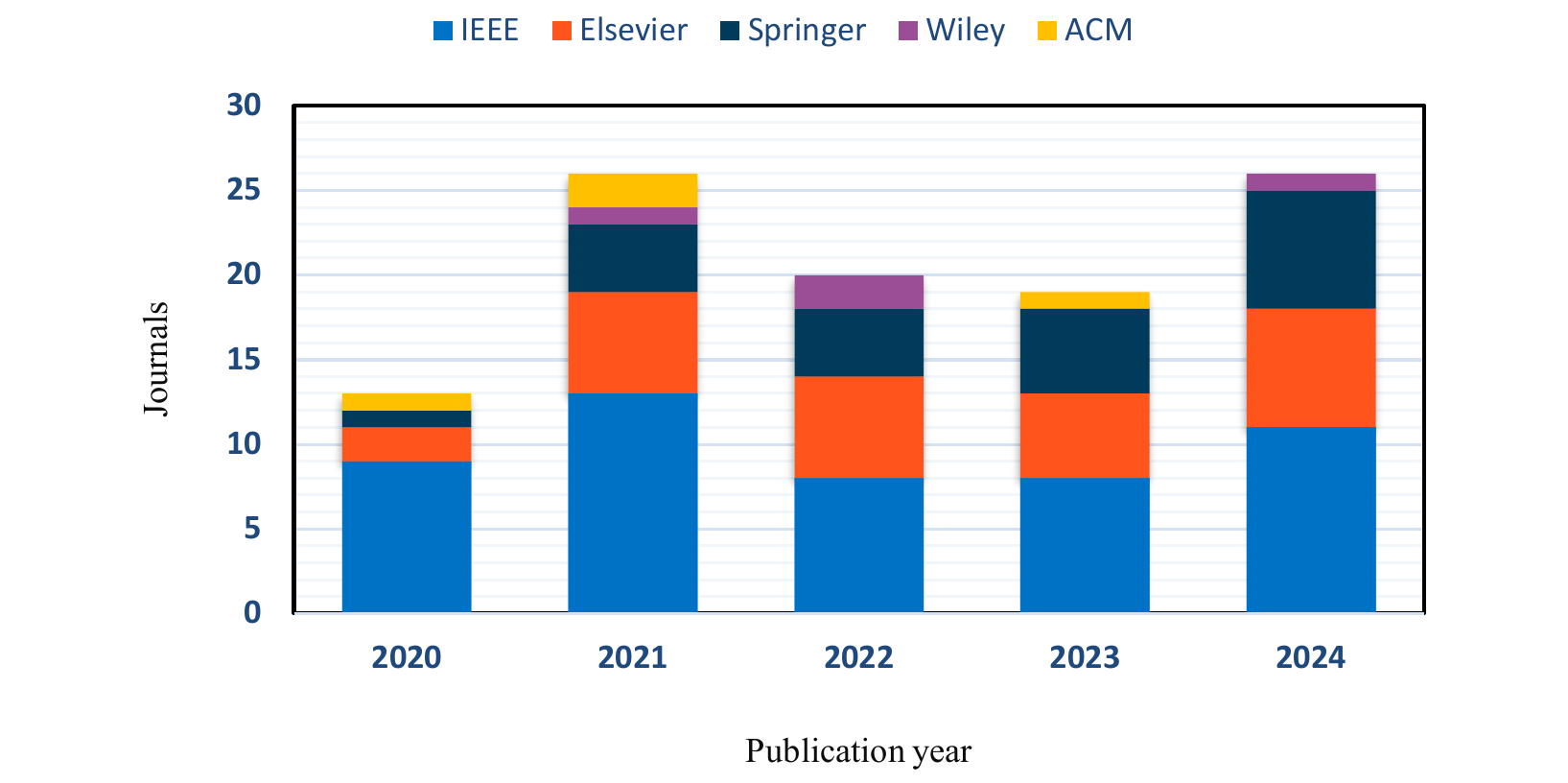}
        \caption{Yearly distribution of selected studies.}
        \label{fig:distribution_journal}
        \Description{Yearly distribution of selected studies and their publishing sources.}
    \end{minipage}
    \hfill
    \begin{minipage}[c]{0.48\textwidth}
        \centering
        \includegraphics[width=\linewidth]{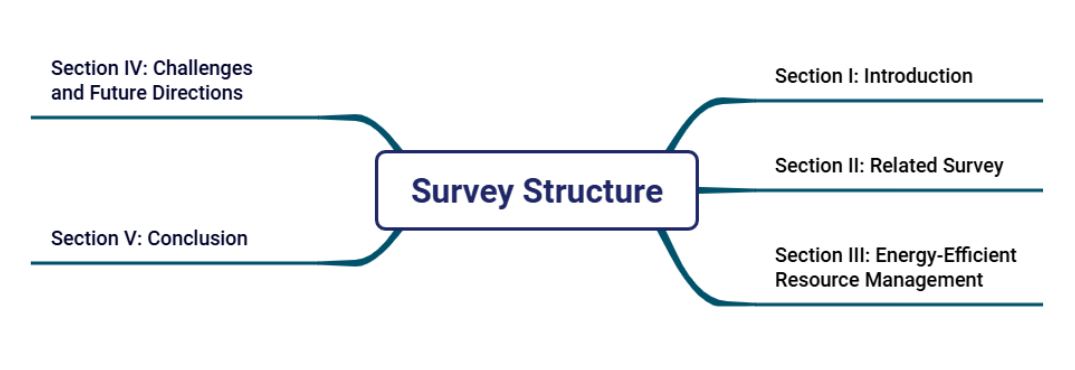}
        \caption{Survey Structure.}
        \label{fig:SurveyStructure}
        \Description{A taxonomy diagram illustrating resource management in fog and edge computing.}
    \end{minipage}
\end{figure}

The distribution of the selected studies by publication year and venue is shown in Figure~\ref{fig:distribution_journal}, illustrating the growing research attention between 2020 and 2024. For consistency, papers published online in late 2024 and assigned print dates in early 2025 are classified under 2024 in our analysis.
 
The rest of this paper is organized as follows: Section~\ref{sec:Related} discusses related surveys, positioning our survey within the broader body of literature. Section~\ref{sec:Energy-Efficient} provides a detailed examination of energy-efficient resource management techniques across the five focus areas. Section~\ref{sec:Challenges} outlines open challenges and potential future research directions, while Section~\ref{sec:Conclusion} concludes the paper with a summary of key findings and their implications for future research, as illustrated in Figure~\ref{fig:SurveyStructure}.
 
\section{Related Surveys}
\label{sec:Related}
This subsection presents a review of related survey papers on resource management issues, specifically focusing on placement, resource provisioning, task scheduling and offloading, resource allocation and replica selection. The advantages and limitations of each related study are discussed and analyzed, as summarized in Table~\ref{table:Tbl01Survey}.

Salaht et al.~\cite{salaht2020overview} provide a comprehensive survey on the Service Placement Problem (SPP) in fog and edge computing, addressing challenges in large-scale, geographically distributed, and heterogeneous environments. They classify research into five scenarios based on deployment objectives, user expectations, and problem descriptions, introducing a new classification scheme that considers problem statements, placement characteristics, optimization strategies, and experimental tools. While the article thoroughly analyzes service placement algorithms, including control models, scheduling approaches, and mobility support, it overlooks key resource management tasks like resource provisioning, task scheduling, and replica selection. Additionally, it offers limited discussion on energy efficiency and excludes MSAs, leaving gaps in understanding their interaction with service placement. The article~\cite{sonkoly2021survey} surveys the optimal placement of computational units like Virtual Machines (VMs), containers, and tasks in edge computing. It introduces a taxonomy that classifies research by platform capabilities, service structure, problem formulation, mathematical methods, and optimization goals, focusing on QoS and cost efficiency. While it provides valuable insights into placement challenges, it overlooks key resource management areas such as resource provisioning, replica selection, and energy efficiency. Its coverage of task scheduling and offloading is limited, and it offers minimal insights into MSA.

Kaur et al.~\cite{kaur2022container} provide a comprehensive survey on container migration techniques and placement algorithms in cloud, fog, and edge computing, with a focus on geographically distributed infrastructures. The article categorizes placement strategies and explores methods for transferring containerized microservices between servers, analyzing trade-offs between resource efficiency and latency requirements. While it thoroughly examines migration and placement, it overlooks key aspects of dynamic resource management, such as resource provisioning, replica selection, and energy efficiency. Coverage of task scheduling and offloading is also limited. Malazi et al.~\cite{malazi2022dynamic} provide a systematic review of dynamic SPP in MEC environments, exploring its links to task scheduling, resource management, and caching. The paper categorizes service placement methods by design objectives and evaluation strategies, identifying key challenges and future directions. While it offers a detailed analysis of service placement, it overlooks crucial aspects like resource provisioning, dynamic scaling, and replica selection, which are vital for workload adaptation and fault tolerance. The article also lacks coverage of energy efficiency and MSA, essential for optimizing resource use and ensuring service reliability.

Smolka et al.~\cite{smolka2022evaluation} present a comprehensive survey of evaluation practices for fog application placement algorithms, analyzing over 100 studies using 40 criteria such as evaluation environments, problem instances, and metrics. While offering valuable insights into placement evaluation methods, the paper overlooks key aspects like resource provisioning, task scheduling, offloading, resource allocation, and replica management, which are vital for optimizing fog system performance. Apat et al.~\cite{apat2023comprehensive} analyze the IoT application placement problem in fog computing, focusing on resource allocation strategies and optimization approaches for single and multiple objectives. While the study addresses placement and resource allocation, it overlooks replica selection, task scheduling, offloading, resource provisioning, MSA, and energy efficiency, leaving gaps in understanding dynamic resource management in fog environments.

Islam et al.~\cite{islam2023optimal} review research on fog application placement in the cloud-to-thing continuum, focusing on strategies for placing applications across cloud, fog, and edge environments. The study categorizes the fog application placement problem (FAPP) and examines related mechanisms, tools, and technologies. While offering valuable insights, it provides limited coverage of task scheduling, offloading, replica selection, resource provisioning, and energy efficiency, leaving gaps in dynamic resource management. The article by Pallewatta et al.~\cite{pallewatta2023placement} provides a comprehensive taxonomy of recent research on the placement of microservices-based IoT applications within fog computing environments. It classifies the placement problem into several categories, emphasizing the challenges and policies associated with deploying microservices in fog settings. By analyzing factors like microservices composition, performance evaluation, and application placement policies, the paper offers valuable insights into the complexities of microservices placement and resource allocation. However, it does not sufficiently address critical aspects such as task scheduling, offloading, replica selection, and energy efficiency, which are essential for ensuring dynamic resource management, optimized computation, and sustainability across fog and edge nodes. 

Pallewatta et al.~\cite{pallewatta2023placement} present a taxonomy of recent research on microservice-based IoT application placement in fog computing. The paper examines challenges, policies, and factors like microservice composition and performance evaluation. While offering valuable insights, it lacks coverage of key aspects such as task scheduling, offloading, replica selection, and energy efficiency, crucial for dynamic resource management and system sustainability. Santos et al.~\cite{santos2023automated} survey automated application deployment in MEC environments, emphasizing Infrastructure as Code (IaC) tools. They explore IaC-enabled deployment methods to tackle MEC’s dynamic nature, improving QoE, QoS, and SLA compliance. Focused on application placement, the study neglects energy efficiency, replica selection, and MSA, with limited coverage of task scheduling and offloading, leaving gaps in MEC resource management.

Sarkohaki et al.~\cite{sarkohaki2024service} survey service placement in fog computing, analyzing optimization strategies. They categorize methods by application, goals, and virtualization, comparing existing approaches’ strengths and weaknesses. While thorough on placement, the paper lacks depth on resource provisioning, replica selection, and MSA, limiting its scope. Qu et al.~\cite{qu2018auto} focus on the challenges and techniques involved in autoscaling web applications in cloud environments, providing a detailed taxonomy based on the Monitoring, Analysis, Planning, and Execution (MAPE) loop. The paper categorizes autoscaling approaches, evaluating their strengths and weaknesses, but focuses primarily on resource provisioning. It largely overlooks critical areas such as service placement, task scheduling, offloading, replica selection, and energy efficiency, while only partially addressing resource allocation and MSA, leaving these topics underexplored.

Duc et al.~\cite{duc2019machine} offers a detailed survey on reliable resource provisioning for distributed applications in joint edge-cloud environments, emphasizing machine learning (ML) techniques. The work is organized around three key themes: workload characterization and prediction, component placement and system consolidation, and application elasticity and remediation. It highlights ML's role in modeling, forecasting, and optimizing resource management tasks, including application placement, migration, provisioning, and remediation. However, the article does not adequately address replica selection, energy efficiency, and MSA, leaving these critical areas underexplored in fog and edge environments. Singh et al.~\cite{singh2019research} survey autoscaling techniques for multi-tier web applications in cloud environments, identifying key challenges and proposing a taxonomy based on approaches, resources, monitoring tools, experiments, workloads, and performance metrics. Their work focuses on dynamic resource scaling in cloud systems, addressing critical aspects of resource provisioning. However, it does not explore service placement, task scheduling, offloading, replica selection, energy efficiency, or MSA, leaving these areas underexamined in resource management.

Verma et al.~\cite{verma2021auto} survey autoscaling, load prediction, and VM migration techniques for IoT-based cloud applications. The study evaluates these methods against QoS parameters, outlines foundational concepts, and identifies challenges in efficient cloud autoscaling. By focusing on load prediction and VM migration, the authors offer insights into resource provisioning for IoT workloads. However, the study is confined to cloud environments, omitting service placement, task scheduling, offloading strategies, replica selection, and MSA—key aspects for managing the constrained, heterogeneous resources of fog/MEC systems. Jawaddi et al.~\cite{jawaddi2022review} conduct a systematic literature review on autoscaling techniques for microservices, focusing on verification aspects. The paper explores scaling challenges and presents a taxonomy of autoscaling and verification approaches, particularly in cloud systems. It emphasizes formal verification methods, like probabilistic model checking, to ensure autoscaling policy effectiveness in dynamic cloud environments. However, the study is limited to cloud systems, excluding edge or MEC challenges. It also overlooks critical areas such as service placement, task scheduling, offloading, replica selection, and offers limited discussion on resource allocation and energy efficiency, leaving these topics underexplored.

Shakarami et al.~\cite{shakarami2022resource} review resource provisioning approaches in edge and fog computing, categorizing them into five classes: framework-based, heuristic/meta-heuristic-based, model-based, ML-based, and game-theoretic mechanisms. The paper compares these approaches based on performance metrics, techniques, case studies, and evaluation tools while proposing a taxonomy to structure them. Although the authors focus on resource provisioning, they do not adequately address task placement, scheduling, offloading, replica selection, and MSA. These underexplored areas limit the paper’s coverage of critical aspects needed for optimizing resource management and fault tolerance in distributed systems. The survey article by Dogani et al.~\cite{dogani2023auto} offers a comprehensive review of autoscaling solutions for container-based virtualization in cloud, edge, and fog computing environments. It classifies various autoscaling techniques based on system design and workload management methods, focusing on both reactive and proactive strategies using orchestration tools like Docker and Kubernetes. The paper discusses the challenges and limitations specific to each environment, providing valuable insights into resource provisioning. However, it does not adequately address service placement strategies, task scheduling, offloading, and replica selection.

In their review, Kaur et al.~\cite{kaur2023systematic} analyze resource provisioning techniques in fog computing, categorizing them by QoS, cost, energy, and latency. They emphasize the cost-efficiency and performance of dynamic over static methods, identify gaps, and suggest future research. However, the article focuses mainly on resource provisioning, offering limited coverage of task placement, replica selection, and MSA. While it briefly addresses task scheduling, offloading, and resource allocation, these areas lack depth. Energy efficiency is also discussed but warrants further exploration. Qassem et al.~\cite{al2024containerized} survey MSA resource management frameworks, focusing on efficient resource utilization in cloud environments for containerized microservices via autoscaling. The study categorizes frameworks into reactive, predictive, and hybrid approaches, assessing their strengths and limitations in preventing SLA violations and enhancing resource efficiency. It also examines their impact on resource utilization and data center costs, providing framework selection guidance. While emphasizing resource provisioning and Allocation, it offers limited coverage of Placement, Task scheduling and Offloading, replica selection, and energy efficiency, which are critical for resource management in distributed and containerized environments.

Tari et al.~\cite{tari2024auto} provide a comprehensive review of autoscaling strategies in serverless computing, focusing on resource provisioning in cloud environments. The paper classifies autoscaling approaches into ML-based, framework-based, and model-based strategies, evaluating their performance and applicability to serverless architectures. While it makes significant contributions to autoscaling for cloud-based serverless applications, it largely overlooks placement strategies and replica selection, and offers limited exploration of task scheduling, offloading, energy efficiency, and MSA. Walia et al.~\cite{walia2023ai} review resource management challenges in fog and edge computing, focusing on computing resource provisioning, task offloading, scheduling, service placement, and load balancing. The paper surveys AI-based and non-AI-based solutions, presenting mathematical models for key areas like placement, provisioning, task scheduling, and allocation. It also explores the integration of emerging technologies such as serverless computing, 5G, Blockchain, Digital Twins, and Quantum Computing. However, it lacks depth in critical areas, notably replica selection. While load balancing is mentioned, dynamic resource allocation and replica selection strategies are not addressed. Additionally, insights into energy efficiency and MSA are limited.

Gasmi et al.~\cite{gasmi2022survey} survey computation offloading and service placement in fog-based IoT systems, analyzing factors influencing offloading decisions and classifying algorithms by types and objectives. It introduces a novel taxonomy for SPP optimization in IoT, focusing on fog node methods. While providing insights into resource utilization and performance, the paper lacks discussion on resource provisioning and dynamic allocation, leaving gaps in scalable management. Additionally, it omits replica selection and MSA, limiting broader resource optimization strategies. The article by Goudarzi et al.~\cite{goudarzi2022scheduling} reviews recent task scheduling works in fog computing for IoT applications. The authors develop a taxonomy to categorize scheduling mechanisms based on application structure, environmental architecture, optimization modeling, decision engine characteristics, and performance evaluation. They identify research gaps and propose guidelines for enhancing scheduling techniques, particularly in task scheduling and offloading. However, the paper offers limited coverage of resource provisioning and replica selection, leaving these critical aspects underexplored. Additionally, the discussion on MSA is incomplete, narrowing its contributions to broader resource optimization strategies.

Jamil et al.~\cite{jamil2022resource} present a comparative study of task scheduling algorithms in fog computing and Internet of Everything (IoE) environments, focusing on various techniques and their optimization metrics, strengths, and weaknesses. The paper reviews task scheduling, workflow scheduling, and resource allocation, exploring intelligent methods like ML, fuzzy logic, and RL. It identifies key research gaps and challenges, providing insights into scheduling algorithms and future directions for fog and IoE. However, it lacks discussion on placement strategies, resource provisioning, and replica selection, and does not address MSA, narrowing its scope in broader system optimization challenges. Kumar et al.~\cite{kumar2022survey} study nature-inspired algorithms, like Genetic Algorithms (GAs) and Deep Neural Network (DNN), for solving service placement and computation offloading in fog and edge computing. The article categorizes approaches into service placement and offloading, comparing various strategies. It highlights challenges in IoT-edge-cloud systems but omits resource provisioning, replica selection, and MSA, leaving gaps in scalable resource management and broader system optimization.

In their systematic review, Tocze et al.~\cite{tocze2018taxonomy} examine AI/ML-based resource management techniques in edge and fog computing, categorizing literature by resource type, management objectives, location, and use. The paper highlights challenges and gaps in research, particularly regarding data, storage, and energy resources, while exploring the impact of mobility and collaboration. It provides a comprehensive taxonomy of AI algorithms for resource management, focusing on resource allocation. However, it lacks coverage of placement strategies, task scheduling, replica selection, and energy efficiency, and does not address MSA, leaving critical gaps in distributed environments. Hong et al.~\cite{hong2019resource} survey resource management in fog and edge computing, focusing on architectures, infrastructures, and algorithms. They classify resource management architectures by data flow, control, and tenancy, and discuss the hardware, system software, and middleware involved. The paper explores algorithms for discovery, benchmarking, load balancing, and placement, emphasizing placement strategies in distributed systems. However, it leaves gaps in critical areas like replica selection and resource provisioning. While task scheduling and resource allocation are well-covered, aspects such as energy efficiency and MSA receive less attention.

Noor et al.~\cite{noor2020survey} survey radio resource allocation in vehicular networks, focusing on DSRC and cellular-based systems, including heterogeneous versions. They categorize RA methods, discuss their pros and cons, and explore ML’s role in optimization. However, the paper misses critical areas like placement, resource provisioning, task scheduling, offloading, replica selection, energy efficiency, and MSA, essential for managing distributed systems and MEC. Afrin et al.~\cite{afrin2021resource} present a comprehensive survey on resource allocation and service provisioning in multi-agent cloud robotics, with a focus on key concepts such as resource pooling, computation offloading, and task scheduling. The paper offers a comprehensive taxonomy for resource allocation and addresses several key challenges, yet it falls short in exploring critical areas such as replica selection and MSA. Additionally, while it touches on placement strategies and energy efficiency, these topics are not thoroughly examined.

Noor et al.~\cite{noor2020survey} survey radio resource allocation in vehicular networks, focusing on DSRC and cellular-based systems, including heterogeneous versions. They categorize RA methods, discuss their pros and cons, and explore ML’s role in optimization. However, the paper misses critical areas like placement, resource provisioning, task scheduling, offloading, replica selection, energy efficiency, and MSA, essential for managing distributed systems and MEC. In their survey, Gupta et al.~\cite{gupta2023toward} compare RL and Deep Reinforcement Learning (DRL) with traditional resource management methods like graph theory, heuristics, and greedy algorithms in fog computing. They highlight DRL's superiority for managing resources in IoT-based fog systems and propose a microservice model based on DRL to address the dynamic nature of fog networks. The paper introduces the Learning-Based Resource Manager (LBRM) framework, which uses actor-critic DRL algorithms for resource management and service migration. While it makes significant contributions to resource allocation, it leaves areas like replica selection and placement strategies underexplored, and energy efficiency is less emphasized compared to other aspects.

Iftikhar et al.~\cite{iftikhar2023ai} present a systematic review of AI/ML algorithms in fog and edge computing resource management. They analyze ML, deep learning, and RL techniques for tasks like resource estimation, task offloading, load balancing, application placement, and orchestration. The paper offers a detailed taxonomy of AI/ML approaches but misses key areas like replica selection and MSA. While it extensively covers resource provisioning and task scheduling, energy efficiency and some resource management aspects are underexplored. Zhang et al.'s survey~\cite{zhang2023resource} on resource management in MEC reviews key aspects like computation offloading, resource allocation, and system optimization. It covers offloading models, allocation strategies, and performance metrics while addressing challenges such as stochastic task arrivals and fluctuating channel quality. The article effectively discusses task scheduling and resource allocation but lacks exploration of critical areas like replica selection and MSA. Additionally, while it mentions placement and energy efficiency, these topics are not thoroughly examined, leaving gaps in comprehensive resource management in distributed edge environments.

Alsadie et al.'s review~\cite{alsadie2024comprehensive} on fog computing resource management examines AI techniques like ML, DL, and meta-heuristic algorithms, focusing on latency, energy consumption, and QoS. While it thoroughly covers task scheduling, offloading, and resource allocation, critical areas like replica selection are overlooked, and MSA receives minimal attention. Additionally, resource provisioning and placement—essential for dynamic optimization—are only partially addressed. Baidya et al.'s survey~\cite{baidya2024comprehensive} examines resource-allocation strategies in edge-computing-enabled metaverse environments, focusing on latency, energy consumption, and QoE. The article evaluates 35 strategies and benchmarks 19 algorithms to optimize these factors for real-time, scalable metaverse applications. While it covers resource allocation and task scheduling extensively, it lacks focus on placement, resource provisioning, replica selection, energy efficiency, and MSA, which are essential for effective resource management and system optimization in distributed environments. Zolghadri et al.'s article~\cite{zolghadri2024resource} reviews and classifies resource allocation, application placement, and scheduling strategies in fog computing, focusing on optimizing resource allocation and bandwidth management. While it provides valuable insights into resource allocation and task scheduling, it lacks exploration of replica selection and MSA and offers limited discussion on resource provisioning, especially in dynamic and real-time contexts.\\
Although numerous surveys have addressed different aspects of resource management in fog and edge computing, each has focused only on a subset of the problem—such as service placement, autoscaling, task scheduling, or resource allocation. To the best of our knowledge, none of the existing studies provides a unified and energy-aware perspective that simultaneously covers all five fundamental dimensions of resource management: service placement, resource provisioning, task scheduling and offloading, resource allocation, and instance selection. In contrast, our survey offers a holistic taxonomy and a comprehensive discussion of these dimensions under microservices-based architectures, highlighting their interdependencies and associated open challenges.\\
Studies that focus on architectural and coordination mechanisms rather than on resource-optimization aspects. For instance, Velasquez et al. (2018)~\cite{velasquez2018fog} examined fog orchestration for the Internet of Everything, identifying architectural challenges such as scheduling, interoperability, and security. Svorobej et al. (2020)~\cite{svorobej2020orchestration} provided an in-depth overview of orchestration frameworks and standards (e.g., NFV-MANO, OpenFog, and Kubernetes) for managing distributed infrastructures, while Ullah et al. (2023)~\cite{ullah2023orchestration} offered a comprehensive taxonomy of orchestration systems in the Cloud-to-Things continuum and proposed a conceptual framework for cross-layer coordination.
Although these works provide valuable insights into system-level orchestration, their focus lies primarily on architecture design, standardization, and service coordination, rather than on energy-aware resource management in microservices-based environments, which is the central theme of the present survey. Therefore, while this study acknowledges the significance of orchestration research within fog and edge computing, a detailed exploration of orchestration architectures is considered beyond its intended scope. 

\begin{table}
\centering
\caption{Comparison of articles in various areas of cloud, fog, and mobile edge computing resource management}
\label{table:Tbl01Survey}
\tiny
\begin{tabular}{lllllllllllll}
row & Reference & Year & 
PL & 
RP & 
TS & 
RA & 
RS & 
EE & 
MS & 
EN & 
TX & 
OI \\ \hline\hline
1 & Salaht et al.~\cite{salaht2020overview} & 2020 & \circlefill{100} & \circlefill{25} & \circlefill{50} & \circlefill{50} & \circlefill{25} & \circlefill{25} & \circlefill{0} & Fog,Edge & \checkmark & . \\
2 & Sonkoly et al.~\cite{sonkoly2021survey} & 2021 & \circlefill{100} & \circlefill{25} & \circlefill{75} & \circlefill{75} & \circlefill{25} & \circlefill{50} & \circlefill{25} & Fog,Edge & \checkmark & . \\
3 & Kaur et al.~\cite{kaur2022container} & 2022 & \circlefill{100} & \circlefill{25} & \circlefill{50} & \circlefill{75} & \circlefill{0} & \circlefill{25} & \circlefill{75} & Cloud,Fog,Edge & \checkmark & \checkmark \\
4 & Malazi et al.~\cite{malazi2022dynamic} & 2022 & \circlefill{100} & \circlefill{25} & \circlefill{50} & \circlefill{75} & \circlefill{25} & \circlefill{25} & \circlefill{25} & Edge & \checkmark & \checkmark \\
5 & Smolka et al.~\cite{smolka2022evaluation} & 2022 & \circlefill{100} & \circlefill{0} & \circlefill{0} & \circlefill{25} & \circlefill{0} & \circlefill{25} & \circlefill{0} & Fog,Edge & . & . \\
6 & Apat et al.~\cite{apat2023comprehensive} & 2023 & \circlefill{100} & \circlefill{25} & \circlefill{50} & \circlefill{75} & \circlefill{0} & \circlefill{50} & \circlefill{25} & Fog,Edge & . & \checkmark \\
7 & Islam et al.~\cite{islam2023optimal} & 2023 & \circlefill{100} & \circlefill{50} & \circlefill{25} & \circlefill{75} & \circlefill{25} & \circlefill{50} & \circlefill{75} & Fog & . & \checkmark \\
8 & Pallewatta et al.~\cite{pallewatta2023placement} & 2023 & \circlefill{100} & \circlefill{50} & \circlefill{25} & \circlefill{75} & \circlefill{25} & \circlefill{25} & \circlefill{100} & Fog,Edge & \checkmark & \checkmark \\
9 & Santos et al.~\cite{santos2023automated} & 2023 & \circlefill{100} & \circlefill{75} & \circlefill{50} & \circlefill{75} & \circlefill{25} & \circlefill{0} & \circlefill{25} & Edge &  & . \\
10 & Sarkohaki et al.~\cite{sarkohaki2024service} & 2024 & \circlefill{100} & \circlefill{25} & \circlefill{50} & \circlefill{75} & \circlefill{0} & \circlefill{50} & \circlefill{25} & Cloud,Fog & \checkmark & \checkmark \\
11 & Qu et al.~\cite{qu2018auto} & 2018 & \circlefill{25} & \circlefill{100} & \circlefill{0} & \circlefill{75} & \circlefill{25} & \circlefill{25} & \circlefill{50} & Cloud & \checkmark & \checkmark \\
12 & Duc et al.~\cite{duc2019machine} & 2019 & \circlefill{75} & \circlefill{100} & \circlefill{50} & \circlefill{75} & \circlefill{25} & \circlefill{50} & \circlefill{25} & Edge,Cloud & . & \checkmark \\
13 & Singh et al.~\cite{singh2019research} & 2019 & \circlefill{0} & \circlefill{100} & \circlefill{0} & \circlefill{75} & \circlefill{0} & \circlefill{25} & \circlefill{25} & Cloud & \checkmark & \checkmark \\
14 & Verma et al.~\cite{verma2021auto} & 2021 & \circlefill{25} & \circlefill{100} & \circlefill{0} & \circlefill{50} & \circlefill{0} & \circlefill{25} & \circlefill{0} & Cloud & . & \checkmark \\
15 & Jawaddi et al.~\cite{jawaddi2022review} & 2022 & \circlefill{0} & \circlefill{100} & \circlefill{0} & \circlefill{25} & \circlefill{0} & \circlefill{25} & \circlefill{50} & Cloud &  & \checkmark \\
16 & Shakarami et al.~\cite{shakarami2022resource} & 2022 & \circlefill{25} & \circlefill{100} & \circlefill{50} & \circlefill{75} & \circlefill{25} & \circlefill{50} & \circlefill{25} & Edge,Fogs & \checkmark & \checkmark \\
17 & Dogani et al.~\cite{dogani2023auto} & 2023 & \circlefill{25} & \circlefill{100} & \circlefill{25} & \circlefill{75} & \circlefill{25} & \circlefill{25} & \circlefill{50} & Cloud,Edge,Fogs & \checkmark & \checkmark \\
18 & Kaur et al.~\cite{kaur2023systematic} & 2023 & \circlefill{25} & \circlefill{100} & \circlefill{50} & \circlefill{75} & \circlefill{25} & \circlefill{50} & \circlefill{25} & Fog & . & \checkmark \\
19 & Al et al.~\cite{al2024containerized} & 2024 & \circlefill{25} & \circlefill{100} & \circlefill{25} & \circlefill{100} & \circlefill{50} & \circlefill{25} & \circlefill{75} & Cloud &  & . \\
20 & Tari et al.~\cite{tari2024auto} & 2024 & \circlefill{0} & \circlefill{100} & \circlefill{25} & \circlefill{75} & \circlefill{0} & \circlefill{25} & \circlefill{25} & Cloud & \checkmark & \checkmark \\
21 & Walia et al.~\cite{walia2023ai} & 2023 & \circlefill{100} & \circlefill{100} & \circlefill{100} & \circlefill{100} & \circlefill{0} & \circlefill{25} & \circlefill{25} & Fog,Edges,IoT & \checkmark & \checkmark \\
22 & Gasmi et al.~\cite{gasmi2022survey} & 2022 & \circlefill{100} & \circlefill{25} & \circlefill{100} & \circlefill{75} & \circlefill{0} & \circlefill{100} & \circlefill{0} & Fog & . & \checkmark \\
23 & Goudarzi et al.~\cite{goudarzi2022scheduling} & 2022 & \circlefill{75} & \circlefill{25} & \circlefill{100} & \circlefill{75} & \circlefill{0} & \circlefill{75} & \circlefill{50} & Fog,Edge & \checkmark & \checkmark \\
24 & Jamil et al.~\cite{jamil2022resource} & 2022 & \circlefill{25} & \circlefill{0} & \circlefill{100} & \circlefill{100} & \circlefill{0} & \circlefill{75} & \circlefill{0} & Fog,IoE & \checkmark & \checkmark \\
25 & Kumar et al.~\cite{kumar2022survey} & 2022 & \circlefill{100} & \circlefill{25} & \circlefill{100} & \circlefill{75} & \circlefill{0} & \circlefill{75} & \circlefill{25} & Fog,Edge & . & \checkmark \\
26 & Tocze et al.~\cite{tocze2018taxonomy} & 2018 & \circlefill{75} & \circlefill{25} & \circlefill{50} & \circlefill{100} & \circlefill{25} & \circlefill{50} & \circlefill{0} & Edge & \checkmark & \checkmark \\
27 & Hong et al.~\cite{hong2019resource} & 2019 & \circlefill{50} & \circlefill{25} & \circlefill{75} & \circlefill{100} & \circlefill{0} & \circlefill{50} & \circlefill{25} & Fog,Edge & . & \checkmark \\
28 & Noor et al.~\cite{noor2020survey} & 2020 & \circlefill{0} & \circlefill{0} & \circlefill{0} & \circlefill{100} & \circlefill{0} & \circlefill{0} & \circlefill{0} & Edge &  &  \\
29 & Afrin et al.~\cite{afrin2021resource} & 2021 & \circlefill{25} & \circlefill{50} & \circlefill{75} & \circlefill{100} & \circlefill{0} & \circlefill{25} & \circlefill{0} & Cloud & \checkmark & \checkmark \\
30 & Naren et al.~\cite{naren2021survey} & 2021 & \circlefill{0} & \circlefill{25} & \circlefill{75} & \circlefill{100} & \circlefill{0} & \circlefill{50} & \circlefill{0} & Edge &  & \checkmark \\
31 & Gupta et al.~\cite{gupta2023toward} & 2023 & \circlefill{50} & \circlefill{75} & \circlefill{75} & \circlefill{100} & \circlefill{25} & \circlefill{50} & \circlefill{75} & Fog & . & \checkmark \\
32 & Iftikhar et al.~\cite{iftikhar2023ai} & 2023 & \circlefill{75} & \circlefill{50} & \circlefill{75} & \circlefill{100} & \circlefill{0} & \circlefill{50} & \circlefill{25} & Cloud,Edge,Fogs & \checkmark & \checkmark \\
33 & Zhang et al.~\cite{zhang2023resource} & 2023 & \circlefill{50} & \circlefill{25} & \circlefill{100} & \circlefill{100} & \circlefill{0} & \circlefill{50} & \circlefill{0} & Edge & . & \checkmark \\
34 & Alsadie et al.~\cite{alsadie2024comprehensive} & 2024 & \circlefill{75} & \circlefill{50} & \circlefill{100} & \circlefill{100} & \circlefill{0} & \circlefill{75} & \circlefill{25} & Fog & . & \checkmark \\
35 & Baidya et al.~\cite{baidya2024comprehensive} & 2024 & \circlefill{25} & \circlefill{25} & \circlefill{75} & \circlefill{100} & \circlefill{0} & \circlefill{50} & \circlefill{0} & Edge &  & \checkmark \\
36 & Zolghadri et al.~\cite{zolghadri2024resource} & 2024 & \circlefill{50} & \circlefill{25} & \circlefill{75} & \circlefill{100} & \circlefill{0} & \circlefill{50} & \circlefill{0} & Fog & \checkmark & \checkmark \\
37 & \textbf{This Survey (Our work)} &  & \circlefill{100} & \circlefill{100} & \circlefill{100} & \circlefill{100} & \circlefill{100} & \circlefill{100} & \circlefill{100} & \textbf{Fog,Edge} & \checkmark & \checkmark
\end{tabular}
    \captionsetup{font=footnotesize}  
    \caption*{\footnotesize 
    \textbf{Abbreviations Guide:} 
    PL: Placement, 
    RP: Resource Provisioning, 
    TS: Task Scheduling, 
    RA: Resource Allocation, 
    RS: Replica Selection, 
    EE: Energy Efficiency, 
    MS: Microservice, 
    EN: Environment, 
    TX: Taxonomy, 
    OI: Open Issue.
    }
\end{table}

\begin{figure}
    \centering
    \includegraphics[width=0.9\linewidth]{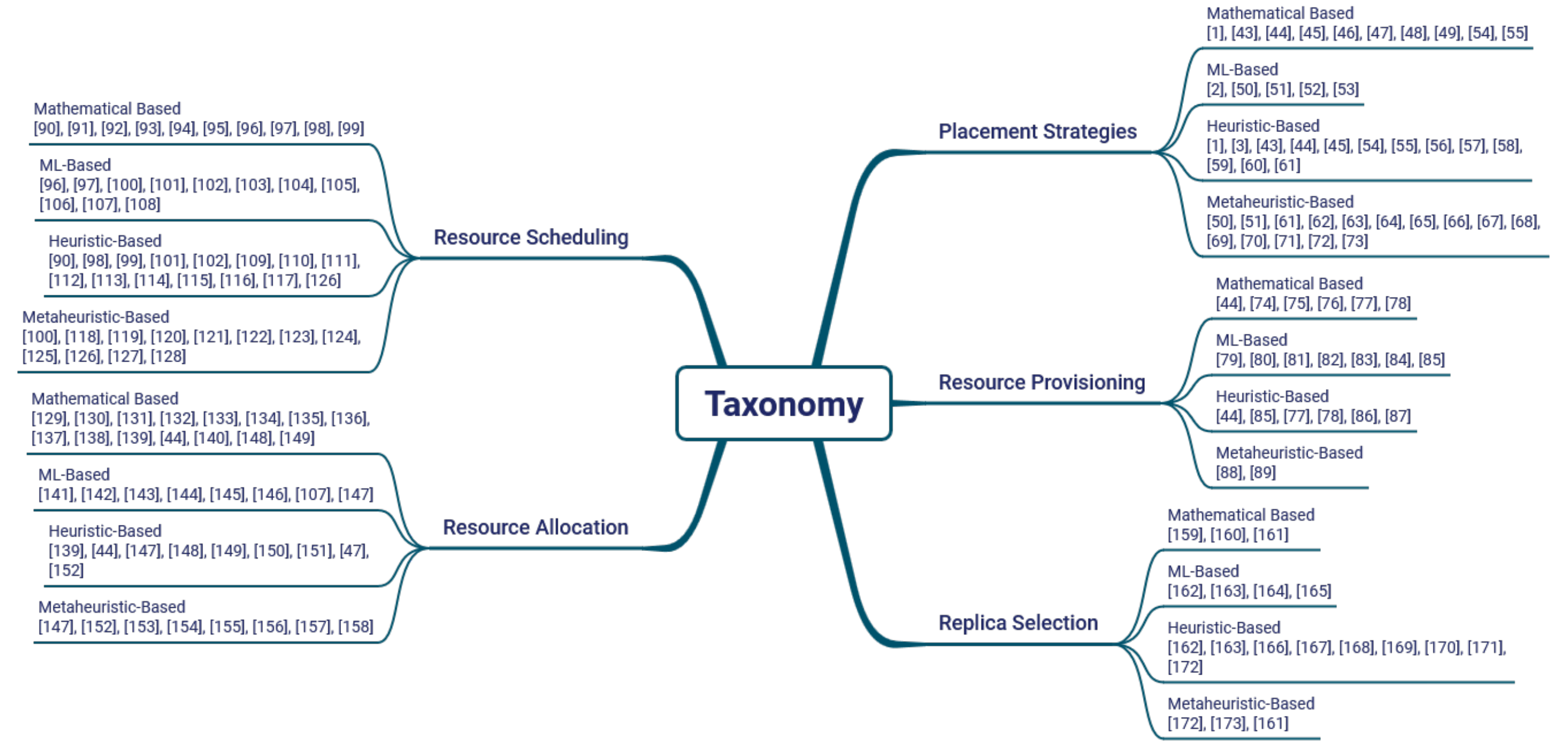} 
    \caption{Taxonomy of Resource Management in fog and edge computing.} 
    \label{fig:Taxonomy} 
    \Description{A taxonomy diagram illustrating resource management in fog and edge computing.}
\end{figure}

\section{Energy-Efficient Resource Management in Microservices-based Fog and Edge Computing}
\label{sec:Energy-Efficient}
The proliferation of edge and fog computing architectures has significantly transformed the landscape of distributed computing by bringing computational resources closer to the end-users. However, managing resources in such heterogeneous and dynamic environments is a complex challenge that requires efficient mechanisms to ensure optimal system performance while meeting the QoS demands of various applications. Typically, resource management encompasses critical tasks such as placement of services, provisioning of resources (autoscaling), scheduling and offloading of tasks, allocation of available resources, and selecting appropriate replicas.

Placement decisions determine the optimal location of microservices or applications across fog and edge nodes to minimize latency and improve resource utilization. Resource provisioning mechanisms, such as autoscaling, dynamically adjust resource allocation based on workload fluctuations, ensuring performance efficiency and energy savings. Task scheduling and offloading distribute computational tasks across edge, fog, and cloud layers, considering constraints like computational capacity, network bandwidth, and energy efficiency for seamless execution. Resource allocation assigns available resources to tasks or services, accounting for real-time availability and QoS parameters, which include service level agreement (SLA), deadline, number of satisfied requests, and failed request rate. Replica selection directs requests to the most suitable service replica, balancing load and reducing latency. This work comprehensively examines these aspects using a structured taxonomy, as illustrated in Fig.\ref{fig:Taxonomy}, providing a detailed framework for efficient resource management in fog and edge computing environments.
Energy optimization in fog/edge systems manifests in three primary forms—computation, communication, and storage energy. Throughout this survey, each reviewed work is categorized according to its dominant energy domain, enabling a fine-grained understanding of trade-offs between energy efficiency and other objectives such as latency and QoS.
\subsection{Placement}
\label{sub:Placement}
Placement in MEC and fog computing is a cornerstone of efficient resource management, ensuring the optimal allocation of services and microservices across distributed nodes. In these environments, the placement process directly impacts critical performance metrics such as energy consumption, latency, network utilization, and system reliability. By strategically assigning computational tasks to appropriate nodes, placement mechanisms enable the seamless execution of services close to the data source, thereby minimizing communication overhead and enhancing real-time responsiveness.
The rise of MSAs has added another layer of complexity to placement strategies. Microservices, characterized by their modular and distributed nature, require meticulous mapping to ensure low latency and effective inter-service communication, all while balancing resource constraints. Furthermore, the dynamic and heterogeneous nature of edge and fog environments presents additional challenges, such as fluctuating workloads, mobility, and energy efficiency.
Given the complexities of placement strategies in MEC and fog computing environments, it is essential to explore diverse algorithmic approaches tailored to specific challenges. In the following sections, we will review various research papers, categorizing them based on their underlying methodologies: Mathematical-based, ML-based, Heuristic-based, and Metaheuristic-based, as shown in Table~\ref{table:tb01}. Additionally, a detailed analysis of the advantages and disadvantages of each approach is provided in Table~\ref{table:add01}.

\begin{table}
\centering
\caption{Optimization objectives and algorithms in placement strategies}
\label{table:tb01}
\tiny
\begin{tabular}{lllccccclccccll}
\multicolumn{1}{c}{\multirow{2}{*}{Row}} & 
\multicolumn{1}{c}{\multirow{2}{*}{CatName}} &
\multicolumn{1}{c}{\multirow{2}{*}{Year}} &
\multicolumn{5}{c}{Objectives} &  & 
\multicolumn{4}{c}{Algorithm} & 
\multirow{2}{*}{Envrionment} &  \\ \cline{4-8}\cline{10-13}
\multicolumn{1}{c}{} & \multicolumn{1}{c}{} & 
\multicolumn{1}{c}{} & 
LT & 
CT & 
LB & 
QoS & 
TP &  & 
MB & 
ML & 
HB & 
MH &  &   \\ \hline\hline
1 & Ahvar et al.~\cite{ahvar2021deca} & 2021 & - & \checkmark & - & - & - &  & \checkmark & - & \checkmark & - & Edge &  \\ \hline
2 & Qu et al.~\cite{qu2021service} & 2021 & - & - & - & - & - &  & \checkmark & - & \checkmark & - & Edge &  \\ \hline
3 & Omer et al.~\cite{omer2021priority} & 2021 & - & \checkmark & \checkmark & - & - &  & \checkmark & - & \checkmark & - & Fog &  \\ \hline
4 & Ahmed et al.~\cite{ahmed2023ecq} & 2023 & \checkmark & \checkmark & - & \checkmark & - &  & \checkmark & - & - & - & Cloud &  \\ \hline
5 & Kumar et al.~\cite{kumar2024auction} & 2024 & - & \checkmark & - & - & - &  & \checkmark & - & - & - & Fog &  \\ \hline
6 & Arzo et al.~\cite{arzo2024softwarized} & 2024 & \checkmark & - & - & - & \checkmark &  & \checkmark & - & - & - & Edge &  \\ \hline
7 & Liang et al.~\cite{liang2024sustainable} & 2024 & - & \checkmark & - & - & - &  & \checkmark & - & - & - & Edge &  \\ \hline
8 & Singh et al.~\cite{singh2023integrated} & 2023 & - & \checkmark & - & - & - &  & - & \checkmark & - & \checkmark & Fog &  \\ \hline
9 & Srichandan et al.~\cite{srichandan2024secure} & 2024 & \checkmark & \checkmark & - & \checkmark & - &  & - & \checkmark & - & \checkmark & Hybrid &  \\ \hline
10 & Rajagopal et al.~\cite{rajagopal2023fedsdm} & 2023 & \checkmark & \checkmark & - & - & - &  & - & \checkmark & - & - & Hybrid &  \\ \hline
11 & Afachao et al.~\cite{afachao2024efficient} & 2024 & \checkmark & - & - & - & - &  & - & \checkmark & - & - & Hybrid &  \\ \hline
12 & Sharma et al.~\cite{sharma2024intelligent} & 2024 & \checkmark & - & - & - & - &  & - & \checkmark & - & - & Fog &  \\ \hline
13 & Goudarzi et al.~\cite{goudarzi2021distributed} & 2021 & - & \checkmark & - & - & - &  & \checkmark & - & \checkmark & - & Hybrid &  \\ \hline
14 & Premsankar et al.~\cite{premsankar2022energy} & 2022 & \checkmark & - & - & - & - &  & \checkmark & - & \checkmark & - & Edge &  \\ \hline
15 & Ray et al.~\cite{ray2023adaptive} & 2023 & \checkmark & - & - & - & - &  & \checkmark & - & \checkmark & - & Edge &  \\ \hline
16 & Hassan et al.~\cite{hassan2020priority} & 2020 & \checkmark & - & - & \checkmark & - &  & - & - & \checkmark & - & Fog &  \\ \hline
17 & Zeng et al.~\cite{zeng2020towards} & 2020 & - & - & \checkmark & - & - &  & - & - & \checkmark & - & Fog &  \\ \hline
18 & Thanh et al.~\cite{thanh2021energy} & 2021 & - & - & - & - & - &  & - & - & \checkmark & - & Hybrid &  \\ \hline
19 & Tzenetopoulos et al.~\cite{tzenetopoulos2021fade} & 2021 & \checkmark & - & - & \checkmark & - &  & - & - & \checkmark & - & Edge &  \\ \hline
20 & Song et al.~\cite{song2023sd} & 2023 & - & - & - & - & - &  & - & - & \checkmark & - & Edge &  \\ \hline
21 & Taleb et al.~\cite{taleb2024energy} & 2024 & - & - & \checkmark & - & - &  & - & - & \checkmark & - & Hybrid &  \\ \hline
22 & Fang et al.~\cite{fang2020iot} & 2020 & \checkmark & - & - & - & - &  & - & - & \checkmark & \checkmark & Edge-Cloud &  \\ \hline
23 & Natesha et al.~\cite{natesha2021adopting} & 2021 & \checkmark & \checkmark & - & - & - &  & - & - & - & \checkmark & Fog &  \\ \hline
24 & Wang et al.~\cite{wang2021effective} & 2021 & - & \checkmark & - & - & - &  & - & - & - & \checkmark & Edge &  \\ \hline
25 & Ghobaei-Arani et al.~\cite{ghobaei2022cost} & 2022 & \checkmark & - & - & \checkmark & \checkmark &  & - & - & - & \checkmark & Fog &  \\\hline
26 & Mortazavi et al.~\cite{mortazavi2022discrete} & 2022 & - & - & - & \checkmark & - &  & - & - & - & \checkmark & Fog &  \\ \hline
27 & Zhao et al.~\cite{zhao2022qos} & 2022 & \checkmark & \checkmark & - & - & - &  & - & - & - & \checkmark & Fog &  \\ \hline
28 & Natesha et al.~\cite{natesha2022meta} & 2022 & \checkmark & \checkmark & - & \checkmark & - &  & - & - & - & \checkmark & Fog &  \\ \hline
29 & Liu et al.~\cite{liu2022solving} & 2022 & \checkmark & - & - & \checkmark & - &  & - & - & - & \checkmark & Hybrid &  \\ \hline
30 & e\{HU2023 et al.~\cite{HU2023} & 2023 & - & - & - & \checkmark & - &  & - & - & - & \checkmark & Hybrid &  \\ \hline
31 & Asghari et al.~\cite{asghari2023energy} & 2023 & \checkmark & - & - & - & - &  & - & - & - & \checkmark & Edge &  \\ \hline
32 & Zare et al.~\cite{zare2024imperialist} & 2024 & \checkmark & \checkmark & - & - & \checkmark &  & - & - & - & \checkmark & Fog &  \\ \hline
33 & Dogani et al.~\cite{dogani2024two} & 2024 & \checkmark & \checkmark & - & - & - &  & - & - & - & \checkmark & Fog &  \\ \hline
34 & Ghasemi et al.~\cite{ghasemi2024mohho} & 2024 & \checkmark & - & - & - & - &  & - & - & - & \checkmark & Fog &  \\ \hline

\end{tabular}
    \captionsetup{font=footnotesize}  
     \caption*{\footnotesize 
    \textbf{Abbreviations Guide:} 
    LT: Latency, 
    CT: Cost, 
    LB: Load Balancing,  
    TP: Throughput, 
    MB: Mathematical-Based, 
    ML: ML-Based, 
    HB: Heuristic-Based, 
    MH: Metaheuristic-Based.
    }
\end{table}

\subsubsection{Mathematical-based:}

Ahvar et al.~\cite{ahvar2021deca} propose dynamic application component placement in edge clouds to minimize energy costs and carbon emissions. Their model considers geographic edge computing distribution, energy price variability, and carbon emission rates, aiming to reduce energy consumption, emissions, and SLA violations. The Dynamic Energy Cost and Carbon Emission-efficient Application placement method (DECA) integrates a prediction-based A* algorithm with Fuzzy Sets for intelligent, adaptive placement decisions, leveraging regional energy and carbon rate variations. In ~\cite{qu2021service}, Qu et al. optimize service placement in UAV-enabled MEC to minimize user energy consumption under latency and resource constraints, enhancing efficiency in the SPUN problem.

Kumar et al.~\cite{kumar2024auction} introduce a distributed fog service placement (DFSP) algorithm to optimize microservices allocation in fog computing environments, targeting reductions in energy consumption and data transfer costs. The model utilizes an iterative combinatorial auction with two variants: DFSP-GF for global cost optimization and DFSP-NF for scalable local decision-making, ensuring energy efficiency and resource constraint adherence. In ~\cite{arzo2024softwarized}, Arzo et al. advocate for MSA in network management systems to improve performance in edge/cloud environments. Their research focuses on decomposing monolithic systems into scalable, containerized microservices using the Microservice-based SDN (MSN) framework, optimizing latency, throughput, resource utilization, and energy consumption through a mathematical model that evaluates end-to-end performance and energy efficiency.

In ~\cite{liang2024sustainable}, Liang et al. emphasize the optimization of Virtual Network Function (VNF) placement and traffic routing in Green Mobile Edge Networks (GMENs) to reduce energy costs while maintaining service availability. Their model addresses dynamic green energy and wireless link availability, formulating the problem as an NP-hard integer nonlinear programming (INLP) task, employing a two-stage online scheme for real-time adaptation. In ~\cite{omer2021priority}, Omer et al. address VM placement in cloud data centers (CDCs), highlighting inefficiencies in power, network, and resource utilization. They formulate the VM placement problem as a MILP model, aiming to minimize power consumption, network usage, and resource wastage. Their priority, power, and traffic-aware approach optimizes VM placement, enhancing resource allocation and energy efficiency.

The articles ~\cite{goudarzi2021distributed,premsankar2022energy,ray2023adaptive,hassan2020priority} primarily rely on their respective core algorithms but also incorporate mathematical methods, such as optimization formulations, Lagrangian relaxation, probabilistic model checking, and mixed-integer linear programming (MILP), to enhance decision-making and achieve energy efficiency, latency reduction, service placement optimization, and QoS enhancement.

\subsubsection{ML-based:}
Singh et al.~\cite{singh2023integrated} propose a secure two-step service placement framework to enhance QoS in fog-cloud ecosystems. This framework addresses latency and security challenges by utilizing fog computing, employing an improved Adaptive Neuro-Fuzzy Inference System (ANFIS) for service classification, and a hybrid CGOA-GA algorithm for efficient fog-tier scheduling, focusing on minimizing makespan, computational cost, and energy consumption. Srichandan et al.~\cite{srichandan2024secure} introduce a novel framework for QoS-aware IoT request processing in fog-cloud environments, addressing latency and resource constraints due to increasing data from Internet-enabled devices. The model integrates a edge layer between devices and CDCs, utilizing ANFIS for request classification and an Improved Honey Badger Algorithm (IHBA) for task scheduling, demonstrating significant QoS improvements through simulations.

Rajagopal et al.~\cite{rajagopal2023fedsdm} introduce FedSDM, a Federated Learning-based Smart Decision Making module, addressing privacy, data ownership, and application heterogeneity in IoT-integrated edge–fog–cloud smart healthcare systems. The approach minimizes energy consumption, network usage, cost, execution time, and latency while ensuring privacy, employing the Federated Averaging (FedAvg) algorithm for decentralized edge-layer processing, outperforming fog and cloud deployments across all metrics. Afachao et al.~\cite{afachao2024efficient} emphasize efficient microservice deployment in edge-cloud networks for user-centric design. Their study optimizes resource allocation and service placement to reduce execution time, network usage, migration delay, and energy consumption. The proposed Bi-Generic Advantage Actor-Critic for Microservice Placement Policy (BAMPP) dynamically adapts to edge-cloud conditions, ensuring optimal resource management.

Sharma et al.~\cite{sharma2024intelligent} present an intelligent service placement algorithm that employs a Double Deep Q-Network with Prioritized Experience Replay (DDQN-PER) to improve service execution time and energy efficiency in IoT-FC environments. The model addresses stochastic service demand and heterogeneous fog nodes, aiming to reduce service latency and energy consumption while dynamically optimizing fog node selection. This DRL approach effectively balances QoS and energy efficiency, contributing significantly to IoT-FC research.

\subsubsection{Heuristic-based:}

Goudarzi et al.~\cite{goudarzi2021distributed} propose a distributed technique for managing application placement and migration in hierarchical fog and edge computing environments, addressing user mobility and diverse IoT resource demands that exceed single-server capabilities. They utilize a weighted cost model to optimize response time and energy consumption, dynamically allocating modules across servers while employing clustering to enhance energy efficiency and reduce communication overhead. Premsankar et al.~\cite{premsankar2022energy} advocate for edge computing as an effective strategy for deploying latency-sensitive AI applications, particularly those utilizing DNN models, while alleviating energy demands from resource overprovisioning. Their model formulates a multiperiod optimization problem to optimize placement and request scheduling, employing a heuristic One-step ahead algorithm and Lagrangian relaxation to enhance service consolidation and scheduling efficiency. 

Ray et al.~\cite{ray2023adaptive} propose an adaptive static-dynamic service placement policy for MEC that optimizes the trade-offs between latency and energy consumption. Their model employs Probabilistic Model Checking to establish a static policy that ensures probabilistic guarantees on energy-latency trade-offs, prioritizing renewable energy availability and resource contention while adapting dynamically to runtime variability in user requirements. Hassan et al.~\cite{hassan2020priority} examine service placement policies in fog-cloud ecosystems, emphasizing the maintenance of high QoS for IoT applications while optimizing energy efficiency. They introduce MinRE, a priority-, network-, and energy-aware placement policy that classifies services as critical or normal, employing heuristic algorithms to minimize response time for critical services and reduce energy consumption for normal services.

Zeng et al.~\cite{zeng2020towards} examine energy-efficient service composition in green energy-powered Cyber–Physical edge Systems (CPFS), aiming to minimize brown energy use through optimized source rate control, load balancing, and service replica deployment. Formulated as an NP-hard MILP model, the approach seeks to maximize green energy utilization while ensuring QoS. A heuristic algorithm yields near-optimal solutions, with simulations confirming substantial energy savings. This study contributes a practical framework for sustainable service composition in fog computing environments. Thanh et al.~\cite{thanh2021energy} propose an energy-aware Service Function Chaining (SFC) embedding strategy to enhance resource and energy efficiency for IoT applications in edge-cloud environments. Their Resource and Energy-Aware Service Chain Embedding (RE-SCE) algorithm addresses dynamic SFC requests and resource constraints through VNF consolidation and offloading strategies, optimizing workload distribution between edge devices and cloud servers while leveraging energy-efficient infrastructure.

Tzenetopoulos et al.~\cite{tzenetopoulos2021fade} highlight the need for innovative approaches to reduce energy consumption in heterogeneous, power-constrained edge-computing devices while maintaining QoS. They propose FADE, a framework that decomposes applications into fine-grained serverless functions, optimizing energy-aware function placement on edge devices through a heuristic-based algorithm that predicts energy consumption and execution time while adhering to user-defined thresholds. Song et al.~\cite{song2023sd} propose an adaptive edge task offloading scheme (SD-AETO) aimed at optimizing energy utilization and processing latency in MEC. Their model integrates service deployment and task offloading to enhance Quality of Experience (QoE) for 5G/B5G, IIoT, and computing networks, utilizing an approximate deployment graph (AD-graph) to minimize redundancy and storage.

Taleb et al.~\cite{taleb2024energy} introduce a model for microservices placement in cloud-fog-edge infrastructures, addressing challenges such as node failures, resource limitations, and network congestion. Their optimization goals include reducing energy consumption and network load while ensuring acceptable response times. The approach combines the Louvain algorithm for community detection with a greedy algorithm to strategically place microservices, enhancing resource utilization and sustainability. Ahmed et al.~\cite{ahmed2023ecq}, identify the challenge of optimizing service migration in MEC systems, where energy consumption and migration costs need to be minimized amidst heterogeneous server configurations and non-universal server presence across base stations. They formulate this issue as a MILP model, aiming to reduce system energy use and migration costs while treating delay as a key constraint. To address this, they propose ECQ, a heuristic-based method designed to deliver an energy-efficient, cost-effective, and QoS-aware solution tailored to the dynamic and time-sensitive nature of real-world MEC environments.

The articles ~\cite{ahvar2021deca,qu2021service} primarily rely on their core algorithms but also incorporate heuristic methods, such as Fuzzy Sets in DECA for optimizing energy cost and carbon emissions, and adaptive strategies in UAV-enabled MEC for dynamic service provisioning, to enhance decision-making and system performance.

\subsubsection{Metaheuristic-based:}
In the research conducted by Fang et al.~\cite{fang2020iot}, the authors tackle the challenges associated with real-time processing of extensive IoT data within conventional cloud networks by introducing an edge-cloud computing architecture. Their optimization objectives focus on minimizing task latency and reducing energy consumption, utilizing a heuristic dynamic task processing algorithm alongside an enhanced discrete particle swarm optimization (PSO) algorithm for module placement and task scheduling. Natesha et al.~\cite{natesha2021adopting} propose the Elitism-based GA (EGA) as an effective solution for the challenges of IoT service placement in fog computing environments. This issue arises from the geographically distributed nature of fog nodes and their limited resources. The optimization aims to minimize service time, cost, and energy consumption while maintaining QoS, employing a two-level resource provisioning framework utilizing Docker and containers. 

Wang et al.~\cite{wang2021effective} propose an effective edge-intelligent service placement algorithm (EISPA) to enhance the performance of 5G-and-beyond Industrial IoT (IIoT) by ensuring service continuity and optimizing system costs under energy constraints. Their model utilizes Multi-Access edge computing to overcome traditional cloud computing limitations in low-latency sensor networks, addressing service discontinuity caused by device mobility. Ghobaei-Arani et al.~\cite{ghobaei2022cost} highlight the increased demand for efficient data processing in fog computing environments due to the rapid growth of IoT applications and the emergence of 5G networks. They introduce a cost-efficient IoT service placement approach utilizing the Whale Optimization Algorithm (WOA) to optimize resource usage, minimize energy consumption, and reduce service delay while satisfying QoS requirements.

Mortazavi et al.~\cite{mortazavi2022discrete} highlight the benefits of deploying IoT applications in fog computing environments, particularly for time-sensitive tasks, by conserving network bandwidth and cloud resources. To address inefficiencies in service deployment, such as resource waste and increased power consumption, they introduce a Discrete Cuckoo Search Algorithm (CSA-SP). This metaheuristic approach optimizes service placement across fog and cloud nodes, minimizing power consumption while ensuring reliability and avoiding single points of failure. Zhao et al.~\cite{zhao2022qos} emphasize the need for efficient service placement mechanisms in fog computing due to the rapid growth of IoT applications. They propose FSP-ODMA, a QoS-aware mechanism utilizing the Open-source Development Model Algorithm (ODMA) to optimize service deployment on fog nodes. This approach minimizes service cost, energy consumption, response time, and latency while maximizing resource utilization, addressing the challenges of energy efficiency and latency in fog computing.

According to ~\cite{natesha2022meta}, Natesha et al. propose that meta-heuristic-based hybrid service placement strategies can significantly enhance the efficiency of fog computing architectures for IIoT applications. The study develops two hybrid algorithms, MGAPSO and EGAPSO, integrating GA and PSO with Elitism-based GA (EGA) to optimize service placement in a two-level fog computing framework, aiming to minimize service time, cost, and energy consumption while ensuring QoS. The NP-hard problem of service placement is addressed, with EGAPSO outperforming state-of-the-art strategies in efficiency and resource utilization. Energy consumption is minimized through optimized load distribution, resource utilization, and explicit consideration of energy during data processing and transmission. In their work ~\cite{liu2022solving}, Liu et al. propose a conceptual framework for IoT service placement in fog computing using the Cuckoo Search Algorithm (CSA) to optimize resource utilization, energy consumption, and QoS. The problem involves efficiently mapping IoT applications to fog cells, modeled as a multi-objective optimization problem considering resource heterogeneity and constraints like CPU, RAM, and storage. The solution leverages CSA to minimize energy consumption, delay, response time, SLA violations, and cost while maximizing fog utilization.

In ~\cite{HU2023}, HU et al. highlight the energy consumption challenges arising from increased communication among IoT devices, wearable sensors, and cloud providers, particularly in cloud-edge data centers. To address this, they propose an energy-aware service placement strategy using a hybrid meta-heuristic algorithm, GA-SSO, which combines GA and Social Spider Optimization (SSO). This approach optimizes VM placement to minimize energy consumption while maintaining QoS, ensuring reliability and safety for IoT applications. According to ~\cite{asghari2023energy}, Asghari et al. propose an energy-aware edge server placement strategy using an improved Butterfly Optimization Algorithm (BOA) combined with the Coral Reefs Optimization (CRO) algorithm. This approach addresses the challenges of reducing energy consumption and minimizing network latency in edge computing environments. The BOA identifies optimal local server locations, while the CRO ensures global convergence, avoiding local optima. Dynamic Voltage and Frequency Scaling (DVFS) is integrated to further minimize energy usage based on task requirements.

In ~\cite{zare2024imperialist}, Zare et al. investigate the deployment of IoT services in fog computing using an Imperialist Competitive Algorithm (ICA)-based approach to address the SPP. The study highlights the limitations of centralized cloud computing for real-time IoT applications due to latency, advocating fog computing to bring processing closer to the edge. The optimization goals include improving QoS, resource utilization, energy consumption, service cost, delay cost, and throughput. The proposed SPP-ICA leverages ICA to prioritize and distribute IoT services across fog nodes, optimizing energy usage and performance. In ~\cite{dogani2024two}, Dogani et al. propose a two-tier multi-objective optimization approach for dynamic service placement in container-based fog-cloud computing environments, addressing the challenges of latency-sensitive IoT applications in traditional cloud computing. The system model integrates fog computing to reduce network traffic and service delays, optimizing for latency, cost-efficiency, and energy consumption. The solution employs Kubernetes for resource management and the NSGA-II algorithm to reconcile conflicting objectives, achieving optimal service placement. As outlined in ~\cite{ghasemi2024mohho}, Ghasemi et al. propose the multi-objective Harris hawks optimization (MOHHO) algorithm to address service placement challenges in fog computing. The system model optimizes energy consumption, end-to-end delay, and network utilization by transforming the multi-objective problem into single-objective subproblems. MOHHO reduces energy usage by turning off idle devices, which consume approximately 70\% of their active state energy. CloudSim simulations demonstrate that MOHHO outperforms existing algorithms by balancing workloads and optimizing resource utilization, showcasing the efficacy of metaheuristic approaches in fog computing.

Both articles ~\cite{singh2023integrated,srichandan2024secure} primarily rely on their core algorithms but also incorporate metaheuristic methods, such as a hybrid chaotic grasshopper optimization algorithm (CGOA) with GA in the first article and an improved honey badger algorithm (IHBA) with chaos theory in the second article, to enhance scheduling, resource utilization, and QoS optimization in fog-cloud environments.

\begin{center}

\tiny
\begin{longtable}{
    >{\RaggedRight}m{0.5cm} 
    >{\RaggedRight}m{0.5cm} 
    >{\RaggedRight}m{0.5cm} 
    >{\RaggedRight}m{0.5cm} 
    >{\RaggedRight}m{4cm}
    >{\RaggedRight}m{4cm} 
    >{\RaggedRight}m{4cm}
}
\caption{Advantages \& disadvantages for placement reviews}
\label{table:add01}\\ 
\hline
\multirow{2}{*}{Article} & 
\multicolumn{3}{c}{\textbf{Energy}} & 
\multirow{2}{*}{Main Idea} & 
\multirow{2}{*}{Advantages} & 
\multirow{2}{*}{Disadvantages} \\ 
 & CE & CoE & SE & & & \\ \hline

\hline
\endhead 
\hline
\multicolumn{7}{r}{{Continued on next page}} \\ \hline
\endfoot 
\hline
\multicolumn{7}{l}{
\footnotesize
\textbf{Abbreviations:} 
CE = Computation Energy, 
CoE = Communication Energy, 
SE = Storage Energy.
} \\
\endlastfoot

~\cite{ahvar2021deca} & \circlefill{100} & \circlefill{75} & \circlefill{25} & DECA optimizes application placement in edge clouds for energy cost and carbon emissions & Combines A* algorithm and Fuzzy Sets for optimizing energy cost and carbon emissions, supporting dynamic re-optimization via live migration & Limited to batch applications, ineffective for interactive environments, and does not fully address real-time workload variations\\ \hline
~\cite{qu2021service} & \circlefill{100} & \circlefill{75} & \circlefill{0} & UAV-enabled mobile edge computing optimizes service provisioning to reduce energy consumption & Efficient algorithms with proven convergence significantly reduce energy consumption through a comprehensive optimization approach & High computational complexity in worst-case scenarios, lacking theoretical performance guarantees and scalability considerations\\ \hline
~\cite{omer2021priority} & \circlefill{100} & \circlefill{25} & \circlefill{0} & Efficient VM placement reduces power, network, and resource waste& Total Cloud Data Center power consumption is minimized by efficient Physical Machines being prioritized and active PMs being reduced & Limited to cloud data centers, may not scale well to integrated IoT-Fog-Cloud environments without modification\\ \hline
~\cite{ahmed2023ecq} & \circlefill{100} & \circlefill{25} & \circlefill{0} & optimizing service migration in MEC systems for energy, cost, and QoS& Reduces energy consumption, cuts migration costs compared to FullMig, improves response time over E-ware, ensures high QoS.& Response time is higher than FullMig; limited to heuristic efficiency, may not fully adapt to extreme dynamic conditions.\\ \hline
~\cite{kumar2024auction} & \circlefill{100} & \circlefill{25} & \circlefill{0} & DFSP algorithm minimizes power and data transfer costs in fog computing through auctions & Offers fast convergence, privacy preservation, and scalable, cost-optimized solutions & High communication overhead (DFSP-GF) with unaddressed dynamic performance issues and limited exploration of global vs\\ \hline
~\cite{arzo2024softwarized} & \circlefill{100} & \circlefill{25} & \circlefill{0} & A mathematical model analyzes and optimizes microservices in edge/cloud environments & Provides a detailed mathematical formulation and experimental validation for scalable network architecture design & Limited analysis of microservice placement for energy efficiency and trade-offs between energy and performance metrics\\ \hline
~\cite{liang2024sustainable} & \circlefill{100} & \circlefill{75} & \circlefill{0} & VNF placement and traffic routing in GMENs minimize energy costs while ensuring service availability & An online strategy ensures service availability and reduces energy costs amid real-time green energy changes & Simplified green energy model that ignores real-world energy fluctuations, limiting practical applicability\\ \hline
~\cite{singh2023integrated} & \circlefill{100} & \circlefill{0} & \circlefill{0} & A secure two-step service placement framework uses ML and metaheuristics to optimize QoS in fog-cloud ecosystems & ML-based classification with CGOA-GA improves QoS metrics using chaos theory and opposition-based learning & Limited testing in dynamic environments\\ \hline
~\cite{srichandan2024secure} & \circlefill{75} & \circlefill{25} & \circlefill{0} & A distributed framework for QoS-aware IoT requests in Fog-Cloud optimizes resource use and service delay & Novel IHBA enhances QoS metrics with ANFIS for intelligent classification & Lacks privacy/security considerations, load prediction, and real-world validation\\ \hline
~\cite{rajagopal2023fedsdm} & \circlefill{100} & \circlefill{75} & \circlefill{0} & FedSDM is a Federated Learning module for ECG data management in Edge-Fog-Cloud environments & Edge deployment reduces energy, cost, and latency while ensuring privacy in healthcare through Federated Learning & Incomplete real-time energy analysis, limited exploration of aggregation techniques, and no blockchain integration for security\\ \hline
~\cite{afachao2024efficient} & \circlefill{100} & \circlefill{25} & \circlefill{0} & BAMPP is a novel RL algorithm for resource allocation and microservice placement in Edge-Cloud & Demonstrates superior performance and adaptability in dynamic environments with comprehensive metrics & Limited comparison with nature-inspired algorithms, lack of scalability exploration, and no real-world validation\\ \hline
~\cite{sharma2024intelligent} & \circlefill{100} & \circlefill{0} & \circlefill{0} & A DDQN-PER-based algorithm optimizes service execution time and energy in IoT-Fog environments & Minimizes service latency and energy consumption while optimizing task execution time & Lacks fault tolerance mechanisms, ignores resource contention risks, and does not provide replication for redundancy\\ \hline
~\cite{goudarzi2021distributed} & \circlefill{100} & \circlefill{75} & \circlefill{0} & A distributed application placement technique minimizes response time and energy in hierarchical fog/edge computing & Improves placement deployment time and reduces task execution costs and migrations & Ignores server energy consumption and monetary costs, limited to pre-copy migration models\\ \hline
~\cite{premsankar2022energy} & \circlefill{100} & \circlefill{25} & \circlefill{75} & Energy-efficient service placement for AI applications optimizes DNN model placement and scheduling & Provides an energy-efficient heuristic solution with low latency maintenance, outperforming baselines & Assumes predictable demand, limiting applicability in dynamic environments and ignoring demand uncertainty\\ \hline
~\cite{ray2023adaptive} & \circlefill{100} & \circlefill{0} & \circlefill{0} & An adaptive service placement policy for MEC optimizes latency-energy trade-offs & Guarantees on latency-energy trade-offs with dynamic adaptation to runtime variability & Ignores heterogeneous MEC servers and relies on simulated data, with limited real-world validation\\ \hline
~\cite{hassan2020priority} & \circlefill{100} & \circlefill{25} & \circlefill{0} & MinRE is a priority-aware service placement policy for IoT applications in fog-cloud environments & Enhances energy consumption, deadline satisfaction, and response time & No support for dependent services, excludes streaming applications, and ignores resource cost factors\\ \hline
~\cite{zeng2020towards} & \circlefill{100} & \circlefill{75} & \circlefill{0} & Energy-efficient service composition in CPFS considers source rate control and load balancing & Achieves high energy efficiency and leverages green energy with a comprehensive MILP formulation & Scalability issues in large deployments, computationally intensive MILP, and limited real-world validation\\ \hline
~\cite{thanh2021energy} & \circlefill{100} & \circlefill{75} & \circlefill{0} & An energy-aware SFC embedding strategy optimizes resource efficiency for IoT applications & Resource-efficient SFC embedding algorithm improves acceptance ratio and handles dynamic loads effectively & Focuses on homogeneous SFCs, limiting applicability to complex topologies and increasing complexity from frequent VNF migrations\\ \hline
~\cite{tzenetopoulos2021fade} & \circlefill{100} & \circlefill{25} & \circlefill{0} & FADE decomposes applications into serverless functions for energy-aware placement on edge devices & Offers a tool-flow for profiling, energy prediction, and efficient function placement, optimizing resource use & Simplistic communication overhead modeling, limited regression model features, and requires evaluation on larger edge clusters\\ \hline
~\cite{song2023sd} & \circlefill{100} & \circlefill{25} & \circlefill{0} & SD-AETO optimizes energy and latency in mobile edge computing through task offloading & Increases edge offloading rates and lowers resource consumption in massive task scenarios & Does not differentiate between user equipment attributes and relies on simulations rather than real-world testing\\ \hline
~\cite{taleb2024energy} & \circlefill{100} & \circlefill{75} & \circlefill{0} & A heuristic model optimizes microservice placement in the Cloud-Fog-Edge continuum & Reduces energy consumption and improves sustainability through community-based placement & Focuses on RAM, neglecting other resource constraints, with limited exploration of dynamic resource changes\\ \hline
~\cite{fang2020iot} & \circlefill{100} & \circlefill{75} & \circlefill{0} & A dynamic task processing algorithm reduces latency and power consumption in edge-cloud computing & Decreases task latency and power consumption while enhancing application service quality & Lacks dynamic modification of application mappings during execution and does not optimize real-time resource utilization\\ \hline
~\cite{natesha2021adopting} & \circlefill{100} & \circlefill{25} & \circlefill{0} & EGA optimizes IoT service placement in fog computing to minimize time, cost, and energy & A formal energy model is used and empirical validation is performed with a physical power meter & No consideration of interdependent IoT applications and limited evaluation in complex scenarios\\ \hline
~\cite{wang2021effective} & \circlefill{100} & \circlefill{50} & \circlefill{0} & EISPA uses PSO and SA for service continuity and cost optimization in Industrial IoT & Lowers system cost and energy consumption, excelling in industrial applications with PSO and SA & Longer decision-making times in large systems, potential scalability issues, and limited focus on time complexity optimization\\ \hline
~\cite{ghobaei2022cost} & \circlefill{100} & \circlefill{50} & \circlefill{0} & A cost-efficient IoT service placement approach uses WOA to optimize resource use and service delay & Improves resource usage and service acceptance while reducing delay and energy consumption & Lack of scalability validation in real-world systems, absence of privacy mechanisms, and limited exploration of optimization techniques\\ \hline
~\cite{mortazavi2022discrete} & \circlefill{100} & \circlefill{25} & \circlefill{0} & CSA-SP balances power consumption and reliability in IoT application deployment in fog environments & Significantly reduces power consumption with high reliability for IoT applications, outperforming other algorithms & Ignores fog pricing models, limited focus on dynamic resource provisioning, and lacks detailed evaluation of scheduling strategies\\ \hline
~\cite{zhao2022qos} & \circlefill{100} & \circlefill{50} & \circlefill{0} & ODMA optimizes service cost, energy, and response time in QoS-aware IoT service placement & Enhances resource usage and service acceptance with a multi-objective optimization approach & Does not address reliability and safety in fog interactions, with limited exploration of dynamic resource management\\ \hline
~\cite{natesha2022meta} & \circlefill{100} & \circlefill{50} & \circlefill{0} & Hybrid strategies MGAPSO and EGAPSO minimize service time, cost, and energy in fog computing & A detailed mathematical model for processing and transmission energy is provided & Ignores interdependent IIoT requests, leading to increased energy consumption with more fog nodes and limited scalability evaluation\\ \hline
~\cite{liu2022solving} & \circlefill{100} & \circlefill{25} & \circlefill{0} & A conceptual framework uses CSA for IoT service placement to optimize resource use and QoS & Multi-objective optimization improves fog utilization and reduces energy consumption and service delay & Incomplete trade-off analysis between energy consumption and delay, lacking dynamic workload variation considerations\\ \hline
~\cite{HU2023} & \circlefill{100} & \circlefill{25} & \circlefill{0} & A hybrid GA-SSO strategy optimizes energy consumption and QoS in IoT environments & Achieves 24\% energy reduction and 88\% QoS fitness with an innovative hybrid approach & No security-aware placement, limited evaluation scope, and simulation-based results\\ \hline
~\cite{asghari2023energy} & \circlefill{100} & \circlefill{25} & \circlefill{0} & An energy-aware edge server placement method combines BOA and CRO to reduce energy and latency & Reduces energy and latency while improving scalability with DVFS techniques & No real-world validation, scalability challenges not discussed, and limited focus on resource management\\ \hline
~\cite{zare2024imperialist} & \circlefill{50} & \circlefill{0} & \circlefill{0} & An ICA-based approach optimizes QoS and resource use in IoT service placement & Improves performance by reducing delay, energy, and cost using the CBR-MADE-k framework & Metaheuristic delays requests, no exploration of DRL, and limited dynamic adaptability\\ \hline
~\cite{dogani2024two} & \circlefill{100} & \circlefill{0} & \circlefill{0} & Dynamic service placement is achieved through two-tier multi-objective optimization in fog-cloud & Enhances cost, latency, and energy consumption through Kubernetes and NSGA-II optimization & No consideration of security and privacy, limited execution time evaluation, and lacks real-world implementation validation\\ \hline
~\cite{ghasemi2024mohho} & \circlefill{100} & \circlefill{0} & \circlefill{0} & MOHHO algorithm focuses on reducing energy consumption and delay in fog computing service placement & Effectively reduces energy consumption and end-to-end delay while improving network utilization & Slight increase in execution time compared to some algorithms, does not address security, and limited focus on load balancing\\ \hline 
\end{longtable}
\end{center}

\subsection{Resource Provisioning}

Resource provisioning, particularly autoscaling, is a critical component in MEC and fog computing, enabling dynamic adjustment of computational resources to meet fluctuating demand. In these environments, autoscaling mechanisms ensure that resources are neither underutilized nor overburdened, thereby optimizing cost, energy efficiency, and system performance. By automatically scaling resources up or down based on real-time workload conditions, these mechanisms support the seamless execution of applications while maintaining QoS.

The dynamic nature of edge and fog environments, characterized by varying user demands, mobility, and resource heterogeneity, poses significant challenges for autoscaling strategies. Effective resource provisioning must account for factors such as latency sensitivity, energy constraints, and network bandwidth, all while adapting to unpredictable workload patterns. Furthermore, the integration of MSAs adds complexity, as each microservice may have unique resource requirements and scaling behaviors. In this section, we review and categorize existing research on autoscaling techniques, focusing on Mathematical-based, ML-based, Heuristic-based, and Metaheuristic-based, as shown in Table~\ref{table:tb02}. The advantages and disadvantages of each method are further explored in Table~\ref{table:add02}.

\subsubsection{Mathematical-based:}

Qu et al.~\cite{qu2021service} discussed in Section ~\ref{sub:Placement} for placement, also optimize resource provisioning in UAV-enabled MEC by addressing latency and energy constraints. According to ~\cite{canete2022proactive}, Cañete et al. propose a proactive horizontal autoscaling framework to address energy consumption and QoS challenges in heterogeneous edge infrastructures. The optimization goals include minimizing global energy consumption, reducing failed request rates, and maintaining reasonable execution times. The solution integrates a workload predictor, an Essential Node Identifier (ENI) module, and a constraint satisfaction problem solved using the Z3 SMT solver to optimize node activation and task allocation. 

In ~\cite{mays2022intent}, Mays et al. highlight the challenges of transitioning to fog-native applications due to heterogeneous capacities, energy prices, and supply. They introduce a decentralized algorithm, iADMM, based on the Alternating Direction Method of Multipliers (ADMM), for intent-based workflow mapping and admission. The optimization goals include minimizing operational costs while considering compute-network capacity, energy prices, and green energy supply ratios. The solution leverages green energy and smart load distribution to reduce energy consumption without compromising workflow intents, such as latency and resource constraints. In their work ~\cite{santos2021towards}, Santos et al. propose integrating fog computing with LPWANs to address IoT challenges in Smart Cities, highlighting the limitations of centralized cloud architectures in meeting low latency and high mobility requirements. They introduce a MILP model to optimize end-to-end resource provisioning, incorporating SFC, LPWAN technologies, and objectives like minimizing active nodes, reducing latency, and maximizing user requests. The model focuses on energy efficiency through strategies such as Minimizing Active Nodes (MIN N), reducing operational costs while maintaining service quality. Evaluated using Smart City use cases, the MILP formulation serves as a scalable benchmark for resource provisioning in edge-cloud environments.

Both articles ~\cite{li2021cost,das2022all} primarily rely on their main algorithms but also incorporate mathematical methods, such as linear programming and Tabu Search in the first article for automatic scaling and replica placement optimization, and multi-criteria optimization in the second article for context-aware data filtration and forwarding, to enhance resource utilization and service efficiency.

\begin{table}
\centering
\caption{Optimization goals and methods in resource provisioning}
\label{table:tb02}
\tiny
\begin{tabular}{lllccccclccccll}
\multicolumn{1}{c}{\multirow{2}{*}{Row}} & 
\multicolumn{1}{c}{\multirow{2}{*}{CatName}} &
\multicolumn{1}{c}{\multirow{2}{*}{Year}} &
\multicolumn{5}{c}{Objectives} &  & 
\multicolumn{4}{c}{Algorithm} & 
\multirow{2}{*}{Envrionment} &  \\ \cline{4-8}\cline{10-13}
\multicolumn{1}{c}{} & \multicolumn{1}{c}{} & 
\multicolumn{1}{c}{} & 
LT & 
CT & 
LB & 
QoS & 
TP &  & 
MB & 
ML & 
HB & 
MH &  &   \\ \hline\hline
1 & Qu et al.~\cite{qu2021service} & 2021 & - & - & - & - & - &  & \checkmark & - & \checkmark & - & Edge &  \\ \hline
2 & Cañete et al.~\cite{canete2022proactive} & 2022 & - & - & - & \checkmark & - &  & \checkmark & - & - & - & Edge &  \\ \hline
3 & Mays et al.~\cite{mays2022intent} & 2022 & - & \checkmark & - & - & - &  & \checkmark & - & - & - & Fog &  \\ \hline
4 & Santos et al.~\cite{santos2021towards} & 2021 & \checkmark & - & - & - & - &  & \checkmark & - & - & - & Fog &  \\ \hline
5 & Hazra et al.~\cite{hazra2021collaborative} & 2021 & \checkmark & - & - & - & - &  & - & \checkmark & - & - & Fog &  \\ \hline
6 & Xiang et al.~\cite{xiang2023dynamic} & 2023 & \checkmark & - & - & - & - &  & - & \checkmark & - & - & Edge &  \\ \hline
7 & Faraji-Mehmandar et al.~\cite{faraji2022self} & 2022 & \checkmark & \checkmark & \checkmark & - & - &  & - & \checkmark & - & - & Fog &  \\ \hline
8 & Mekki et al.~\cite{mekki2023xai} & 2023 & - & - & \checkmark & - & - &  & - & \checkmark & - & - & Hybrid &  \\ \hline
9 & Nandhakumar et al.~\cite{nandhakumar2024edgeaisim} & 2024 & - & - & \checkmark & - & - &  & - & \checkmark & - & - & Edge &  \\ \hline
10 & Nizamis et al.~\cite{nizamis2024enact} & 2024 & \checkmark & - & - & - & - &  & - & \checkmark & - & - & Hybrid &  \\ \hline
11 & Choupani et al.~\cite{choupani2025joint} & 2024 & - & \checkmark & - & \checkmark & - &  & - & \checkmark & \checkmark & - & Edge &  \\ \hline

12 & Li et al.~\cite{li2021cost} & 2021 & \checkmark & - & \checkmark & - & - &  & \checkmark & - & \checkmark & - & Edge &  \\ \hline
13 & Das et al.~\cite{das2022all} & 2022 & - & - & \checkmark & - & - &  & \checkmark & - & \checkmark & - & Edge &  \\ \hline
14 & Siasi et al.~\cite{siasi2020delay} & 2020 & \checkmark & \checkmark & - & \checkmark & - &  & - & - & \checkmark & - & Hybrid &  \\ \hline
15 & Adeppady et al.~\cite{adeppady2023energy} & 2023 & \checkmark & - & - & \checkmark & - &  & - & - & \checkmark & - & Edge &  \\ \hline
16 & Xu et al.~\cite{xu2022dynamic} & 2022 & \checkmark & - & - & - & - &  & - & - & - & \checkmark & Edge &  \\ \hline
17 & Rajagopal et al.~\cite{rajagopal2023resource} & 2023 & \checkmark & \checkmark & - & - & - &  & - & - & - & \checkmark & Hybrid &  \\ \hline
\end{tabular}
    \captionsetup{font=footnotesize}  
     \caption*{\footnotesize 
    \textbf{Abbreviations Guide:} 
    LT: Latency, 
    CT: Cost, 
    LB: Load Balancing,  
    TP: Throughput, 
    MB: Mathematical-Based, 
    ML: ML-Based, 
    HB: Heuristic-Based, 
    MH: Metaheuristic-Based.
    }
\end{table}

\subsubsection{ML-based:}

As delineated in ~\cite{hazra2021collaborative}, Hazra et al. advocate for the incorporation of DRL within green industrial fog networks to enhance energy efficiency and mitigate latency for latency-sensitive IIoT applications. Their system model confronts challenges related to effective service provisioning and energy optimization by implementing a DRL-enabled partial service provisioning strategy, which allocates tasks among fog devices and cloud servers to reduce centralized offloading. In the research conducted by Xiang et al.~\cite{xiang2023dynamic}, the authors propose that MEC systems can be dynamically reconfigured to markedly enhance energy efficiency, stability, and service response times. Their system model processes service requests from mobile devices within a MEC framework, addressing the escalating issue of energy consumption. The optimization objectives emphasize minimizing average service response time while ensuring stability and energy efficiency through a reinforcement learning-based algorithm, specifically the Deep Deterministic Policy Gradient (DDPG).

Faraji et al.~\cite{faraji2022self} introduce a self-learning framework utilizing DRL to improve resource and service provisioning in fog computing environments. The primary optimization goals encompass minimizing energy consumption, latency, and operational costs while ensuring effective resource utilization. Their solution method employs a DRL-based framework integrated with a DNN to adaptively allocate resources, leveraging historical strategies to enhance overall efficiency. Mekki et al.~\cite{mekki2023xai} propose a Zero-touch Service Management (ZSM) framework that employs ML and eXplainable AI (XAI) to tackle the complexities of fine-grained resource management in cloud-native environments. The optimization objectives include minimizing resource over-provisioning, maintaining service quality, and enhancing energy efficiency, particularly in edge computing contexts. Their solution method integrates XGBoost for performance prediction and SHAP for explainable AI, facilitating precise vertical resource autoscaling and root-cause analysis during service degradation.

In their study, Nandhakumar et al.~\cite{nandhakumar2024edgeaisim} present EdgeAISim, a lightweight Python-based toolkit designed to address the increasing challenges of resource management in edge computing environments. Their system model extends EdgeSimPy by incorporating AI models such as Multi-Armed Bandit with UCB, DQN, DQN-GNN, and Actor-Critic Network to optimize power consumption and task migration. The optimization goals focus on minimizing energy usage and enhancing energy efficiency through improved task scheduling and service migration.

Nizamis et al.~\cite{nizamis2024enact} propose the ENACT framework, which signifies a substantial advancement in managing hyper-distributed data-intensive applications across the edge-to-cloud continuum. The problem statement centers on optimizing resource orchestration, dynamic scaling, and energy efficiency in distributed environments, addressing challenges such as latency and resource utilization. Their optimization goals include minimizing execution time and maximizing resource utilization, employing AI techniques, specifically Graph Neural Networks (GNNs) and reinforcement learning (RL), for adaptive scheduling and predictive resource management. Choupani et al.~\cite{choupani2025joint} identify the challenges associated with resource autoscaling and request scheduling in serverless edge computing environments, which may result in inefficiencies and suboptimal QoS for users. To address this issue, they propose a self-adaptive approach that integrates a Q-learning algorithm for dynamic scaling of function instances with a heuristic algorithm for optimal request scheduling. This solution aims to enhance resource efficiency while maintaining high QoS by accurately predicting the required number of active instances and selecting the most suitable instances for incoming requests.

\subsubsection{Heuristic-based:}

In the research conducted by Li et al.~\cite{li2021cost}, the authors tackle challenges inherent in edge-cloud environments, particularly concerning latency-sensitive applications and escalating data volumes. They propose cost-aware automatic scaling and workload-aware replica management strategies. The optimization objectives encompass reducing energy consumption, minimizing total costs, enhancing response times, and balancing workloads through a dynamic threshold-based data migration strategy that utilizes historical data and dynamic predictions. Das et al.~\cite{das2022all} address the complexities of efficient data management in shared IoT infrastructures, exacerbated by the proliferation of IoT devices and edge computing. They introduce CaDGen, a framework designed to minimize energy consumption and optimize resource utilization. This framework employs context-aware data filtration to mitigate irrelevant data transmission and a selective forwarding mechanism that dynamically routes data based on load, energy, and buffer states.

In their study, Siasi et al.~\cite{siasi2020delay} propose a hybrid fog-cloud architecture that adeptly balances resource utilization, network delay, and energy consumption for both delay-sensitive and delay-tolerant requests. Their system model features a delay-aware SFC provisioning scheme that optimizes VNF placement across fog and cloud nodes, prioritizing fog nodes for delay-sensitive requests while utilizing cloud nodes for delay-tolerant ones. Adeppady et al.~\cite{adeppady2023energy} assert that serverless edge computing can substantially diminish the energy footprint of edge data centers through optimized microservice provisioning. Their study addresses the management of container states (cold, warm, running) to balance energy efficiency with QoS requirements. The optimization goal focuses on minimizing energy consumption while adhering to service delay constraints through a threshold-based algorithm that dynamically manages container lifecycle based on request queue thresholds.
This article~\cite{qu2021service} primarily relies on its main algorithm but also incorporates heuristic methods, such as relaxation and randomized rounding techniques, to address the MINLP problem and optimize service provisioning in UAV-enabled MEC.

\subsubsection{Metaheuristic-based:}

As delineated in ~\cite{xu2022dynamic}, Xu et al. posit that edge computing, as an extension of cloud computing, can effectively mitigate challenges associated with uncertainty in workflow scheduling while optimizing energy consumption and task completion time. Their system model employs a dynamic resource provisioning method (UARP) within an SDN-based edge computing framework to address uncertainties such as performance degradation and service failures. The optimization goals focus on minimizing energy consumption and task completion time using the NSGA-III algorithm, with the solution method integrating SAW and MCDM techniques for optimal scheduling, validated through comparative experiments. In the research conducted by Rajagopal et al.~\cite{rajagopal2023resource}, the authors propose a meta-heuristic-based resource provisioning model with mobility management to enhance the efficiency of IoT microservices in smart healthcare systems. The study addresses resource allocation and task scheduling challenges in fog and edge computing environments, aiming to optimize energy consumption, network usage, cost, execution time, and latency. The solution employs modified genetic and flower pollination algorithms, demonstrating significant improvements in a simulated smart healthcare application.

\begin{center}

\tiny
\begin{longtable}{
    >{\RaggedRight}m{0.5cm} 
    >{\RaggedRight}m{0.5cm} 
    >{\RaggedRight}m{0.5cm} 
    >{\RaggedRight}m{0.5cm} 
    >{\RaggedRight}m{4cm}
    >{\RaggedRight}m{4cm} 
    >{\RaggedRight}m{4cm}
}
\caption{Advantages \& disadvantages for autoscaling reviews}
\label{table:add02}\\ 
\hline
\multirow{2}{*}{Article} & 
\multicolumn{3}{c}{\textbf{Energy}} & 
\multirow{2}{*}{Main Idea} & 
\multirow{2}{*}{Advantages} & 
\multirow{2}{*}{Disadvantages} \\ 
 & CE & CoE & SE & & & \\ \hline
\endhead 
\hline
\multicolumn{7}{r}{{Continued on next page}} \\ \hline
\endfoot 
\hline
\multicolumn{7}{l}{
\footnotesize
\textbf{Abbreviations:} 
CE = Computation Energy, 
CoE = Communication Energy, 
SE = Storage Energy.
} \\
\endlastfoot 

~\cite{qu2021service} & \circlefill{100} & \circlefill{75} & \circlefill{0} & UAV-MEC energy optimization via joint service, trajectory, and resource allocation & Two optimization algorithms reduce UE energy consumption, ensure convergence, and manage resources comprehensively & Theoretical performance beyond convergence is not thoroughly analyzed, leaving gaps in understanding efficiency\\ \hline
~\cite{canete2022proactive} & \circlefill{100} & \circlefill{50} & \circlefill{25} & Proactive horizontal autoscaling for edge energy efficiency & Achieves significant energy savings, maintains high request success rates, and adapts to various edge configurations and workloads & autoscaling can be time-consuming for larger infrastructures, requires adjustments for optimal performance, and depends on accurate workload predictions\\ \hline
~\cite{mays2022intent} & \circlefill{100} & \circlefill{25} & \circlefill{0} & Decentralized iADMM for fog microservice optimization & Minimizes operational costs, offers flexibility and scalability in fog environments, and incorporates green energy for sustainability & Algorithm complexity hinders implementation in dynamic or large-scale fog networks, decentralized coordination causes inefficiencies, and green energy variability is not fully addressed\\ \hline
~\cite{santos2021towards} & \circlefill{25} & \circlefill{75} & \circlefill{0} & MILP model for fog/edge energy-efficient resource provisioning & Optimizes cloud and wireless needs, provides trade-off insights, and integrates multiple objectives for flexible resource management & MILP model has high computational demands, limited scalability for real-time deployments, and performance may degrade for latency-sensitive objectives\\ \hline
~\cite{hazra2021collaborative} & \circlefill{100} & \circlefill{75} & \circlefill{0} & DRL-based partial service provisioning in green fog networks & Reduces energy consumption and latency, efficiently handles large-scale industrial applications & Does not address scalability challenges or focus on real-time adaptability in dynamic environments\\ \hline
~\cite{xiang2023dynamic} & \circlefill{75} & \circlefill{25} & \circlefill{0} & RL for MEC dynamic reconfiguration to optimize response time and energy & Balances response time, energy efficiency, and stability, improves system stability, and adapts dynamically via reinforcement learning & Uses a simplistic linear resource-performance model, lacks queue theory, and neglects non-functional performance metrics\\ \hline
~\cite{faraji2022self} & \circlefill{75} & \circlefill{25} & \circlefill{0} & DRL framework for fog resource and service provisioning & Reduces response time, cost, and energy, enhances resource utilization, and adapts to workload changes & Lacks predictive techniques for workload estimation, DRL implementation is complex, and horizontal scaling strategies are not comprehensively addressed\\ \hline
~\cite{mekki2023xai} & \circlefill{75} & \circlefill{0} & \circlefill{0} & ZSM with ML/XAI for vertical autoscaling in cloud-native environments & Scales CPU/memory based on XAI insights, improves service quality with fewer resources & Focuses on vertical scaling, lacks broader applicability, and ML/XAI reliance increases complexity\\ \hline
~\cite{nandhakumar2024edgeaisim} & \circlefill{100} & \circlefill{0} & \circlefill{0} & EdgeAISim: AI-driven edge resource management simulation & Reduces power consumption using advanced AI models, promotes eco-friendly edge computing, and supports diverse AI models & Lacks focus on practical implementation challenges, security/privacy concerns, and edge device diversity\\ \hline
~\cite{nizamis2024enact} & \circlefill{100} & \circlefill{25} & \circlefill{0} & ENACT: AI for edge-to-cloud resource and energy optimization & Optimizes app deployment, enhances energy efficiency with dynamic resource modeling, and supports hyper-distributed applications & No robust solutions for cloud-to-edge continuum, challenges in interoperability and security, and requires further real-world validation\\ \hline
~\cite{choupani2025joint} & \circlefill{25} & \circlefill{0} & \circlefill{0} & optimizing serverless edge computing with Q-learning and heuristics& High request success rate, reduced cold starts, efficient resource utilization, improved QoS& Limited discussion on adaptive warm time, may not handle highly diverse QoS needs\\ \hline
~\cite{li2021cost} & \circlefill{100} & \circlefill{25} & \circlefill{25} & Cost-aware scaling and replica management for edge-cloud efficiency & Reduces total cost, improves CPU utilization, balances workloads, and minimizes response time and energy consumption & Lacks evaluation of network latency/bandwidth, scalability to dynamic workloads, and security/fault-tolerance in replica management\\ \hline
~\cite{das2022all} & \circlefill{100} & \circlefill{75} & \circlefill{0} & CaDGen: Context-aware edge microservices for energy conservation & Reduces data transmission, improves energy efficiency through dynamic path selection, and enhances network scalability & No security/privacy considerations in data sharing, and limited focus on task scheduling\\ \hline
~\cite{siasi2020delay} & \circlefill{0} & \circlefill{25} & \circlefill{0} & Delay-aware SFC provisioning in hybrid fog-cloud architectures & Balances fog and cloud solutions, reduces network delays, and improves energy and cost efficiency & Ignores failure probabilities, lacks multi-tier fog architecture exploration, and does not address resource management optimizations\\ \hline
~\cite{adeppady2023energy} & \circlefill{100} & \circlefill{0} & \circlefill{0} & Serverless edge microservice optimization using container management & Reduces energy and memory consumption in edge data centers while maintaining service quality & Limited scalability in small-scale scenarios, NP-hard optimization complicates implementation, and lacks comprehensive testing\\ \hline
~\cite{xu2022dynamic} & \circlefill{100} & \circlefill{25} & \circlefill{0} & UARP: Dynamic resource provisioning for SDN-based edge workflows & Addresses performance uncertainties, reduces energy and completion time, and uses SDN for dynamic resource management & Not validated in real-world scenarios, incomplete uncertainty coverage, and scalability concerns in large-scale deployments\\ \hline
~\cite{rajagopal2023resource} & \circlefill{100} & \circlefill{25} & \circlefill{0} & Meta-heuristic IoT resource provisioning for smart healthcare & Outperforms in energy, network usage, cost, execution time, and latency, integrates IoT mobility, and uses microservices for scalability & Does not address regulatory compliance, ignores real-world user behavior impact, and lacks scalability solutions for large-scale deployments\\ \hline

\end{longtable}
\end{center}

\subsection{Task Scheduling and Offloading}
\label{subsec:taskschedul}
Task scheduling and offloading are pivotal in MEC and fog computing, as they determine how and where computational tasks are executed to maximize efficiency and performance. These processes involve deciding whether to process tasks locally on edge devices, offload them to nearby edge or fog nodes, or send them to centralized cloud servers. The goal is to minimize latency, reduce energy consumption, and optimize resource utilization while meeting application-specific requirements.

The complexity of task scheduling and offloading arises from the dynamic and heterogeneous nature of edge and fog environments, where factors such as device mobility, network conditions, and resource availability must be carefully considered. Additionally, the rise of latency-sensitive applications, such as augmented reality (AR) and autonomous vehicles, has heightened the need for efficient scheduling and offloading strategies. In this section, we explore various research contributions, as well as examining the role of Mathematical-based, ML-based, Heuristic-based, and Metaheuristic-based techniques in addressing these challenges, as shown in Table~\ref{table:tb03}. Additionally, a detailed analysis of the advantages and disadvantages of each approach is provided in Table~\ref{table:add03}.

\subsubsection{Mathematical-based:}
In ~\cite{hazra2020joint}, Hazra et al. highlight fog computing as a transformative technology for enhancing IoT application performance by decentralizing cloud services to the network edge. However, escalating energy consumption in fog networks presents a critical efficiency challenge. To mitigate this, the authors propose an Energy-Efficient Task Offloading (EETO) policy, integrating hierarchical fog networks with priority-aware scheduling and data offloading. Utilizing Lyapunov optimization, EETO minimizes energy consumption, stabilizes virtual queues, and ensures QoS compliance, optimizing the energy-performance trade-off. In ~\cite{zhao2021dynamic}, Zhao et al. explore the integration of MEC with energy harvesting devices to reduce energy consumption while maintaining high QoS for IoT applications. Addressing the challenge of limited computing resources, they formulate a stochastic optimization problem to minimize energy and resource usage, leveraging energy harvesting and grid optimization. The authors introduce an online Dynamic Offloading and Resource Scheduling (DORS) algorithm based on Lyapunov optimization, transforming the stochastic problem into a deterministic framework for real-time decision-making, thereby enhancing efficiency.

As described in ~\cite{luo2021optimization}, Luo et al. propose a multi-UAV-enabled MEC system to reduce ground users' energy consumption through optimized task scheduling, bit allocation, and UAV trajectory. The system employs a two-layer optimization strategy: the upper layer uses dynamic programming for task scheduling, while the lower layer applies ADMM for bit allocation and trajectory optimization, reducing computational complexity. A re-optimization strategy prevents UAV path conflicts, ensuring operational safety and energy efficiency. In ~\cite{samanta2020dyme}, Samanta et al. introduce a dynamic microservice scheduling framework for MEC-enabled IoT platforms, optimizing network delay, energy efficiency, QoS, and total network cost. The framework dynamically allocates resources and schedules microservices, integrating energy consumption models and utility functions to balance energy efficiency with performance. Through simulations, the authors demonstrate significant improvements in delay, cost, satisfaction, ECR, failure rate, and throughput, providing a scalable solution to energy efficiency and resource allocation challenges in edge computing and IoT.

In ~\cite{mays2023federated}, Mays et al. propose iBADMM, a federated optimization algorithm for scheduling fog-native microservice workflows across multi-domain edge-to-cloud ecosystems. The algorithm optimizes workflow placement, mapping, routing, and admission while ensuring QoS thresholds and efficient data supply. By integrating Bender’s decomposition with ADMM, iBADMM decomposes the problem into sub-problems for distributed decision-making, achieving a 15\% improvement in workflow greenness compared to iADMM, thereby enhancing energy efficiency and environmental sustainability. According to ~\cite{yin2024joint}, Yin et al. advocate for integrating energy harvesting techniques into MEC systems to address mobile device battery limitations. Their system model minimizes the weighted sum of energy consumption and computational latency while ensuring stability, leveraging energy harvesting and DVFS for energy optimization. The authors develop an online Lyapunov optimization-based dynamic task allocation (LODTA) algorithm, which adjusts CPU frequency and transmission power without requiring future state predictions. Simulations demonstrate LODTA’s effectiveness in balancing energy consumption, latency, and battery stability, ensuring efficient task backlogs and green computing in MEC systems.

The articles ~\cite{raju2023delay,zhao2024ets,zou2024fine,syrigos2023eelas} not only focus on their main algorithms but also explore mathematical methods. For instance, the first article employs non-convex MINLP and fuzzy-based RL, while the second article utilizes an M/G/1/PR queuing model combined with a DDPG algorithm to optimize task scheduling, energy consumption, and QoS in fog and edge computing environments. Additionally, the third article incorporates Lyapunov theory, Lagrangian relaxation, and dynamic programming, whereas the fourth article leverages integer linear programming (ILP) to enhance energy efficiency, latency, and resource allocation in edge-edge collaboration and cloud-to-things continuum environments.

\begin{table}
\centering
\caption{Key optimization objectives and techniques in task scheduling}
\label{table:tb03}
\tiny
\begin{tabular}{lllccccclccccll}
\multicolumn{1}{c}{\multirow{2}{*}{Row}} & 
\multicolumn{1}{c}{\multirow{2}{*}{CatName}} &
\multicolumn{1}{c}{\multirow{2}{*}{Year}} & 
\multicolumn{5}{c}{Objectives} &  & 
\multicolumn{4}{c}{Algorithm} & 
\multirow{2}{*}{Envrionment} &  \\ 
\cline{4-8}\cline{10-13}
\multicolumn{1}{c}{} & 
\multicolumn{1}{c}{} & 
\multicolumn{1}{c}{} & 
LT & 
CT & 
LB & 
QoS & 
TP &  & 
MB & 
ML & 
HB & 
MH &  &   \\ \hline\hline
1 & Hazra et al.~\cite{hazra2020joint} & 2020 & \checkmark & - & - & \checkmark & - &  & \checkmark & - & \checkmark & - & Fog &  \\ \hline
2 & Zhao et al.~\cite{zhao2021dynamic} & 2021 & - & \checkmark & - & - & - &  & \checkmark & - & - & - & Edge &  \\ \hline
3 & Luo et al.~\cite{luo2021optimization} & 2021 & - & - & - & - & - &  & \checkmark & - & - & - & Edge &  \\ \hline
4 & Samanta et al.~\cite{samanta2020dyme} & 2020 & \checkmark & \checkmark & - & \checkmark & \checkmark &  & \checkmark & - & - & - & Edge &  \\ \hline
5 & Mays et al.~\cite{mays2023federated} & 2023 & - & \checkmark & - & - & - &  & \checkmark & - & - & - & Fog &  \\ \hline
6 & Yin et al.~\cite{yin2024joint} & 2024 & \checkmark & - & - & - & - &  & \checkmark & - & - & - & Edge &  \\ \hline
7 & Raju et al.~\cite{raju2023delay} & 2023 & \checkmark & - & - & \checkmark & - &  & \checkmark & \checkmark & - & - & Fog &  \\ \hline
8 & Zhao et al.~\cite{zhao2024ets} & 2024 & \checkmark & - & - & - & - &  & \checkmark & \checkmark & - & - & Edge &  \\ \hline
9 & Shruthi et al.~\cite{shruthi2022mayfly} & 2022 & - & \checkmark & - & \checkmark & - &  & - & \checkmark & - & \checkmark & Hybrid &  \\ \hline
10 & Lakhan et al.~\cite{lakhan2022federated} & 2022 & \checkmark & - & - & - & - &  & - & \checkmark & \checkmark & - & Fog &  \\ \hline
11 & Vijayalakshmi et al.~\cite{vijayalakshmi2023reinforcement} & 2023 & - & - & - & \checkmark & - &  & - & \checkmark & \checkmark & - & Fog &  \\ \hline
12 & Murtaza et al.~\cite{murtaza2020qos} & 2020 & \checkmark & - & - & - & - &  & - & \checkmark & - & - & Fog &  \\ \hline
13 & Tuli et al.~\cite{tuli2021cosco} & 2021 & \checkmark & - & - & \checkmark & - &  & - & \checkmark & - & - & Fog &  \\ \hline
14 & Seo et al.~\cite{seo2021slo} & 2021 & - & - & - & \checkmark & \checkmark &  & - & \checkmark & - & - & Edge &  \\ \hline
15 & Sellami et al.~\cite{sellami2022deep} & 2022 & \checkmark & - & \checkmark & - & \checkmark &  & - & \checkmark & - & - & Edge &  \\ \hline
16 & Wang et al.~\cite{wang2023cloud} & 2023 & \checkmark & - & - & - & - &  & - & \checkmark & - & - & Edge &  \\ \hline
17 & Filinis et al.~\cite{filinis2024intent} & 2024 & - & \checkmark & - & \checkmark & - &  & - & \checkmark & - & - & Edge &  \\ \hline
18 & Zou et al.~\cite{zou2024fine} & 2024 & \checkmark & - & - & - & - &  & \checkmark & - & \checkmark & - & Edge &  \\ \hline
19 & Syrigos et al.~\cite{syrigos2023eelas} & 2023 & \checkmark & - & - & - & - &  & \checkmark & - & \checkmark & - & Edge &  \\ \hline
20 & Ijaz et al.~\cite{ijaz2021energy} & 2021 & - & - & - & - & - &  & - & - & \checkmark & - & Edge &  \\ \hline
21 & Hosseini et al.~\cite{hosseini2022optimized} & 2022 & \checkmark & - & - & \checkmark & - &  & - & - & \checkmark & - & Fog &  \\ \hline
22 & Azizi et al.~\cite{azizi2022deadline} & 2022 & \checkmark & - & - & \checkmark & - &  & - & - & \checkmark & - & Fog &  \\ \hline
23 & Aslanpour et al.~\cite{aslanpour2022energy} & 2022 & \checkmark & - & - & \checkmark & \checkmark &  & - & - & \checkmark & - & Edge &  \\ \hline
24 & Singh et al.~\cite{singh2022energy} & 2022 & \checkmark & - & - & - & - &  & - & - & \checkmark & - & Fog &  \\ \hline
25 & Hosseini et al.~\cite{hosseini2023energy} & 2023 & \checkmark & - & - & - & - &  & - & - & \checkmark & - & Fog &  \\ \hline
26 & Verma et al.~\cite{verma2024lease} & 2024 & \checkmark & - & - & \checkmark & - &  & - & - & \checkmark & - & Edge &  \\ \hline
27 & Huang et al.~\cite{huang2023satedge} & 2023 & \checkmark & - & - & \checkmark & - &  & - & - & \checkmark & - & Edge &  \\ \hline
28 & Ali et al.~\cite{ali2024edgebus} & 2024 & \checkmark & - & - & - & - &  & - & - & \checkmark & - & Edge &  \\ \hline
29 & Krishnan et al.~\cite{krishnan2024reliability} & 2024 & \checkmark & \checkmark & - & \checkmark & - &  & - & - & - & \checkmark & Fog &  \\ \hline
30 & Abdel-Basset et al.~\cite{abdel2020energy} & 2020 & - & - & - & - & - &  & - & - & - & \checkmark & Fog &  \\ \hline
31 & Abdel-Basset et al.~\cite{abdel2020energy2} & 2020 & - & \checkmark & - & - & - &  & - & - & - & \checkmark & Fog &  \\ \hline
32 & Hoseiny et al.~\cite{hoseiny2021pga} & 2021 & \checkmark & - & - & \checkmark & - &  & - & - & - & \checkmark & Cloud-Fog &  \\ \hline
33 & Dabiri et al.~\cite{dabiri2022optimizing} & 2022 & \checkmark & - & - & - & - &  & - & - & - & \checkmark & Cloud-Fog &  \\ \hline
34 & Bali et al.~\cite{bali2023effective} & 2023 & \checkmark & - & - & - & - &  & - & - & - & \checkmark & Edge &  \\ \hline
35 & Karami et al.~\cite{karami2024bi} & 2024 & \checkmark & \checkmark & - & \checkmark & - &  & - & - & - & \checkmark & Cloud-Fog &  \\ \hline
36 & Salimi et al.~\cite{salimi2024greedy} & 2024 & \checkmark & - & - & \checkmark & \checkmark &  & - & - & - & \checkmark & Cloud-Fog &  \\ \hline
37 & Shukla et al.~\cite{shukla2024motors} & 2024 & \checkmark & \checkmark & - & - & - &  & - & - & \checkmark & \checkmark & Fog &  \\ \hline
38 & Mahapatra et al.~\cite{mahapatra2024energy} & 2024 & \checkmark & - & \checkmark & - & - &  & - & - & \checkmark & \checkmark & Fog &  \\ \hline
39 & Ijaz et al.~\cite{ijaz2025energy} & 2024 & \checkmark & \checkmark & - & - & - &  & - & - & - & \checkmark & Fog &  \\ \hline
\end{tabular}
    \captionsetup{font=footnotesize}  
     \caption*{\footnotesize 
    \textbf{Abbreviations Guide:} 
    LT: Latency, 
    CT: Cost, 
    LB: Load Balancing,  
    TP: Throughput, 
    MB: Mathematical-Based, 
    ML: ML-Based, 
    HB: Heuristic-Based, 
    MH: Metaheuristic-Based.
    }
\end{table}
\subsubsection{ML-based:}
In their study, Raju et al.~\cite{raju2023delay} propose a fuzzy RL-based task scheduling mechanism aimed at significantly reducing service delay and energy consumption in fog-enabled IoT applications. They formulate the scheduling challenge as a MINLP model, focusing on minimizing energy consumption and service time while adhering to deadline and resource constraints through a fuzzy-based RL approach. Zhao et al.~\cite{zhao2024ets} introduce an energy-efficient and QoS-guaranteed edge task scheduling method, termed ETS-DDPG, which optimizes task response time and energy consumption in edge computing environments. By integrating a DDPG algorithm with an M/G/1/PR queuing model, the authors aim to minimize energy costs while ensuring QoS, demonstrating superior performance in reducing both response time and energy consumption.

As detailed by Shruthi et al.~\cite{shruthi2022mayfly}, the integration of the Mayfly Taylor Optimization Algorithm (MTOA) with Deep Q-Network (DQN) addresses dynamic scheduling challenges in fog-cloud computing environments. The optimization focuses on minimizing energy consumption, reducing SLA violations, and optimizing computation costs, employing DQN for energy prediction and VM energy optimization to ensure efficient task scheduling and load distribution. Lakhan et al.~\cite{lakhan2022federated} propose a Federated Learning-Aware Multi-Objective Modeling and Blockchain-Enabled system for IIoT applications, targeting energy consumption and application delay in decentralized fog-cloud networks. Their optimization strategies, including Deadline-Efficient Task Sequencing and Scheduling (DETS), Latency-Efficient Task Scheduling (LETS), and Energy-Efficient Task Scheduling (EETS), are supported by the Blockchain-Enabled Federated Learning Algorithm Framework (DLEBAF) to enhance system performance.

Vijayalakshmi et al.~\cite{vijayalakshmi2023reinforcement} explore a RL-based Multi-Objective Energy-Efficient Task Scheduling (MEETS-RL) model for optimizing energy efficiency and task scheduling in fog-cloud IIoT systems. The model classifies tasks using the ID3 algorithm and employs the First Fit algorithm for task offloading to CDCs, focusing on minimizing energy consumption and maximizing task completion rates through a dynamic RL-based Task Scheduling (RLTS) algorithm. Murtaza et al.~\cite{murtaza2020qos} introduce the Learning Repository fog-cloud (LRFC) technique, an adaptive task scheduling method addressing energy consumption and QoS challenges in fog computing. The system incorporates a smart layer between IoE devices and fog nodes, utilizing rule-based learning and case-based reasoning to allocate resources dynamically, thereby minimizing energy consumption and enhancing QoS through reduced processing delays.

Tuli et al.~\cite{tuli2021cosco} suggest that advanced optimization strategies and simulation frameworks can enhance task placement and management in fog computing environments. They propose a Gradient-based Optimization Strategy using Back-propagation of gradients with respect to Input (GOBI) and a Coupled Simulation and Container Orchestration Framework (COSCO) to optimize QoS parameters, leveraging predictive digital-twin models for improved reactivity and reduced scheduling overhead. Seo et al.~\cite{seo2021slo} address the challenges of scheduling multiple ML inference tasks on edge platforms with heterogeneous processors. Their optimization goals include ensuring bounded response latency, energy efficiency, and high system throughput. The authors employ heterogeneity-aware scheduling policies and a novel model slicing technique to decompose large ML tasks into smaller sub-tasks, facilitating efficient workload allocation to energy-efficient processors.

Sellami et al.~\cite{sellami2022deep} propose a framework integrating DRL with Blockchain and Software-Defined Networking (SDN) to enhance energy efficiency in 5G massive IoT edge networks. Their approach addresses complexity, security, and energy consumption, focusing on reducing energy usage, improving throughput, and minimizing latency through the Asynchronous Actor-Critic Agent (A3C) algorithm combined with Proof-of-Authority (PoA) Blockchain consensus. Wang et al.~\cite{wang2023cloud} advocate for a cloud-native task scheduling strategy to enhance the efficiency of the Internet of Vehicles (IoV). They introduce the cloud-native Vehicular edge computing Framework (CVEF), which optimizes task scheduling using a Multi-Agent DDPG (MADDPG) algorithm within a Markov game framework, thereby minimizing task execution latency and energy consumption. Filinis et al.~\cite{filinis2024intent} propose a transition from centralized to distributed computing continuum systems for managing serverless applications with strict QoS requirements. Their intent-driven orchestration approach utilizes RL for autoscaling and a multi-objective optimization framework for scheduling, aiming to minimize latency, energy consumption, and transformation costs.

\subsubsection{Heuristic-based:}
In their study, Zou et al.~\cite{zou2024fine} address energy-constrained edge-edge collaboration challenges by proposing a Fine-Grained Service Lifetime Optimization (FGSLO) scheme. This approach minimizes system delay while adhering to energy constraints through dynamic control of service lifetimes on edge servers. The problem is decomposed into service placement, lifetime decision, and task scheduling sub-problems, solved using dynamic programming and heuristic methods to optimize energy and delay. Syrigos et al.~\cite{syrigos2023eelas} propose the Energy Efficient and Latency-Aware Scheduling (EELAS) framework to optimize resource allocation across the cloud-to-things continuum. The framework minimizes energy consumption while meeting latency requirements by formulating the problem as an ILP task and employing heuristic methods. Integrated with Kubernetes, EELAS leverages Bender’s decomposition and ADMM for efficient load distribution and near-optimal energy savings.

Ijaz et al.~\cite{ijaz2021energy} introduce the Energy-Makespan Multi-Objective Optimization (EM-MOO) algorithm for workflow scheduling in fog-cloud environments. The algorithm balances energy consumption and application completion time by scheduling latency-sensitive tasks to fog resources and computationally intensive tasks to cloud resources. Using Deadline-Aware Stepwise Frequency Scaling (DAFS) and DVFS, EM-MOO dynamically adjusts processor frequency, reducing energy consumption with minimal impact on makespan. Hosseini et al.~\cite{hosseini2022optimized} propose the Priority Queue, Fuzzy, and Analytical Hierarchy Process (PQFAHP) algorithm for task scheduling in mobile fog computing (MFC). The algorithm prioritizes tasks using a multi-criteria decision-making approach, considering completion time, energy consumption, RAM, and deadlines. PQFAHP minimizes energy consumption, average waiting time, and delay while maximizing service level, addressing dynamic scheduling criteria in MFC environments.

Azizi et al.~\cite{azizi2022deadline} propose two semi-greedy-based algorithms, Priority-aware Semi-Greedy (PSG) and PSG with Multistart Procedure (PSG-M), for task scheduling in fog computing. These algorithms optimize energy consumption and minimize deadline violation times by incorporating controlled randomness to avoid local optima. The approach enhances load distribution, resource utilization, and task completion timeliness in real-time IoT applications. Aslanpour et al.~\cite{aslanpour2022energy} introduce energy-aware resource scheduling techniques for serverless edge computing, focusing on IoT domains powered by renewable energy. The solution employs zone-oriented and priority-based algorithms, incorporating sticky offloading and warm scheduling to minimize energy waste and preserve QoS. A heuristic-based greedy algorithm is evaluated on a solar-powered Raspberry Pi cluster, demonstrating operational availability and energy efficiency.

Singh et al.~\cite{singh2022energy} propose an energy-efficient fuzzy data offloading (EEFO) scheme for Internet of Medical Things (IoMT) systems in cloud-fog environments. The scheme classifies medical data into emergency, significant, and general categories using fuzzy logic, prioritizing processing to minimize energy consumption and latency. Dynamic clustering of fog nodes and adaptive scheduling enhance resource utilization and system performance in IoMT applications. Hosseini et al.~\cite{hosseini2023energy} propose an energy-efficient scheduling model for MFC, integrating mobile network processing with edge computing. The model employs a Greedy Knapsack Offloading Algorithm (GKOA) for task offloading and a High Low Priority Scheduling (HLPS) model using dual priority queues. These techniques minimize energy consumption and improve resource allocation in fifth-generation networks.

Verma et al.~\cite{verma2024lease} introduce LEASE, a dynamic, energy-aware resource scheduling framework for serverless edge computing. LEASE employs a priority-based heuristic algorithm to offload functions from over-provisioned to under-provisioned nodes, minimizing energy consumption while meeting microservice deadlines. The framework addresses workload variability and strict latency constraints in edge environments. Huang et al.~\cite{huang2023satedge} propose SatEdge, a satellite edge cloud platform, and the OMS-AM microservice scheduling algorithm. OMS-AM optimizes task processing latency, energy consumption, and task failure rates in satellite-terrestrial networks. Leveraging KubeEdge, the algorithm adapts to dynamic network topologies, enhancing satellite edge computing efficiency and reliability. Ali et al.~\cite{ali2024edgebus} propose EdgeBus, a framework for resource management in heterogeneous MEC environments. The framework employs MANGO, a lightweight scheduler, to minimize latency, energy consumption, and container migrations. Integrated with Kubernetes, EdgeBus uses co-simulation to evaluate container migration impacts, balancing performance and resource utilization in MEC environments.

The articles~\cite{hazra2020joint},~\cite{lakhan2022federated},~\cite{vijayalakshmi2023reinforcement},~\cite{shukla2024motors}, and~\cite{mahapatra2024energy} primarily utilize their main algorithms while integrating various heuristic methods to enhance performance in fog-cloud systems. For instance,~\cite{hazra2020joint} employs a priority-aware queueing assignment policy to improve energy efficiency and task offloading. The first two articles~\cite{lakhan2022federated} and~\cite{vijayalakshmi2023reinforcement} incorporate techniques such as deadline-efficient task sequencing (DETS), latency-efficient task scheduling (LETS), energy-efficient task scheduling (EETS), the ID3 algorithm for task classification, and the first fit algorithm for task offloading to optimize resource management. Additionally,~\cite{shukla2024motors} combines a hybrid Harmony Search Algorithm (HSA) with a GA, while~\cite{mahapatra2024energy} utilizes a Binary Linear-Weight JAYA (BLWJAYA) algorithm, both aimed at optimizing task offloading, resource scheduling, and load balancing.

\subsubsection{Metaheuristic-based:}
In ~\cite{krishnan2024reliability}, Krishnan et al. propose enhancing IoT integration with cloud and fog computing through dynamic optimization scheduling. The study introduces the Dynamic Mayfly Optimization Scheduling (DMOS) algorithm, combining PSO, GA, and Firefly Algorithm (FA) to optimize resource allocation and task scheduling. The objectives include minimizing energy consumption, meeting task deadlines, and ensuring system reliability in multi-agent microservice cloud-fog architectures. In ~\cite{abdel2020energy}, Abdel et al. investigate an energy-aware task scheduling model using the Marine Predators Algorithm (MPA) to improve QoS in IoT-based fog computing. The study focuses on optimizing energy efficiency, reducing makespan, flow time, and carbon emissions. Two modified versions of MPA—Modified MPA (MMPA) and Improved MMPA (IMMPA)—are proposed, with IMMPA demonstrating superior performance in energy efficiency and workload distribution for IoT-based fog computing systems.

According to ~\cite{abdel2020energy2}, Abdel et al. highlight fog computing as a solution to enhance QoS in IIoT applications by addressing high latency and bandwidth demands in cloud computing. The authors propose the HHOLS algorithm, integrating local search and swap mutation for efficient task scheduling in fog computing. This approach employs a virtualized layered fog computing model, normalization, and scaling to optimize energy consumption and makespan, improving workload distribution and system performance. As outlined in ~\cite{bali2023effective}, Bali et al. propose the Priority-aware Task Scheduling (PaTS) algorithm to manage data from industrial sensors in IoT applications. The algorithm prioritizes tasks based on urgency, leveraging edge and cloud servers to minimize queue delay, computation time, and energy consumption, thereby optimizing performance in IoT-based systems. In ~\cite{shukla2024motors}, Shukla et al. introduce the Multi-Objective Task Offloading and Resource Scheduling (MOTORS) algorithm for heterogeneous fog-cloud environments. The algorithm employs fuzzy dominance-based task clustering and offloading (FDTCO) and a hybrid optimization-based resource scheduling algorithm (HORSA) to minimize makespan, cost, and energy consumption while maximizing resource utilization.

According to ~\cite{mahapatra2024energy}, Mahapatra et al. propose an energy-aware task offloading and load balancing strategy for latency-sensitive IoT applications in the fog-cloud continuum. The study introduces a Fuzzy logic algorithm for task offloading and the Binary Linear-Weight JAYA (BLWJAYA) algorithm for scheduling, dynamically adjusting weights to balance exploration and exploitation, thereby improving resource utilization and energy efficiency. In ~\cite{ijaz2025energy}, Ijaz et al. propose an energy-efficient time and cost (ETC) constraint scheduling algorithm for fog computing integrated with IoT. The algorithm leverages an improved multi-objective differential evolution (I-MODE) meta-heuristic to minimize execution time, energy consumption, and cost, incorporating deadline-aware frequency scaling to optimize energy usage without compromising deadlines.

In ~\cite{hoseiny2021pga}, Hoseiny et al. address task scheduling in heterogeneous fog-cloud environments by formulating a multi-objective optimization problem. Their priority-aware GA (PGA) integrates task classification and genetic optimization to minimize computation time and energy consumption while maximizing deadline-satisfied tasks. In ~\cite{dabiri2022optimizing}, Dabiri et al. propose nature-inspired optimization techniques—grey wolf optimization (GWO) and grasshopper optimization algorithm (GOA)—for job scheduling in hybrid fog-cloud systems. These algorithms aim to minimize deadline violation time and energy consumption by efficiently assigning tasks to computing nodes.

In ~\cite{karami2024bi}, Karami et al. address workflow scheduling in virtualized fog-cloud computing (VFCC) systems by formulating a bi-objective optimization model. The proposed two-phase method combines NSGA-II with a semi-greedy initialization and an energy-aware heuristic algorithm to balance makespan and energy consumption. In ~\cite{salimi2024greedy}, Salimi et al. propose a Greedy Randomized Adaptive Search Procedure (GRASP) for scheduling IoT tasks in virtualized fog-cloud environments. The approach aims to enhance deadline satisfaction and minimize energy consumption by optimizing task assignment to virtualized resources.

The article~\cite{shruthi2022mayfly} primarily relies on its main algorithm but also incorporates a metaheuristic method, the Mayfly Taylor Optimisation Algorithm (MTOA), to enhance dynamic scheduling and optimize energy consumption, SLA violation, and computation cost in fog-cloud computing environments.

\begingroup
\centering
\tiny
\begin{longtable}{
    >{\RaggedRight}m{0.5cm} 
    >{\RaggedRight}m{0.5cm} 
    >{\RaggedRight}m{0.5cm} 
    >{\RaggedRight}m{0.5cm} 
    >{\RaggedRight}m{4cm}
    >{\RaggedRight}m{4cm} 
    >{\RaggedRight}m{4cm}
}
\caption{Advantages \& disadvantages for scheduling reviews}
\label{table:add03}\\ 
\hline
\multirow{2}{*}{Article} & 
\multicolumn{3}{c}{\textbf{Energy}} & 
\multirow{2}{*}{Main Idea} & 
\multirow{2}{*}{Advantages} & 
\multirow{2}{*}{Disadvantages} \\ 
 & CE & CoE & SE & & & \\ \hline
\endhead 
\hline
\multicolumn{7}{r}{{Continued on next page}} \\ \hline
\endfoot 
\hline
\multicolumn{7}{l}{
\footnotesize
\textbf{Abbreviations:} 
CE = Computation Energy, 
CoE = Communication Energy, 
SE = Storage Energy.
} \\
\endlastfoot 

~\cite{hazra2020joint} & \circlefill{100} & \circlefill{75} & \circlefill{0} & Proposes EETO policy for energy-performance trade-offs in IoT using Lyapunov optimization & Achieves significant energy savings, reduces queue waiting time, and improves throughput in fog networks & Does not optimize multiple QoS parameters, lacks reward-penalty strategies for resource configurations\\ \hline
~\cite{zhao2021dynamic} & \circlefill{100} & \circlefill{75} & \circlefill{0} & Introduces DORS algorithm for MEC with energy harvesting, minimizing energy and resource use while meeting QoS & DORS algorithm balances cost and performance, ensures queue stability, and adapts to varying MEC workloads & Assumes limitless cloud resources, ignores user interference, and lacks real-world applicability\\ \hline
~\cite{luo2021optimization} & \circlefill{25} & \circlefill{100} & \circlefill{0} & Optimizes task scheduling, bit allocation, and UAV trajectory in multi-UAV MEC to reduce user energy consumption & Two-layer optimization reduces user energy, avoids UAV conflicts, and outperforms greedy and random strategies & Scalability issues with users and UAVs, assumes fixed UAV altitude, and ignores other QoS metrics\\ \hline
~\cite{samanta2020dyme} & \circlefill{25} & \circlefill{50} & \circlefill{25} & Proposes microservice scheduling for MEC IoT platforms to optimize delay, energy, QoS, and network price & Improves network throughput, resource use, QoS, and reduces energy and failure rates, outperforming baselines & Ignores privacy concerns, lacks real-world prototype for validation\\ \hline
~\cite{mays2023federated} & \circlefill{100} & \circlefill{25} & \circlefill{0} & Introduces iBADMM for fog-native microservice workflows, focusing on green energy and intent-awareness & Minimizes data sharing, improves workflow greenness, and incorporates actor intents into scheduling & Depends on green energy variability, struggles in high-cost edge environments\\ \hline
~\cite{yin2024joint} & \circlefill{100} & \circlefill{50} & \circlefill{0} & Proposes joint task allocation and offloading in MEC with EH to minimize energy and latency while ensuring stability & Introduces LODTA for dynamic task allocation, reduces task loss and latency, and balances energy and delay & Focuses on single mobile device, lacks multi-device energy and latency optimization, and scalability for large systems\\ \hline
~\cite{raju2023delay} & \circlefill{100} & \circlefill{0} & \circlefill{0} & Combines fuzzy logic and SARSA reinforcement learning to minimize delay and energy under deadlines & Reduces latency and energy, uses fuzzy logic for prioritization, and ensures deadline compliance & Ignores device mobility and fog server expansion, focuses only on task scheduling\\ \hline
~\cite{zhao2024ets} & \circlefill{100} & \circlefill{0} & \circlefill{0} & Introduces ETS-DDPG for edge task scheduling using DDPG and M/G/1/PR queuing to optimize response time and energy & Dynamic server power modeling, integration of real-time electricity prices, simultaneous optimization of energy and performance using deep reinforcement learning & Does not address failure rates, ignores task offloading and migration success rates\\ \hline
~\cite{shruthi2022mayfly} & \circlefill{100} & \circlefill{25} & \circlefill{0} & Develops MTOA with DQN for fog-cloud resource management, focusing on energy, SLA violations, and cost & MTOA-based DQN improves energy, SLA violations, and cost, suitable for real-time data centers & Lacks data privacy/security, focuses only on scheduling, ignoring placement and provisioning\\ \hline
~\cite{lakhan2022federated} & \circlefill{100} & \circlefill{25} & \circlefill{0} & Proposes federated learning-aware multi-objective modeling with blockchain for IIoT energy and delay optimization & Improves energy efficiency, reduces training delay, and ensures secure data sharing via blockchain & No mobility support, limited application scope, and increased complexity with blockchain and federated learning\\ \hline
~\cite{vijayalakshmi2023reinforcement} & \circlefill{100} & \circlefill{50} & \circlefill{0} & Introduces MEETS-RL for fog-cloud IIoT to optimize task completion, resource use, and energy efficiency & Achieves high task scheduling accuracy, reduces energy, and outperforms FCFS, GfE, SJF, RR, and LBP-ACS & Ignores VM reliability, introduces overhead in real-time applications, and lacks dynamic adaptation\\ \hline
~\cite{murtaza2020qos} & \circlefill{100} & \circlefill{25} & \circlefill{0} & Proposes LRFC for adaptive task scheduling in fog computing to enhance QoS and energy efficiency & Introduces smart layer for efficient resource allocation, reduces power consumption, and uses formal verification & Limited scalability testing, ignores network conditions, and lacks real-world deployment\\ \hline
~\cite{tuli2021cosco} & \circlefill{100} & \circlefill{25} & \circlefill{0} & Introduces GOBI and COSCO for fog task placement, optimizing energy and response time & Improves QoS (energy, response time, scheduling time), adapts quickly, and combines simulation with AI & High computational overhead, lacks formal guarantees, and requires advanced infrastructure\\ \hline
~\cite{seo2021slo} & \circlefill{100} & \circlefill{0} & \circlefill{0} & Proposes SLO-aware inference scheduling for edge platforms to optimize ML model mapping, latency, and energy & Reduces latency and SLO violations, achieves performance speedup, and lowers violation rates & Limited fairness between tasks, effectiveness drops in high request rates or overloaded systems\\ \hline
~\cite{sellami2022deep} & \circlefill{100} & \circlefill{50} & \circlefill{0} & Integrates DRL, blockchain, and SDN for energy-aware task scheduling in 5G IoT edge networks & Improves energy efficiency, reduces latency, and uses blockchain for secure transactions & Lacks interoperability between distributed ledgers, no federated learning for scalability, and increased complexity\\ \hline
~\cite{wang2023cloud} & \circlefill{50} & \circlefill{50} & \circlefill{0} & Proposes multi-agent RL for IoV task scheduling to minimize latency and energy & Integrates task scheduling, resource allocation, and microservice deployment, reducing time and energy & High computational complexity, andlimited scalability\\ \hline
~\cite{filinis2024intent} & \circlefill{100} & \circlefill{25} & \circlefill{0} & Introduces intent-driven orchestration for serverless apps using RL for autoscaling and multi-objective optimization & Reduces QoS violations, total cost, and power consumption, managing dynamic workloads with RL & Over-provisions resources, and limited scalability evaluation\\ \hline
~\cite{zou2024fine} & \circlefill{100} & \circlefill{25} & \circlefill{0} & Proposes FGSLO for edge-edge collaboration to minimize delay under energy constraints & Reduces task delays, optimizes resource use, and operates without cloud dependency & No scalability analysis, ignores network conditions, and focuses only on energy constraints\\ \hline
~\cite{syrigos2023eelas} & \circlefill{100} & \circlefill{0} & \circlefill{0} & Introduces EELAS for ML inference workload deployment, optimizing energy and latency & Reduces energy, improves latency and resource use, and integrates with Kubernetes for real-world deployment & No ML training support, lacks dynamic workload profiling, and ignores workload migration\\ \hline
~\cite{ijaz2021energy} & \circlefill{100} & \circlefill{25} & \circlefill{0} & Proposes EM-MOO for fog-cloud workflow scheduling, balancing energy and makespan & Reduces energy, uses deadline-aware frequency scaling, and balances energy and makespan & No exploration of advanced meta-heuristics\\ \hline
~\cite{hosseini2022optimized} & \circlefill{100} & \circlefill{75} & \circlefill{0} & Introduces PQFAHP for mobile fog computing to optimize cost-latency trade-offs using fuzzy AHP & Reduces waiting time, delay, and energy, optimizes resource use, and combines multiple criteria for prioritization & Scalability issues, lacks forecasting system, and limited scope to task scheduling\\ \hline
~\cite{azizi2022deadline} & \circlefill{100} & \circlefill{0} & \circlefill{0} & Proposes PSG and PSG-M for deadline-aware, energy-efficient task scheduling in fog computing & Improves deadline compliance, reduces energy, and minimizes violation time & No support for dependent tasks, limited task types, and single-task execution assumption\\ \hline
~\cite{aslanpour2022energy} & \circlefill{100} & \circlefill{25} & \circlefill{0} & Introduces energy-aware resource scheduling for serverless edge computing to enhance renewable energy use & Increases bottleneck node availability, reduces energy variance, and introduces sticky offloading and warm scheduling & Limited scalability, no heterogeneity consideration, and narrow focus on CPU-intensive tasks\\ \hline
~\cite{singh2022energy} & \circlefill{100} & \circlefill{75} & \circlefill{25} & Proposes EEFO in cloud-fog-IoMT using fuzzy logic and adaptive scheduling for energy, latency, and load balancing & Reduces energy, improves latency, and ensures efficient load balancing & Ignores network failures, multiple protocols cause delays, and high architectural complexity\\ \hline
~\cite{hosseini2023energy} & \circlefill{100} & \circlefill{75} & \circlefill{0} & Introduces GKOA and HLPS for energy-efficient scheduling in mobile fog computing & Improves energy efficiency, reduces overhead and delay, and optimizes task prioritization and resource allocation & No data security, lacks quick offloading system, and limited focus on broader resource management\\ \hline
~\cite{verma2024lease} & \circlefill{100} & \circlefill{25} & \circlefill{0} & Proposes LEASE for serverless edge computing to optimize latency-sensitive IoT services & Reduces completion time, failure rates, and ensures deadlines are met while optimizing energy & No deadline variation consideration, limited exploration of real-time scheduling thresholds\\ \hline
~\cite{huang2023satedge} & \circlefill{100} & \circlefill{75} & \circlefill{0} & Introduces SatEdge and OMS-AM for satellite-terrestrial networks to optimize latency, energy, and task failure rates & SatEdge provides scalable edge cloud platform, OMS-AM reduces latency, and offers comprehensive architecture & Limited experimental validation, struggles with complex mobility, and reliability concerns\\ \hline
~\cite{ali2024edgebus} & \circlefill{100} & \circlefill{0} & \circlefill{0} & Proposes EdgeBus for MEC resource management, optimizing container placement, latency, and migration costs & Addresses user mobility and resource heterogeneity, improves latency, and reduces container migrations & Ignores variable workloads, no stateful migration, and limited real-world application testing\\ \hline
~\cite{krishnan2024reliability} & \circlefill{100} & \circlefill{50} & \circlefill{25} & Introduces DMOS for IoT task scheduling in fog-cloud, minimizing energy while maintaining reliability & DMOS outperforms in energy efficiency, reduces latency, and optimizes resource use in cloud-fog environments & Scalability challenges, resource overhead, and integration complexity in IoT systems\\ \hline
~\cite{abdel2020energy} & \circlefill{100} & \circlefill{0} & \circlefill{0} & Proposes MPA for energy-aware task scheduling in IoT fog computing, optimizing energy, makespan, and emissions & IMMPA outperforms in energy efficiency, makespan, and environmental impact, reducing $CO_2$ emissions & No task dependency consideration, lacks real-world testing, and narrow focus on task scheduling\\ \hline
~\cite{abdel2020energy2} & \circlefill{100} & \circlefill{0} & \circlefill{0} & Introduces HHOLS for task scheduling in fog computing to enhance IIoT QoS & Outperforms in energy, makespan, cost, flow time, and $CO_2$ emission rate, offering economic and environmental benefits & Limited to independent tasks, time-consuming local search, and no task migration support\\ \hline
~\cite{hoseiny2021pga} & \circlefill{100} & \circlefill{0} & \circlefill{0} & optimizing task scheduling in fog-cloud computing with PGA& A holistic modeling of server energy and a joint optimization with migration cost and delay are performed & Does not consider task dependencies or network topology (noted as future work)\\ \hline
~\cite{dabiri2022optimizing} & \circlefill{100} & \circlefill{0} & \circlefill{0} & optimizing job scheduling in fog-cloud using GWO and GOA. & Deadline violation time and energy consumption are reduced, with the aid of a detailed computational energy model & High running time, especially for latency-sensitive jobs
\\ \hline
~\cite{bali2023effective} & \circlefill{100} & \circlefill{50} & \circlefill{0} & Introduces PaTS for task offloading and scheduling in industrial IoT, minimizing delay and energy & Reduces queue delay, computation time, and energy, prioritizes tasks, and uses NSGA-II for optimization & No real-time dynamic offloading, and limited scalability testing\\ \hline
~\cite{karami2024bi} & \circlefill{100} & \circlefill{0} & \circlefill{0} & optimizing workflow scheduling in VFCC using NSGA-II.& Reduces makespan and energy consumption, outperforms HEFT, Green-HEFT, MOHEFT, and MOACO in quality metrics& complexity of NSGA-II and heuristic combination may increase computational overhead.
\\ \hline
~\cite{salimi2024greedy} & \circlefill{100} & \circlefill{0} & \circlefill{0} & optimizing IoT task scheduling in VFCC using GRASP& Outperforms existing methods in deadline satisfaction, response time, energy consumption, and makespan; simple and easy to implement.& Limited realism (e.g., no diverse task initiation times or varying CPU power scales)\\ \hline
~\cite{shukla2024motors} & \circlefill{100} & \circlefill{25} & \circlefill{0} & Proposes MOTORS for multi-objective task offloading in fog-cloud, optimizing makespan, cost, and energy & Balances makespan, cost, and resource use, improves performance metrics, and uses hybrid GA + HS approach & No deadline or cost constraint consideration, limited optimization for communication-intensive tasks, and scalability issues\\ \hline
~\cite{mahapatra2024energy} & \circlefill{100} & \circlefill{50} & \circlefill{25} & Introduces energy-aware offloading and load balancing for IoT in fog-cloud, reducing latency and energy & Improves resource use, service rate, latency, energy, and load balancing, uses fuzzy logic and JAYA algorithm & Ignores task dependency, limited scalability discussion, and fuzzy logic overhead\\ \hline
~\cite{ijaz2025energy} & \circlefill{100} & \circlefill{0} & \circlefill{0} & Proposes ETC using I-MODE for fog workflow scheduling, optimizing time, energy, and cost & Reduces energy, execution time, and cost, uses DVFS for optimization, and scales well with workflow size & Limited energy optimization scope, no real-time adaptation, and high computational complexity\\ \hline
 
\end{longtable}
\endgroup

\subsection{Resource Allocation}
Resource allocation in MEC and fog computing is a fundamental process that ensures the fair and efficient distribution of computational, storage, and network resources among competing applications and users. Effective resource allocation is essential for maintaining high system performance, meeting QoS requirements, and ensuring energy efficiency in resource-constrained edge and fog environments.

The challenges of resource allocation are compounded by the dynamic and distributed nature of these environments, where resources are often limited, heterogeneous, and subject to frequent changes in demand. Moreover, the need to support diverse applications with varying latency, bandwidth, and computational requirements further complicates the allocation process. In this section, we review and classify existing research on resource allocation strategies, focusing on Mathematical-based, ML-based, Heuristic-based, and Metaheuristic-based approaches, as well as their effectiveness in addressing the unique challenges of MEC and fog computing, as shown in Table~\ref{table:tb04}. The advantages and disadvantages of each method are further explored in Table~\ref{table:add04}.

\begin{table}
\centering
\caption{Summary of optimization approaches in resource allocation}
\label{table:tb04}
\tiny
\begin{tabular}{lllccccclccccll}
\multicolumn{1}{c}{\multirow{2}{*}{Row}} & 
\multicolumn{1}{c}{\multirow{2}{*}{CatName}} & 
\multicolumn{1}{c}{\multirow{2}{*}{Year}} &
\multicolumn{5}{c}{Objectives} &  & \multicolumn{4}{c}{Algorithm} & \multirow{2}{*}{Envrionment} &  \\ \cline{4-8}\cline{10-13}
\multicolumn{1}{c}{} & \multicolumn{1}{c}{} & 
\multicolumn{1}{c}{} & 
LT & 
CT & 
LB & 
QoS & 
TP &  & 
MB & 
ML & 
HB & 
MH &  &  
\\ \hline\hline
1 & Zhao et al.~\cite{zhao2021energy} & 2021 & \checkmark & - & - & \checkmark & - &  & \checkmark & - & - & - & Edge &  \\ \hline
2 & Chang et al.~\cite{chang2020dynamic} & 2020 & \checkmark & \checkmark & - & - & - &  & \checkmark & - & - & - & Fog &  \\ \hline
3 & Zhang et al.~\cite{zhang2020dynamic} & 2020 & \checkmark & - & - & - & - &  & \checkmark & - & - & - & Edge &  \\ \hline
4 & Li et al.~\cite{li2020energy} & 2020 & - & - & - & - & - &  & \checkmark & - & - & - & Edge &  \\ \hline
5 & Alharbi et al.~\cite{alharbi2022energy} & 2022 & - & \checkmark & - & - & - &  & \checkmark & - & - & - & Cloud-Fog &  \\ \hline
6 & Mohajer et al.~\cite{mohajer2022heterogeneous} & 2022 & - & - & - & - & \checkmark &  & \checkmark & - & - & - & Edge &  \\ \hline
7 & Zhou et al.~\cite{zhou2022service} & 2022 & \checkmark & - & - & - & - &  & \checkmark & - & - & - & Edge &  \\ \hline
8 & Bebortta et al.~\cite{bebortta2023optimal} & 2023 & \checkmark & - & - & - & - &  & \checkmark & - & - & - & Fog &  \\ \hline
9 & Vahabi et al.~\cite{vahabi2023energy} & 2023 & - & - & - & \checkmark & - &  & \checkmark & - & - & - & Edge &  \\ \hline
10 & Fé et al.~\cite{fe2023model} & 2023 & - & - & - & - & - &  & \checkmark & - & - & - & Cloud-Fog &  \\ \hline
11 & Sun et al.~\cite{sun2020energy} & 2020 & \checkmark & \checkmark & - & - & - &  & \checkmark & - & \checkmark & - & Fog &  \\ \hline
12 & Qu et al.~\cite{qu2021service} & 2021 & - & - & - & - & - &  & \checkmark & - & \checkmark & - & Edge &  \\ \hline
13 & Bellal et al.~\cite{bellal2024gas} & 2024 & \checkmark & - & - & - & - &  & \checkmark & - & - & - & Edge &  \\ \hline
14 & Tong et al.~\cite{tong2020adaptive} & 2020 & \checkmark & - & - & - & - &  & - & \checkmark & - & - & Edge &  \\ \hline
15 & Zhu et al.~\cite{zhu2021energy} & 2021 & - & - & - & - & - &  & - & \checkmark & - & - & Fog &  \\ \hline
16 & Seid et al.~\cite{seid2021multi} & 2021 & \checkmark & \checkmark & - & \checkmark & - &  & - & \checkmark & - & - & Edge &  \\ \hline
17 & Santos et al.~\cite{santos2021reinforcement} & 2021 & - & \checkmark & - & - & - &  & - & \checkmark & - & - & Fog &  \\ \hline
18 & Santos et al.~\cite{santos2021resource} & 2021 & \checkmark & \checkmark & - & \checkmark & - &  & - & \checkmark & - & - & Fog &  \\ \hline
19 & Mekala et al.~\cite{mekala2022drl} & 2022 & - & \checkmark & - & \checkmark & - &  & - & \checkmark & - & - & Edge &  \\ \hline
20 & Wang et al.~\cite{wang2023cloud} & 2023 & \checkmark & - & - & - & - &  & - & \checkmark & - & - & Edge &  \\ \hline
21 & Lingayya et al.~\cite{lingayya2024dynamic} & 2024 & \checkmark & - & - & - & - &  & - & \checkmark & \checkmark & \checkmark & Edge &  \\ \hline
22 & Xu et al.~\cite{xu2020energy} & 2020 & \checkmark & - & - & \checkmark & - &  & \checkmark & - & \checkmark & - & Edge &  \\ \hline
23 & Fan et al.~\cite{fan2021cloud} & 2021 & \checkmark & \checkmark & - & - & - &  & \checkmark & - & \checkmark & - & Edge &  \\ \hline
24 & Yadav et al.~\cite{yadav2020energy} & 2020 & \checkmark & - & - & - & - &  & - & - & \checkmark & - & Edge &  \\ \hline
25 & Mahmud et al.~\cite{mahmud2021pi} & 2021 & \checkmark & - & - & - & - &  & - & - & \checkmark & - & Fog &  \\ \hline
26 & Kumar et al.~\cite{kumar2024auction} & 2024 & - & \checkmark & - & - & - &  & - & - & \checkmark & - & Fog &  \\ \hline
27 & Yuan et al.~\cite{yuan2020profit} & 2020 & \checkmark & \checkmark & \checkmark & \checkmark & - &  & - & - & \checkmark & \checkmark & Edge &  \\ \hline
28 & Li et al.~\cite{li2021computation} & 2021 & \checkmark & - & - & - & - &  & - & - & - & \checkmark & Edge &  \\ \hline
29 & Abbasi et al.~\cite{abbasi2021intelligent} & 2021 & \checkmark & - & - & - & - &  & - & - & - & \checkmark & Fog &  \\ \hline
30 & Cao et al.~\cite{cao2021large} & 2021 & \checkmark & \checkmark & \checkmark & \checkmark & - &  & - & - & - & \checkmark & Edge &  \\ \hline
31 & Apat et al.~\cite{apat2024hybrid} & 2024 & - & \checkmark & - & - & - &  & - & - & - & \checkmark & Fog &  \\ \hline
32 & Liu et al.~\cite{liu2024multi} & 2024 & \checkmark & - & \checkmark & - & - &  & - & - & - & \checkmark & Edge &  \\ \hline
33 & Ghasemzadeh et al.~\cite{ghasemzadeh2025optimizing} & 2024 & \checkmark & - & \checkmark & - & - &  & - & - & - & \checkmark & Edge &  \\ \hline
\end{tabular}

    \captionsetup{font=footnotesize}  
     \caption*{\footnotesize 
    \textbf{Abbreviations Guide:} 
    LT: Latency, 
    CT: Cost, 
    LB: Load Balancing,  
    TP: Throughput, 
    MB: Mathematical-Based, 
    ML: ML-Based, 
    HB: Heuristic-Based, 
    MH: Metaheuristic-Based.
    }
\end{table}
\subsubsection{Mathematical-based:}
In ~\cite{zhao2021energy}, Zhao et al. address energy minimization in MEC systems while meeting latency constraints. They decompose the problem into offloading ratio selection, transmission power optimization, and resource allocation, solving them iteratively using the Block Coordinate Descent (BCD) method. Advanced techniques like parametric convex programming and the primal-dual method ensure energy efficiency through closed-form solutions. In ~\cite{chang2020dynamic}, Chang et al. propose a fog computing system to enhance IoT performance by offloading tasks to fog nodes (FNs). The authors introduce a dynamic optimization scheme coordinating radio and computational resources, minimizing system costs, including latency and energy consumption. energy harvesting is integrated to extend mobile device lifetimes, with evaluations validating the scheme’s efficiency in improving system performance.

Zhang et al. in ~\cite{zhang2020dynamic} propose dynamic task offloading and resource allocation in MEC-enabled dense cloud Radio Access Networks (C-RAN) to optimize energy efficiency and service delay. They formulate a stochastic mixed-INLP problem, decomposing it into four subproblems using Lyapunov optimization theory. The solution employs convex decomposition methods and matching game theory, achieving superior performance in energy efficiency and service delay. In ~\cite{li2020energy}, Li et al. improve energy efficiency in UAV-assisted MEC through joint optimization of UAV trajectory, user transmit power, and computation load allocation. The study employs the Dinkelbach algorithm, SCA, and ADMM to address the nonconvex problem, integrating spatial distribution estimation to adapt to limited user mobility information, ensuring adaptability under uncertainty.

As outlined in ~\cite{alharbi2022energy}, Alharbi et al. propose a multi-objective optimization framework for energy-efficient UAV-based service offloading over cloud-fog architectures. Using a MILP approach, the model minimizes cost functions such as end-to-end network power consumption and UAV flight distance, prioritizing energy-saving strategies in resource allocation and trajectory planning. In ~\cite{mohajer2022heterogeneous}, Mohajer et al. introduce a dynamic optimization model to enhance energy efficiency and user fairness in NOMA-based MEC systems. The model maximizes uplink/downlink energy efficiency while adhering to QoS constraints, employing a subgradient method for resource allocation and SCA with dual decomposition, achieving significant improvements in energy efficiency and fairness.

Zhou et al. in ~\cite{zhou2022service} integrate dense small cell networks with MEC to address user equipment computational limitations and 5G latency requirements. Their optimization approach minimizes service delay under long-term energy constraints, utilizing a two-layer method combining Gibbs sampling for user equipment assignment and Lyapunov optimization for resource allocation, balancing energy consumption and delay. In ~\cite{bebortta2023optimal}, Bebortta et al. propose an optimal fog-cloud offloading framework to mitigate energy consumption and service latency in heterogeneous IoT networks. The study formulates the problem as a dynamic ILP technique, focusing on task offloading optimization while ensuring timely task completion, demonstrating superior energy efficiency and reduced latency.

As outlined in ~\cite{vahabi2023energy}, Vahabi et al. propose an energy-efficient resource management technique for Function-as-a-Service (FaaS) edge computing platforms. The approach minimizes energy consumption and SLA violations through a system model that dynamically schedules edge nodes based on demand, utilizing linear programming to optimize resource allocation and maintain QoS. In ~\cite{fe2023model}, Fé et al. explore a model-driven approach to evaluate the dependability and energy consumption of Kubernetes-based cloud-fog systems. The authors propose a framework balancing dependability and energy efficiency, employing stochastic modeling and sensitivity analysis to quantify energy usage and assess redundancy impacts, improving resource allocation and reducing environmental impact.

According to ~\cite{sun2020energy}, Sun et al. propose a general IoT-fog-cloud architecture to optimize energy consumption and task completion time. They introduce the Energy and Time Efficient Computation Offloading and Resource Allocation (ETCORA) algorithm, employing dynamic transmission power allocation using a Newton iteration method, demonstrating superior performance in reducing energy consumption and completion time. In ~\cite{qu2021service}, Qu et al. introduce two alternating optimization methods (BnB with SCA, and relaxation with randomized rounding) for efficient computation resource allocation in SPUN, focusing on energy minimization in UAV-enabled MEC systems. According to ~\cite{bellal2024gas}, Bellal et al. propose the GAS framework, enhancing energy efficiency in edge computing microservices (ECM) while meeting stringent latency requirements. The framework leverages DVFS to dynamically adjust CPU frequency and voltage, minimizing energy consumption without compromising latency, ensuring sustainable edge computing in energy-constrained environments.

\subsubsection{ML-based:}
As outlined in ~\cite{tong2020adaptive}, Tong et al. address the increasing demand for data processing in mobile environments by proposing an adaptive task offloading and resource allocation algorithm within a MEC framework. Utilizing DRL, the algorithm aims to minimize average task response time and total system energy consumption, validated through simulations that demonstrate its scalability and efficiency. In their work ~\cite{zhu2021energy}, Zhu et al. introduce the Energy-Efficient Deep Reinforced Traffic Grooming (EDTG) algorithm to mitigate uneven traffic distribution in elastic optical networks (EONs) for cloud-FC. The optimization goal is to minimize energy consumption by reducing the utilization of network resources. The method employs a DRL-based framework, converting network states into images for feature extraction and optimizing grooming strategies through a reward-punishment mechanism.

Seid et al. in ~\cite{seid2021multi} propose a multi-agent DRL (MADRL) approach to enhance task offloading and resource allocation in multi-UAV-enabled IoT edge networks. This method addresses the limitations of computational capacity and energy availability in edge user equipment. The optimization goal is to minimize overall network computation costs while ensuring QoS, utilizing the Multi-Agent DDPG (MADDPG) algorithm for cooperative optimization. In the study presented in ~\cite{santos2021reinforcement}, Santos et al. demonstrate that RL can optimize SFC allocation in fog computing environments. The authors leverage RL to address the dynamic nature of modern networks, focusing on minimizing system costs and enhancing energy efficiency. The solution employs a MILP formulation and Q-learning for dynamic resource allocation, ensuring low latency and high service acceptance rates.

As outlined in ~\cite{santos2021resource}, Santos et al. address the inadequacy of traditional cloud systems in managing the scalability and heterogeneity of IoT devices. They propose a novel DRL approach, enhanced with double Q-learning and Prioritized Experience Replay (PER), to optimize SFC allocation. The method achieves a 95\% request acceptance rate and demonstrates superior adaptability compared to traditional MILP formulations. In ~\cite{mekala2022drl}, Mekala et al. propose a DRL-based service offloading (DSO) approach utilizing a directed acyclic graph (DAG) to optimize resource sharing and improve service reliability in edge computing. The study addresses inefficiencies in Social IoT (SIoT) applications by minimizing edge server costs through a DRL-influenced resource and execution time analysis model, effectively reducing energy consumption.

Wang et al. ~\cite{wang2023cloud}, discussed in Section ~\ref{subsec:taskschedul} for Task Scheduling, also
propose a cloud-native resource allocation strategy to enhance the efficiency of the IoV. They introduce the CVEF, which optimizes resource allocation energy consumption, considering various network characteristics. In their work ~\cite{lingayya2024dynamic}, Lingayya et al. propose a dynamic task offloading framework for resource allocation and privacy preservation in KubeEdge-based edge computing. The framework employs ML, blockchain, and RL to enhance efficiency and security. The solution integrates a privacy-preserving blockchain mechanism and a Multi-Agent Collaborative-RL approach, demonstrating improved offloading rates and reduced resource consumption while balancing latency and throughput for IoT applications.

\subsubsection{Heuristic-based:}
In their work ~\cite{xu2020energy}, Xu et al. propose a joint optimization framework to minimize energy consumption in NOMA-based HetNets by optimizing task offloading, local CPU frequency, power control, and resource allocation while meeting QoS requirements. The problem is decoupled into sub-problems, solved iteratively using SCA and CCP techniques alongside heuristic algorithms like swap matching, demonstrating superior energy efficiency in NOMA-based MEC networks. In the study presented in ~\cite{fan2021cloud}, Fan et al. propose optimizing cloud-EC resource allocation and pricing for mobile blockchain environments using an iterative greedy-and-search-based algorithm (IGS). The system model addresses high energy consumption in IoT devices by offloading tasks to cloud/edge servers, formulated as a Stackelberg game to balance revenue and minimize delays and energy consumption, employing IGS for efficient resource allocation and pricing.

According to ~\cite{yadav2020energy}, Yadav et al. propose Vehicular fog computing (VFC) to alleviate energy consumption and service latency challenges in dynamic computation offloading. The system model introduces ECOS, an energy-efficient scheme that minimizes energy and latency costs by leveraging underutilized vehicular resources, employing a heuristic-based algorithm to prioritize high CPU, low memory tasks for offloading, demonstrating superior performance in simulations. In the study presented in ~\cite{mahmud2021pi}, Mahmud et al. propose Con-Pi, a distributed container-based framework for resource management in edge and fog computing, particularly for latency-sensitive IoT applications on resource-constrained SBCs. The framework enhances resource sharing, energy efficiency, and system longevity through containerization, dynamic task offloading, and renewable energy integration, validated via real-world experiments demonstrating improved response time and energy usage. Building on their DFSP algorithm, Kumar et al.~\cite{kumar2024auction} further optimize microservices allocation in fog computing environments. Their iterative combinatorial auction approach, implemented through DFSP-GF and DFSP-NF variants, ensures efficient resource allocation while minimizing energy consumption and data transfer costs, adhering to resource constraints and scalability requirements.

The four articles~\cite{sun2020energy,qu2021service,lingayya2024dynamic,yuan2020profit} primarily rely on their main algorithms but also incorporate heuristic methods, such as the ETCORA algorithm in the first article, relaxation and randomized rounding techniques in the second article, the Salp Swarm Algorithm and Hybrid Greedy Randomized Adaptive Stackelberg-Auction Game Approach in the third article, and a simulated-annealing-based migrating birds optimization procedure in the fourth article, to optimize energy consumption, resource allocation, task offloading, and privacy preservation in IoT-fog-cloud, UAV-enabled MEC, and edge computing environments.

\subsubsection{Metaheuristic-based:}
As outlined in ~\cite{yuan2020profit}, Yuan et al. propose a profit-maximized collaborative computation offloading and resource allocation algorithm to enhance the efficiency of distributed cloud and edge computing systems. The study addresses the challenge of balancing energy consumption and resource allocation between CDCs and edge computing layers, utilizing the Simulated-Annealing-based Migrating Birds Optimization (SMBO) algorithm to optimize resources and load balance. In ~\cite{li2021computation}, Li et al. emphasize the importance of effective computation offloading and resource allocation for improving service performance in MEC environments. The authors introduce a dynamic computation offloading model using a GA to optimize processing delay and energy consumption, alongside a resource allocation model based on utility maximization through a double auction process.

In their work ~\cite{abbasi2021intelligent}, Abbasi et al. propose intelligent workload allocation in IoT-fog-cloud architectures to enhance MEC by reducing delay and power consumption. The study addresses challenges arising from IoT expansion and 5G, introducing a modified energy model that incorporates green energy to optimize workload distribution using the NSGA II multi-objective GA, ensuring quality and security. In the study presented in ~\cite{cao2021large}, Cao et al. optimize the deployment of edge servers in the IoV by considering six key objectives: transmission delay, workload balancing, energy consumption, deployment costs, network reliability, and the number of edge servers. The authors construct a six-objective optimization model and employ the many-objective evolutionary algorithm PCMaLIA to achieve optimal edge server placement, enhancing system efficiency.

In their work ~\cite{apat2024hybrid}, Apat et al. introduce a hybrid meta-heuristic algorithm that combines GA and Simulated Annealing (SA) to address multi-objective IoT service placement in fog computing environments. The proposed hybrid algorithm, FSPGSA, optimizes makespan, cost, and energy consumption under resource constraints, demonstrating superior performance in minimizing multi-objective goals compared to other algorithms. As outlined in ~\cite{liu2024multi}, Liu et al. propose a multi-objective resource allocation method (MRAM) utilizing Pareto Archived Evolution Strategy (PAES) to optimize completion time, load balance, and energy consumption in MEC for IoT applications. The method addresses challenges in minimizing task completion time and enhancing energy efficiency, employing PAES to identify Pareto-optimal strategies and Multiple Criteria Decision Making (MCDM) for optimal resource management. According to ~\cite{ghasemzadeh2025optimizing}, Ghasemzadeh et al. emphasize the importance of optimizing edge server placement and allocation in MEC environments to enhance energy efficiency and reduce latency. The study proposes a multi-objective approach that integrates decision variable classification and elitism, employing a modified NSGA-II algorithm with a novel mutation strategy, achieving significant improvements in latency and energy savings compared to existing methods.

This article~\cite{lingayya2024dynamic} primarily relies on its main algorithm but also incorporates metaheuristic methods, such as the Salp Swarm Algorithm and a Hybrid Greedy Randomized Adaptive Stackelberg-Auction Game Approach, to enhance resource allocation, task offloading, and privacy preservation in edge computing environments.

\begin{center}

\tiny
\begin{longtable}{
    >{\RaggedRight}m{0.5cm} 
    >{\RaggedRight}m{0.5cm} 
    >{\RaggedRight}m{0.5cm} 
    >{\RaggedRight}m{0.5cm} 
    >{\RaggedRight}m{4cm}
    >{\RaggedRight}m{4cm} 
    >{\RaggedRight}m{4cm}
}
\caption{Advantages \& disadvantages for allocation reviews}
\label{table:add04}\\ 
\hline
\multirow{2}{*}{Article} & 
\multicolumn{3}{c}{\textbf{Energy}} & 
\multirow{2}{*}{Main Idea} & 
\multirow{2}{*}{Advantages} & 
\multirow{2}{*}{Disadvantages} \\ 
 & CE & CoE & SE & & & \\ \hline
\endhead 
\hline
\multicolumn{7}{r}{{Continued on next page}} \\ \hline
\endfoot 
\hline
\multicolumn{7}{l}{
\footnotesize
\textbf{Abbreviations:} 
CE = Computation Energy, 
CoE = Communication Energy, 
SE = Storage Energy.
} \\
\endlastfoot 

~\cite{zhao2021energy} & \circlefill{100} & \circlefill{75} & \circlefill{0} & Minimizes energy in MEC by optimizing task offloading, transmission power, and resource allocation & Reduces energy, supports partial offloading, and outperforms reference schemes & No dynamic environment consideration\\ \hline
~\cite{chang2020dynamic} & \circlefill{100} & \circlefill{75} & \circlefill{0} & Proposes dynamic resource allocation for IoT fog with EH, minimizing execution cost using Lyapunov optimization & Jointly optimizes offloading, power, and subcarrier assignment, incorporates energy harvesting, and handles stochastic environments & Limited scalability, and narrow focus on IoT scenarios\\ \hline
~\cite{zhang2020dynamic} & \circlefill{100} & \circlefill{75} & \circlefill{0} & Introduces dynamic offloading and resource allocation for MEC in dense C-RAN to optimize energy and delay & Provides semidistributed algorithm for energy-delay tradeoff, validated through simulations & Assumes unlimited fronthaul capacity, ignores limited fronthaul impact\\ \hline
~\cite{li2020energy} & \circlefill{100} & \circlefill{75} & \circlefill{0} & Optimizes UAV trajectory, user power, and computation load in UAV-assisted MEC for energy efficiency & Uses SCA, Dinkelbach, and ADMM for distributed solution, addresses user mobility uncertainty & Assumes limited user mobility knowledge, may miss real-time environmental changes\\ \hline
~\cite{alharbi2022energy} & \circlefill{100} & \circlefill{75} & \circlefill{0} & Proposes UAV-based service offloading in cloud-fog, optimizing resource allocation and trajectory using MILP & Provides multi-objective optimization for energy efficiency, achieves power savings & No real-time implementation, simplified UAV power modeling, and no multi-UAV scenarios\\ \hline
~\cite{mohajer2022heterogeneous} & \circlefill{75} & \circlefill{100} & \circlefill{0} & Introduces dynamic optimization for NOMA-based MEC, maximizing energy efficiency and user fairness & Improves energy efficiency and throughput, ensures fairness, and supports flexible backhauling & Lacks dynamic energy efficiency, no mechanisms for link capacity fluctuations\\ \hline
~\cite{zhou2022service} & \circlefill{100} & \circlefill{75} & \circlefill{0} & Proposes service-oriented resource allocation for blockchain MEC, optimizing deployment and power under energy constraints & Introduces blockchain-enabled image sharing, uses two-layer optimization for resource management & Ignores complex services and factors like service reliability\\ \hline
~\cite{bebortta2023optimal} & \circlefill{100} & \circlefill{75} & \circlefill{0} & Introduces ILP-based fog-cloud offloading to enhance energy efficiency and reduce latency in IoT & Balances computation and communication overhead, reduces latency, and adapts to real-time task requirements & Limited mobility consideration, and scalability issues\\ \hline
~\cite{vahabi2023energy} & \circlefill{100} & \circlefill{0} & \circlefill{0} & Proposes energy-efficient resource management for FaaS edge, minimizing energy and SLA violations & Reduces SLA violations and energy consumption, powers down idle nodes for sustainability & Longer execution time, limited to periodic invocations, and static edge resource assumption\\ \hline
~\cite{fe2023model} & \circlefill{100} & \circlefill{0} & \circlefill{25} & Presents model-driven approach for dependability and power trade-offs in Kubernetes cloud-fog systems & Provides hierarchical model for dependability and energy analysis, improves system availability & Limited autoscaling analysis, no focus on dynamic resource variability, and practical implementation gaps\\ \hline
~\cite{sun2020energy} & \circlefill{100} & \circlefill{75} & \circlefill{0} & Introduces ETCORA for IoT-fog-cloud task offloading, minimizing energy and completion time & Reduces energy and completion time, leverages fog-cloud synergy, and outperforms baselines & No real-world implementation, lacks resource availability and fog node mobility consideration\\ \hline
~\cite{qu2021service} & \circlefill{100} & \circlefill{75} & \circlefill{0} & Optimizes service provisioning in UAV-enabled MEC, minimizing energy for terrestrial UEs & Provides efficient algorithms with proven convergence, reduces UE energy consumption & No theoretical performance guarantees beyond convergence\\ \hline
~\cite{bellal2024gas} & \circlefill{100} & \circlefill{0} & \circlefill{0} & Introduces GAS for edge microservices, dynamically adjusting CPU frequency to minimize energy & Reduces energy, ensures latency guarantees, and adjusts CPU frequency dynamically & Heterogeneity limitation, frequency transition latency, and incomplete evaluation\\ \hline
~\cite{tong2020adaptive} & \circlefill{100} & \circlefill{50} & \circlefill{0} & Proposes DRL for adaptive task offloading in MEC, minimizing response time and energy & Effective multi-objective optimization, realistic mobility modeling, and entropy-based weighting & No cloud-MEC integration, and narrow scope selection\\ \hline
~\cite{zhu2021energy} & \circlefill{100} & \circlefill{25} & \circlefill{0} & Introduces EDTG for elastic optical networks, optimizing resource management and energy & Automates feature extraction, reduces energy, and uses images for network state representation & Redundant network state information, lacks simplification, and no GNN exploration\\ \hline
~\cite{seid2021multi} & \circlefill{100} & \circlefill{75} & \circlefill{0} & Proposes MADRL for multi-UAV IoT edge, minimizing computation costs while ensuring QoS & Reduces computation costs, improves resource use and QoS, and scales well with IoT devices & No advanced integration, limited focus on network congestion, and single application context\\ \hline
~\cite{santos2021reinforcement} & \circlefill{100} & \circlefill{25} & \circlefill{0} & Uses RL for SFC allocation in fog computing, minimizing system costs and improving energy efficiency & RL provides scalable SFC allocation, comparable to ILP, and adapts to dynamic conditions & Struggles with fluctuating demands, higher costs, and complexity in dynamic scenarios\\ \hline
~\cite{santos2021resource} & \circlefill{100} & \circlefill{25} & \circlefill{0} & Proposes DRL for SFCA in fog computing, focusing on energy efficiency and dynamic provisioning & DRL scales better than MILP, adapts to dynamic changes, and focuses on energy efficiency & Relies on MILP, trade-off between cost efficiency and request acceptance\\ \hline
~\cite{mekala2022drl} & \circlefill{100} & \circlefill{25} & \circlefill{0} & Introduces DSO using DAG for edge computing, optimizing resource sharing and reducing costs & Reduces execution costs, optimizes resource use, and minimizes energy consumption & No dynamic environment handling, and scalability concerns\\ \hline
~\cite{wang2023cloud} & \circlefill{50} & \circlefill{50} & \circlefill{0} & Proposes multi-agent RL for IoV task scheduling, minimizing latency and energy & Integrates task scheduling, resource allocation, and microservice deployment, reduces latency and energy & Limited scalability\\ \hline
~\cite{lingayya2024dynamic} & \circlefill{75} & \circlefill{50} & \circlefill{25} & Introduces dynamic offloading framework for KubeEdge, optimizing efficiency and security with ML and blockchain & Enhances resource efficiency, reduces latency, and improves privacy with blockchain & Scalability concerns, computational overhead, and static evaluation metrics\\ \hline
~\cite{xu2020energy} & \circlefill{100} & \circlefill{75} & \circlefill{0} & Minimizes energy in NOMA-based MEC by optimizing offloading, CPU frequency, and resource allocation & Achieves near-optimal energy savings, outperforms benchmark approaches & No collaborative offloading, ignores MEC server caching\\ \hline
~\cite{fan2021cloud} & \circlefill{100} & \circlefill{25} & \circlefill{0} & Proposes IGS for cloud/edge resource allocation in mobile blockchain, optimizing revenue for CESP and IoT users & Improves revenue, reduces communication delays, and uses Stackelberg game model & No edge-to-edge or edge-to-cloud offloading, limited applicability, and narrow focus on mobile blockchain\\ \hline
~\cite{yadav2020energy} & \circlefill{100} & \circlefill{75} & \circlefill{0} & Introduces ECOS for VFC, minimizing energy and latency using under-utilized vehicle resources & Reduces energy and latency, considers vehicle mobility, and uses low-complexity heuristic & No interoperability, and limited scalability\\ \hline
~\cite{mahmud2021pi} & \circlefill{100} & \circlefill{25} & \circlefill{0} & Proposes Con-Pi for edge-fog resource management, focusing on energy and microservice execution & Improves response time, supports renewable energy, and enables dynamic scaling & No security features, limited policy support, and reliability assumption\\ \hline
~\cite{kumar2024auction} & \circlefill{100} & \circlefill{25} & \circlefill{0} & Introduces DFSP for fog microservice allocation, minimizing energy and data transfer costs & Decentralized and privacy-preserving, fast convergence, and scalable for fog environments & Higher communication overhead, lacks dynamic adaptation, and skewed allocation distribution\\ \hline
~\cite{yuan2020profit} & \circlefill{100} & \circlefill{0} & \circlefill{0} & Proposes profit-maximized offloading for cloud-edge systems, ensuring task response limits and maximizing profit & Jointly optimizes offloading and resource allocation, increases profit and load balancing & High computational complexity and potential scalability limitations\\ \hline
~\cite{li2021computation} & \circlefill{100} & \circlefill{25} & \circlefill{0} & Introduces genetic algorithm for MEC offloading, optimizing delay and energy & Reduces energy and processing delay, maximizes system utility, and adapts dynamically & Limited wireless resource allocation, incomplete bidding strategy, and lack of diverse scenario testing\\ \hline
~\cite{abbasi2021intelligent} & \circlefill{100} & \circlefill{75} & \circlefill{0} & Proposes NSGA II for IoT-fog-cloud workload allocation, reducing delay and power consumption & Reduces power and delay, uses green energy, and balances objectives with NSGA II & Narrow focus, no dynamic environment testing\\ \hline
~\cite{cao2021large} & \circlefill{100} & \circlefill{0} & \circlefill{0} & Optimizes ES deployment in IoV using many-objective evolutionary algorithm, considering delay, energy, and cost & Addresses six objectives for ES deployment, improves workload balancing, and uses PCMaLIA for optimization & No security considerations, and limited scalability discussion\\ \hline
~\cite{apat2024hybrid} & \circlefill{100} & \circlefill{25} & \circlefill{0} & Introduces GA-SA for IoT service placement in fog, optimizing makespan, cost, and energy & GA-SA outperforms in resource allocation, energy reduction, and system performance, combines GA and SA & Scalability issues, and no queuing models\\ \hline
~\cite{liu2024multi} & \circlefill{100} & \circlefill{25} & \circlefill{0} & Proposes PAES for MEC resource allocation, optimizing completion time, load balance, and energy & Reduces task completion time, improves load balance, and minimizes energy consumption & Privacy concerns, no real-world validation, and limited scalability discussion\\ \hline
~\cite{ghasemzadeh2025optimizing} & \circlefill{100} & \circlefill{75} & \circlefill{0} & Optimizes edge server placement in MEC for energy efficiency and latency reduction using multi-objective approach & Reduces energy and latency, outperforms NSGA-II, NSGA-III, and MOPSO, improves convergence and diversity & Static deployment, no real-time adaptability, and limited scope\\ \hline
 
\end{longtable}
\end{center}

\subsection{Replica Selection}
Replica selection is a crucial aspect of MEC and fog computing, particularly in scenarios where data and services are replicated across multiple nodes to enhance availability, reliability, and performance. The process involves selecting the most appropriate replica to serve user requests, taking into account factors such as latency, network conditions, and resource availability. By optimizing replica selection, systems can reduce response times, minimize bandwidth usage, and improve overall user experience. The dynamic and distributed nature of edge and fog environments introduces significant challenges for replica selection, including node mobility, fluctuating network conditions, and varying resource capacities. Additionally, the need to support real-time and latency-sensitive applications further complicates the selection process. In this section, we examine various research contributions on replica selection mechanisms, categorizing them into Mathematical-based, ML-based, Heuristic-based, and Metaheuristic-based approaches, and evaluating their suitability for MEC and fog computing environments, as shown in Table~\ref{table:tb05}. Table~\ref{table:add02} outlines the advantages and disadvantages associated with each approach.

\subsubsection{Mathematical-based:}
According to ~\cite{satar2022best}, Satar et al. emphasize the significance of optimal replica selection in Data Grids to improve performance metrics, including access time, data transmission speed, and congestion management. Their system model employs the A* priori algorithm to analyze real-time indicators such as CPU load, memory size, and bandwidth, ensuring efficient replica selection from geographically distributed servers. While energy consumption is not addressed, the study offers a comprehensive framework for enhancing distributed system performance through advanced replica selection techniques. In their study ~\cite{sun2024minimizing}, Sun et al. propose a method to minimize service latency in MEC through image-based microservice caching and a randomized request routing algorithm. The problem is formulated as an ILP model with multi-condition constraints, addressing storage and computing resource limitations at edge servers. Their solution leverages a randomization-based approximation algorithm, utilizing Chernoff’s theorem to achieve near-optimal solutions in polynomial time, thereby reducing average request latency.

This article~\cite{wang2022service} primarily relies on its main algorithm but also incorporates mathematical methods, such as integer non-linear optimization and matching game theory, to solve the NP-hard service routing problem and optimize service delay and resource cost in multi-tier edge computing environments.

\subsubsection{ML-based:}
In ~\cite{hu2024joint}, Hu et al. propose a joint optimization framework for microservices deployment and routing in edge computing to enhance performance and resource efficiency. The system model addresses CPU and storage constraints through multi-edge collaboration and queuing network analysis. Optimization goals include minimizing user delay, reducing resource consumption, and improving energy efficiency. The solution employs heuristic algorithms and a RL approach (RSPPO) with reward shaping for dynamic adjustments. According to ~\cite{ahmed2024neil}, Ahmed et al. emphasize the importance of intelligent replica selection for optimizing network performance in edge-to-cloud infrastructures. The study addresses the limitations of traditional methods like anycast and DNS redirections in dynamic network conditions. Optimization goals focus on improving latency and throughput using NeIL, a client-side framework combining Multi-Armed Bandit (MAB) algorithms with expert-based prediction models, enabling decentralized, performance-aware replica selection.

In ~\cite{liu2021virtual}, Liu et al. investigate VR streaming challenges in blockchain-enabled F-RANs, focusing on joint resource allocation and replica selection to achieve ultralow latency and energy efficiency. Optimization goals include minimizing energy consumption for VR streaming and blockchain maintenance using DVFS. The proposed DDPG-based scheme leverages edge caching and computing resources on F-APs, enhancing performance and sustainability in decentralized blockchain operations. As outlined in ~\cite{guo2023deep}, Guo et al. propose optimal microservice instance selection in MEC to reduce service access delays. The study addresses the dynamic and heterogeneous nature of cloud-edge environments, formulating the problem as a Markov decision-making process. The authors introduce MSDDPG, a DDPG algorithm, leveraging DRL and an experience pool for adaptability, highlighting its role in improving user experience and network performance.

\begin{table}
\centering
\caption{Optimization criteria and methods in replica selection}
\label{table:tb05}
\tiny
\begin{tabular}{lllccccclccccll}
\multicolumn{1}{c}{\multirow{2}{*}{Row}} & \multicolumn{1}{c}{\multirow{2}{*}{CatName}} &
\multicolumn{1}{c}{\multirow{2}{*}{Year}} & \multicolumn{5}{c}{Objectives} &  & \multicolumn{4}{c}{Algorithm} & \multirow{2}{*}{Envrionment} &  \\ \cline{4-8}\cline{10-13}
\multicolumn{1}{c}{} & \multicolumn{1}{c}{} & 
\multicolumn{1}{c}{} &
LT & 
CT & 
LB & 
QoS & 
TP &  & 
MB & 
ML & 
HB & 
MH &  &   \\ \hline\hline
1 & Satar et al.~\cite{satar2022best} & 2022 & - & - & \checkmark & - & - &  & \checkmark & - & - & - & Hybrid &  \\ \hline
2 & Sun et al.~\cite{sun2024minimizing} & 2024 & \checkmark & - & - & - & - &  & \checkmark & - & - & - & Edge &  \\ \hline
3 & Hu et al.~\cite{hu2024joint} & 2024 & \checkmark & - & \checkmark & - & - &  & - & \checkmark & \checkmark & - & Edge &  \\ \hline
4 & Ahmed et al.~\cite{ahmed2024neil} & 2024 & \checkmark & - & - & - & \checkmark &  & - & \checkmark & \checkmark & - & Hybrid &  \\ \hline
5 & Liu et al.~\cite{liu2021virtual} & 2021 & - & - & - & - & - &  & - & \checkmark & - & - & Fog &  \\ \hline
6 & Guo et al.~\cite{guo2023deep} & 2023 & \checkmark & - & - & - & - &  & - & \checkmark & - & - & Hybrid &  \\ \hline
7 & Xie et al.~\cite{xie2020bandwidth} & 2020 & \checkmark & - & \checkmark & - & - &  & - & - & \checkmark & - & Hybrid &  \\ \hline
8 & Fahs et al.~\cite{fahs2020voila} & 2020 & - & - & \checkmark & \checkmark & - &  & - & - & \checkmark & - & Fog &  \\ \hline
9 & Dias et al.~\cite{dias2021adaptive} & 2021 & - & - & - & - & - &  & - & - & \checkmark & - & Hybrid &  \\ \hline
10 & Zhao et al.~\cite{zhao2021temperature} & 2021 & \checkmark & \checkmark & \checkmark & - & - &  & - & - & \checkmark & - & Hybrid &  \\ \hline
11 & Sethunath et al.~\cite{sethunath2022joint} & 2022 & \checkmark & \checkmark & - & - & - &  & - & - & \checkmark & - & Hybrid &  \\ \hline
12 & Zhao et al.~\cite{zhao2024winning} & 2024 & - & - & - & \checkmark & - &  & - & - & \checkmark & - & Edge &  \\ \hline
13 & Jaradat et al.~\cite{jaradat2020multiple} & 2020 & - & \checkmark & - & \checkmark & - &  & - & - & \checkmark & \checkmark & Hybrid &  \\ \hline
14 & Zou et al.~\cite{zou2021towards} & 2021 & \checkmark & - & - & - & - &  & - & - & - & \checkmark & Edge &  \\ \hline
15 & Wang et al.~\cite{wang2022service} & 2022 & \checkmark & \checkmark & - & - & - &  & \checkmark & - & - & \checkmark & Edge &  \\ \hline
\end{tabular}
    \captionsetup{font=footnotesize}  
     \caption*{\footnotesize 
    \textbf{Abbreviations Guide:} 
    LT: Latency, 
    CT: Cost, 
    LB: Load Balancing,  
    TP: Throughput, 
    MB: Mathematical-Based, 
    ML: ML-Based, 
    HB: Heuristic-Based, 
    MH: Metaheuristic-Based.
    }
\end{table}

\subsubsection{Heuristic-based:}
In the study presented in ~\cite{xie2020bandwidth}, Xie et al. tackle the complexities of managing data-intensive workflows in multi-cloud environments by introducing the Bandwidth and Latency-based Replica Selection Mechanism (BLRS). The primary optimization goal is to maximize concurrent workflow instances while balancing network bandwidth usage and enhancing QoS. The heuristic-based solution outperforms traditional strategies in network performance and data access efficiency, although energy efficiency remains unaddressed, suggesting a potential avenue for future research. Fahs et al. propose Voilà, a tail-latency-aware auto-scaler for fog computing environments, in ~\cite{fahs2020voila}. This system addresses the challenge of maintaining low tail latency and preventing replica overload under non-stationary workloads. Integrated into Kubernetes, Voilà dynamically adjusts replica count and placement to minimize latency and ensure QoS. The heuristic-based algorithms employed enhance performance and resource efficiency for latency-sensitive fog applications.

According to ~\cite{dias2021adaptive}, Dias et al. focus on adaptive replica selection in MEC environments, emphasizing energy efficiency, latency reduction, and load balancing. The proposed solution integrates MECERRA, a heuristic-based replica ranking algorithm, into the WASABI framework to dynamically select replicas while minimizing overhead. A redirection mechanism prioritizes replicas with higher battery capacity, optimizing energy usage and prolonging device availability. Zhao et al. introduce a temperature matrix-based data placement method (TEMPLIH) in edge computing environments in ~\cite{zhao2021temperature}. This framework leverages spatiotemporal data characteristics to optimize placement, incorporating a replica selection algorithm (RSA-TM) to meet latency requirements. An improved Hungarian algorithm (IHA-RM) balances latency, cost, and load distribution, highlighting the potential of advanced algorithms in addressing edge computing challenges and indirectly enhancing energy efficiency through optimized resource utilization.

In their work ~\cite{sethunath2022joint}, Sethunath et al. propose a joint function warm-up and request routing scheme to optimize serverless computing performance in edge and cloud environments. The optimization goals include maximizing the hit ratio of serverless requests, reducing cold-start latency, and adhering to memory and budget constraints. Heuristic strategies—greedy, conservative, and balanced approaches—are employed to optimize request routing and function warm-up, demonstrating superior performance in simulations compared to baseline methods, although energy consumption is not explicitly addressed. In ~\cite{zhao2024winning}, Zhao et al. explore improving edge Data Integrity (EDI) verification efficiency in MEC environments, where indiscriminate inspection of data replicas incurs significant overhead. The authors formalize the Unreliable Data Replica Selection (URS) problem as an NP-hard constrained optimization problem, aiming to enhance verification efficiency by prioritizing unreliable replicas based on cache service QoS and verification success rates. They propose URS-P, a priority-based heuristic algorithm that reduces unnecessary computation and communication overhead, demonstrating effectiveness in scenarios with numerous data replicas and indirectly lowering resource usage.

The four articles~\cite{hu2024joint,ahmed2024neil,jaradat2020multiple,choupani2025joint} primarily rely on their main algorithms but also incorporate heuristic methods, such as a heuristic-based horizontal scaling algorithm in the first article, Multi-Armed Bandit (MAB) algorithms in the second article, a hybrid GA with a user-preference algorithm in the third article, and a warmest policy for request scheduling in the fourth article. 

\subsubsection{Metaheuristic-based:}
 
In ~\cite{jaradat2020multiple}, Jaradat et al. propose a hybrid approach combining GAs and user-preference algorithms to enhance fair replica selection in data grids. The study addresses limited data resources and inequitable user satisfaction in traditional methods, aiming to minimize total user satisfaction (TUPQ) and achieve fair user satisfaction (FUS) by considering QoS attributes such as time, availability, security, cost, and user preferences. The integrated GA and user-preference algorithm simultaneously process all user requests, demonstrating superior performance in replica selection and resource allocation, thereby improving user satisfaction and fairness in data grid management. In their work ~\cite{zou2021towards}, Zou et al. focus on optimizing service instance selection in MEC to minimize response time and enhance user experience. The service instance selection problem (SISP) is modeled by considering user mobility, edge server resource constraints, and user interference, formulated as an optimization problem. The authors introduce GASISMEC, a GA incorporating response time-aware mutation and normalization, to efficiently distribute service requests across edge servers. Evaluated on real-world datasets, GASISMEC outperforms six baseline methods in reducing latency and improving service performance, leveraging metaheuristic techniques for scalability and adaptability. 

In ~\cite{wang2022service}, Wang et al. propose a novel approach to service routing in multi-tier edge computing environments, focusing on optimizing microservice-based service provision. The study addresses operational complexities in service mesh infrastructure, such as multi-tier edge servers, shared microservices, and diverse QoS requirements, formulating the problem as an NP-hard integer non-linear optimization. The authors introduce the Dependency-aware Deferred Acceptance algorithm with Dynamic Quota (DDA-DQ), leveraging matching game theory to minimize service delay and resource cost, providing an efficient solution for complex edge computing environments.

\begin{center}

\tiny
\begin{longtable}{
    >{\RaggedRight}m{0.5cm} 
    >{\RaggedRight}m{0.5cm} 
    >{\RaggedRight}m{0.5cm} 
    >{\RaggedRight}m{0.5cm} 
    >{\RaggedRight}m{4cm}
    >{\RaggedRight}m{4cm} 
    >{\RaggedRight}m{4cm}
}
\caption{Advantages \& disadvantages for replica selection reviews}
\label{table:add05}\\ 
\hline
\multirow{2}{*}{Article} & 
\multicolumn{3}{c}{\textbf{Energy}} & 
\multirow{2}{*}{Main Idea} & 
\multirow{2}{*}{Advantages} & 
\multirow{2}{*}{Disadvantages} \\ 
 & CE & CoE & SE & & & \\ \hline
\endhead 
\hline
\multicolumn{7}{r}{{Continued on next page}} \\ \hline
\endfoot 
\hline
\multicolumn{7}{l}{
\footnotesize
\textbf{Abbreviations:} 
CE = Computation Energy, 
CoE = Communication Energy, 
SE = Storage Energy.
} \\
\endlastfoot

~\cite{satar2022best} & \circlefill{0} & \circlefill{25} & \circlefill{0} & Proposes replica selection in Data Grids using A priori algorithm to optimize transmission time & Selects replicas based on bandwidth, CPU, and memory, reduces transmission time, and scales well & No energy management, limited real-world evaluation, and narrow focus on Data Grids\\ \hline
~\cite{sun2024minimizing} & \circlefill{0} & \circlefill{25} & \circlefill{0} & Aims to minimize latency in MEC through image-based microservice caching and randomized routing & Reduces latency, optimizes storage with container image sharing, and improves edge server performance & No energy efficiency focus\\ \hline
~\cite{hu2024joint} & \circlefill{100} & \circlefill{25} & \circlefill{0} & Proposes joint optimization for microservice deployment in edge computing using RSPPO to minimize delay and resource use & Combines microservice deployment and routing, adapts dynamically, and addresses latency, energy, and load balancing & Limited energy efficiency focus, scalability issues, and no user mobility consideration\\ \hline
~\cite{ahmed2024neil} & \circlefill{25} & \circlefill{0} & \circlefill{0} & Introduces NeIL for replica selection in edge-to-cloud using MAB algorithms to enhance network performance & Adapts to network dynamics, operates decentralized, and reduces tail latency & Narrow scope, no energy management, and limited application to network performance\\ \hline
~\cite{liu2021virtual} & \circlefill{100} & \circlefill{75} & \circlefill{25} & Proposes DDPG for VR streaming in F-RANs, optimizing resource allocation and replica selection for energy efficiency & Balances energy and resource allocation, leverages edge caching, and uses double-layer PBFT for blockchain & No real-world implementation, no privacy protection, and limited scalability exploration\\ \hline
~\cite{guo2023deep} & \circlefill{0} & \circlefill{25} & \circlefill{0} & Introduces MS-DDPG for microservice selection in MEC to minimize service access delay & MS-DDPG reduces service access delay and scales well in dynamic environments & No energy consumption consideration, ignores task dependencies, and lacks workload balancing\\ \hline
~\cite{xie2020bandwidth} & \circlefill{0} & \circlefill{25} & \circlefill{0} & Proposes BLRS for replica selection in multi-cloud, optimizing data access and avoiding network overload & Balances bandwidth usage, increases concurrent instances, and offers decentralized framework & No energy efficiency focus, limited to multi-cloud, and no real-world validation\\ \hline
~\cite{fahs2020voila} & \circlefill{25} & \circlefill{25} & \circlefill{25} & Introduces Voilà for fog autoscaling, ensuring low latency and optimizing resource utilization & Minimizes tail latency, avoids replica overload, and integrates with Kubernetes & No predictive traffic modeling, high resource cost for strict latency\\ \hline
~\cite{dias2021adaptive} & \circlefill{100} & \circlefill{50} & \circlefill{0} & Proposes MECERRA and WASABI for adaptive replica selection in MEC, focusing on energy and load balancing & Improves replica selection quality, optimizes energy, and introduces low-overhead framework & Simulation-based evaluation, limited comparison, and partial system view\\ \hline
~\cite{zhao2021temperature} & \circlefill{0} & \circlefill{25} & \circlefill{0} & Introduces TEMPLIH for data placement in edge computing, optimizing latency, cost, and load balancing & Reduces data placement costs, optimizes latency, and maintains load balance & No energy efficiency, narrow scope, and no task scheduling or provisioning\\ \hline
~\cite{sethunath2022joint} & \circlefill{25} & \circlefill{0} & \circlefill{0} & Proposes joint function warm-up and routing for serverless computing, minimizing latency under constraints & Addresses cold-start latency, allocates resources dynamically, and outperforms greedy approaches & Ignores energy efficiency, and no real-world implementation\\ \hline
~\cite{zhao2024winning} & \circlefill{25} & \circlefill{0} & \circlefill{0} & Enhances EDI verification by pre-selecting unreliable replicas using cache QoS and verification success rate & Targets unreliable replicas, reduces overhead, and scales well for large data scenarios & No energy/resource management, no dynamic adaptability, and limited indicators\\ \hline
~\cite{jaradat2020multiple} & \circlefill{0} & \circlefill{25} & \circlefill{0} & Introduces hybrid GA and user-preference algorithm for fair replica selection in data grids & Addresses multiple user requests, outperforms UPA and AHP, and scales for large data grids & No energy efficiency, and slow computational speed\\ \hline
~\cite{zou2021towards} & \circlefill{0} & \circlefill{25} & \circlefill{0} & Aims to optimize service instance selection in MEC, minimizing response time under resource constraints & Handles dynamic workloads and mobility, reduces response time, and considers interference & No energy efficiency, limited real-world validation, and scalability concerns\\ \hline
~\cite{wang2022service} & \circlefill{25} & \circlefill{0} & \circlefill{0} & Proposes DDA-DQ for service routing in multi-tier edge computing, minimizing delay and resource cost & Handles task and colleague dependencies, outperforms baselines, and uses real-world dataset & No energy efficiency, limited scalability focus, and no fault tolerance discussion\\ \hline

\end{longtable}
\end{center}

 \section{Open Issues and Future Directions}
\label{sec:Challenges}

Future research on energy-efficient resource management in fog and edge computing must evolve toward deeply integrated frameworks that unify sustainability, intelligence, and security across all architectural layers. Building upon the taxonomy and limitations identified in this survey, we outline six promising directions for advancing the field.

\subsection{Energy Efficiency with Security Awareness}
While prior works emphasize energy optimization via intelligent orchestration, system-wide security integration remains underexplored. Federated fog computing enables decentralized and privacy-preserving coordination while balancing computation and communication costs. However, lightweight encryption, key management, and trust propagation across heterogeneous devices introduce practical challenges. Future research should integrate multi-objective reinforcement learning with trust-aware scheduling to jointly optimize latency, energy, and security metrics. Empirical validation in smart city surveillance and healthcare monitoring systems can demonstrate the viability of security-aware energy optimization under realistic workloads and potential cyberattacks~\cite{alsadie2024ai}.

\subsection{Renewable and Sustainable Energy-Aware Edge Systems}
Integrating renewable energy sources into fog and edge infrastructures is vital for long-term sustainability, yet volatility in renewable supply endangers service-level reliability. Beyond simulation-based approaches, real deployments require dynamic switching between renewable and grid power, predictive energy harvesting, and workload migration under uncertain conditions. Hybrid power management frameworks that combine short-term forecasting with adaptive orchestration can mitigate these challenges. Field experiments—such as renewable-powered IoT gateways for smart agriculture or environmental sensing—can provide empirical insights into the trade-offs between energy sustainability and service continuity~\cite{mrabet2025secureedge}.

\subsection{Green and Self-Healing Edge Ecosystems}
Sustainability extends beyond energy reduction to encompass resilience, fault recovery, and lifecycle-aware resource management. Distributed orchestration equipped with eco-aware and self-healing policies can autonomously detect inefficiencies, reallocate workloads, and recover from energy or hardware failures. Yet, existing frameworks rarely unify fault-tolerance, anomaly detection, and energy optimization. Future systems should employ self-healing controllers that combine AI-driven anomaly detection with circular resource utilization strategies (e.g., workload recycling or predictive maintenance). Demonstrations in industrial IoT or vehicular edge testbeds can showcase how self-healing mechanisms maintain QoS under fluctuating resource conditions~\cite{vishwakarma2025distributed}.

\subsection{Energy-Aware Hybrid Serverful–Serverless Management}
Hybrid fog and edge systems increasingly combine long-running serverful microservices with event-driven serverless functions to balance persistence and elasticity. This convergence, however, raises challenges in energy efficiency due to idle serverful instances and cold-start overheads in serverless execution. Future research should focus on unified orchestration frameworks that adaptively distribute workloads between persistent and ephemeral services based on energy and QoS metrics. AI-assisted autoscaling and predictive invocation can further reduce idle and startup energy consumption without compromising responsiveness. Evaluations on mixed workloads—such as real-time video analytics or anomaly detection pipelines—can reveal practical trade-offs between energy, latency, and throughput~\cite{aslanpour2022energy}.

\subsection{Quantum-Enhanced Resource Management}
Quantum computing offers a promising paradigm for addressing NP-hard optimization problems in fog and edge orchestration. Hybrid quantum–classical optimization can accelerate convergence and reduce computational energy. Nonetheless, practical deployment faces hardware access constraints, error rates, and integration complexity with classical schedulers. Future research should explore quantum-assisted decision engines co-executing with edge orchestrators in simulated or emulated testbeds before real-world adoption. Quantum-inspired algorithms such as QPSO or QGA may also serve as feasible intermediate solutions~\cite{GOLEC2024190}.

\subsection{Next-Generation AI-Driven Edge Intelligence}
AI-based orchestration is fundamental for scalable, adaptive, and trustworthy edge ecosystems. Approaches such as swarm intelligence, meta-learning, and explainable reinforcement learning enable self-optimization under highly dynamic workloads. Practical deployment requires interpretable models that ensure fairness, explainability, and low computational overhead on constrained devices. Future research should advance federated and explainable edge AI frameworks capable of runtime transparency and privacy preservation. Evaluations in latency-critical domains such as autonomous driving or smart manufacturing can bridge the gap between theoretical intelligence and operational reliability~\cite{ghafari2025swarm}. For example, while AI-driven optimization provides adaptivity, its energy overhead may offset gains in low-power nodes—unlike quantum-inspired solvers that inherently minimize computation cost. Similarly, serverless paradigms improve scalability but introduce cold-start energy spikes. These comparative insights clarify trade-offs that future research must address~\cite{11222695}.

\subsection*{Summary}
Achieving sustainable, secure, and intelligent fog–edge ecosystems requires a holistic integration of energy optimization, renewable adaptation, security, and adaptive orchestration. Bridging the gap between simulation-based studies and real-world deployment—through cross-layer testbeds, field case studies, and open benchmarking platforms—remains a central challenge. Addressing these intertwined issues will pave the way toward autonomous, green, and trustworthy next-generation edge infrastructures.

\section{Conclusion}
\label{sec:Conclusion}
This survey has provided a comprehensive analysis of resource management strategies in fog and edge computing, emphasizing critical areas such as placement, autoscaling, instance selection, resource allocation, and replica selection. The continuous evolution of these computing paradigms highlights the urgent need for efficient resource management frameworks capable of addressing the unique challenges posed by decentralized and dynamic environments. Our review of over 136 recent studies has identified key methodological trends, research gaps, and emerging opportunities, particularly in the context of energy efficiency, QoS, and the integration of MSAs. Future research should focus on developing AI-driven optimization techniques, enhancing the synergy between energy consumption and QoS, and addressing the complexities introduced by heterogeneous edge environments. By advancing these areas, we can pave the way for more resilient, efficient, and scalable computing solutions, ensuring sustainable progress in the IoT-driven digital landscape.

As we look to the future, it is evident that more research is needed to develop innovative solutions that enhance the efficiency, scalability, and reliability of resource management in fog and edge computing. This includes exploring the potential of AI-driven approaches, optimizing energy-QoS trade-offs, and addressing the complexities of heterogeneous environments. By advancing our understanding and capabilities in these areas, we can pave the way for more robust and sustainable computing solutions that meet the demands of emerging applications in the IoT and beyond. In conclusion, this survey serves as a foundational resource for researchers and practitioners alike, providing insights into the current state of resource management in fog and edge computing and guiding future innovations in this rapidly evolving field.


\bibliographystyle{ACM-Reference-Format}
\bibliography{references}

\appendix
\end{document}